\documentclass[11pt,preprint]{aastex}
\usepackage{amsmath}
\usepackage[]{graphicx}
\usepackage{indentfirst}
\usepackage{lscape}
\usepackage{afterpage}
\usepackage{rotating}

\usepackage{caption}

\usepackage{wrapfig}
\usepackage{multicol}
\usepackage{textcomp}
\newcommand{\textapprox}{\raisebox{0.5ex}{\texttildelow}}

\usepackage[dvips,bookmarks,breaklinks,colorlinks=true,urlcolor=black,linkcolor=black,citecolor=black]{hyperref}

\usepackage{multirow}

\usepackage{xcolor}
\usepackage{ulem}

\newcommand{\bPicb}{\beta\mbox{ Pic b}}
\newcommand{\bPicA}{\beta\mbox{ Pic A}}

\newcommand{\meters}{\mbox{m}}

\newcommand{\microns}{\mu\mbox{m}}

\begin{document}

\title{Magellan Adaptive Optics first-light observations of the exoplanet $\beta$ Pic b. I. 
\\Direct imaging in the far-red optical with MagAO+VisAO 
\\and in the near-IR with NICI.
\footnote{This paper includes data gathered with the 6.5 m Magellan Telescopes located at Las Campanas Observatory, Chile.}
\footnote{Based in part on observations obtained at the Gemini Observatory.}
}
\author{Jared R. Males$^{1,\dagger,\ddagger}$, 
Laird M. Close$^1$, 
Katie M. Morzinski$^{1,\dagger}$, 
Zahed Wahhaj$^2$, 
Michael C. Liu$^3$, 
Andrew J. Skemer$^1$,  
Derek Kopon$^{1,4}$, 
Katherine B. Follette$^1$, 
Alfio Puglisi$^5$, 
Simone Esposito$^5$, 
Armando Riccardi$^5$, 
Enrico Pinna$^5$, 
Marco Xompero$^5$, 
Runa Briguglio$^5$,  
Beth A. Biller$^6$, 
Eric L. Nielsen$^3$,
Philip M. Hinz$^1$,
Timothy J. Rodigas$^{1,7}$,  
Thomas L. Hayward$^8$, 
Mark Chun$^3$,
Christ Ftaclas$^3$,
Douglas W. Toomey$^9$
and Ya-Lin Wu$^1$}
\affil{\small
$^1$Steward Observatory, University of Arizona, Tucson, AZ 85721.
  $^2$European Southern Observatory, Alonso de Cordova 3107,  Vitacura, Casilla 19001, Santiago, Chile. 
  $^3$Institute for Astronomy, University of Hawaii, 2680 Woodlawn Drive, Honolulu, HI 96822, USA.  
  $^4$Max Planck Institute for Astronomy, Koenigstuhl 17, 69117 Heidelberg, Germany. 
  $^5$Arcetri Observatory/INAF, Largo E. Fermi 5, 50125-Firenze, Italy. 
  $^6$Institute for Astronomy, The University of Edinburgh, Royal Observatory, Blackford Hill, Edinburgh EH9 3HJ, United Kingdom. 
  $^7$Department of Terrestrial Magnetism, Carnegie Institution of Washington 5241 Broad Branch Road NW, Washington, DC 20015, USA. 
  $^8$Gemini Observatory, Southern Operations Center, c/o AURA, Casilla 603, La Serena, Chile. 
  $^9$Mauna Kea Infrared, LLC, 21 Pookela St., Hilo, HI 96720. \\
$^\dagger$NASA Sagan Fellow\\
$^\ddagger$jrmales@as.arizona.edu}

\begin{abstract}
We present the first ground-based CCD ($\lambda < 1\mu$m) image of an extrasolar planet.  Using MagAO's VisAO camera we detected the extrasolar giant planet (EGP) $\beta$ Pictoris b in $Y$-short ($Y_S$, 0.985 $\mu$m), at a separation of $0.470 \pm 0.010''$ and a contrast of $(1.63 \pm 0.49) \times 10^{-5}$.   This detection has a signal-to-noise ratio of 4.1, with an empirically estimated upper-limit on false alarm probability of 1.0\%. We also present new photometry from the NICI instrument on the Gemini-South telescope, in $CH_{4S,1\%}$ ($1.58$ $\mu m$), $K_S$ ($2.18\mu m$), and $K_{cont}$ (2.27 $\mu m$).  A thorough analysis of our photometry combined with previous measurements yields an estimated near-IR spectral type of L$2.5\pm1.5$, consistent with previous estimates.  We estimate $\log (L_{bol}/L_\sun ) = -3.86 \pm 0.04$, which is consistent with prior estimates for $\bPicb$ and with field early-L brown dwarfs.  This yields a hot-start mass estimate of $11.9 \pm 0.7$ $M_{Jup}$ for an age of $21\pm4$ Myr, with an upper limit below the deuterium burning mass.  Our $L_{bol}$ based hot-start estimate for temperature is $T_{eff}=1643\pm32$ K (not including model dependent uncertainty).   Due to the large corresponding model-derived radius of $R=1.43\pm0.02$ $R_{Jup}$, this $T_{eff}$ is $\sim$$250$ K cooler than would be expected for a field L2.5 brown dwarf.  Other young, low-gravity (large radius), ultracool dwarfs and directly-imaged EGPs also have lower effective temperatures than are implied by their spectral types. However, such objects tend to be anomalously red in the near-IR compared to field brown dwarfs.  In contrast, $\bPicb$ has near-IR colors more typical of an early-L dwarf despite its lower inferred temperature.  
\end{abstract}

\section{Introduction}

In contrast to the stellar main sequence, brown dwarfs (BDs) form a true evolutionary sequence.  BDs are not massive enough to maintain a constant effective temperature ($T_{eff}$) via hydrogen fusion \citep[$M\lesssim0.075$ $M_\sun$, e.g.][]{1997ApJ...491..856B}.  Thus, a BD cools as it ages, radiating away the gravitational potential energy from its formation \citep{2001RvMP...73..719B}.  BDs are classified into spectral types by comparison to anchor objects.  Various clues to classification were judiciously chosen such that they should correspond to temperature, at least in a relative sense \citep{1999ApJ...519..802K, 2002ApJ...564..421B, 2003ApJ...594..510B}.  Temperature is not a readily observable quantity, however it is well established from theory that substellar objects ranging in mass from $\sim$$1$ to $\sim$$75$ $M_{Jup}$ will have radius in a narrow range of $0.8-1.1$  $R_{Jup}$ \citep[e.g.][]{2011ApJ...736...47B, 2007ApJ...659.1661F}.  This means that bolometric luminosity, $L_{bol} = 4\pi\sigma_B R^2 T_{eff}^4$, is approximately determined by temperature alone.  $L_{bol}$ is observable, so with our theoretical understanding of radius we can infer $T_{eff}$ and find that field brown dwarf spectral types appear to be a well-defined temperature sequence \citep{2004AJ....127.3516G, 2009ApJ...702..154S}, except perhaps for the coolest objects \citep{2013arXiv1309.1422D}.   The result is that the SpT of a brown dwarf is a function of both mass and age.

The situation is even more challenging for young objects, which have not completed post-formation contraction.   The radius of such an object can be significantly larger, depending on how it formed \citep{1997ApJ...491..856B, 2000ApJ...542..464C, 2003A&A...402..701B, 2007ApJ...655..541M, 2012ApJ...745..174S}.  Young objects will also have lower mass than older objects of the same temperature.  With lower mass and larger radius these young objects have lower surface gravity (low-g), which changes their spectral morphology \citep[e.g.][]{2001MNRAS.326..695L, 2006ApJ...639.1120K}, but even so their spectra can generally be classified within the ultracool dwarf sequence \citep{2009AJ....137.3345C, 2013ApJ...772...79A}. 

This population of such low-g BDs is especially interesting because they potentially serve as analogs for young extrasolar giant planets (EGPs).   Many of the best studied low-g BDs are companions, such as AB Pic B \citep{2005A&A...438L..29C}, and 2M0122 B \citep{2013ApJ...774...55B}.  Examples of isolated low-g objects are 2M0355 \citep{2013AJ....145....2F}, and PSO318.5 \citep{2013arXiv1310.0457L}. These objects tend to have fainter near-IR absolute magnitudes \citep{2013AJ....145....2F,2013AN....334...85L}, and have $T_{eff}$ several hundred K cooler than field BDs of the same SpT \citep{2013ApJ...774...55B,2013arXiv1310.0457L}. These low-g BDs also tend to be much redder in near-IR colors, and despite being fainter in the bluer filters and having lower $T_{eff}$,  their bolometric luminosities tend to be consistent with the field for their spectral types \citep{2013arXiv1310.0457L}. 

The first handful of directly imaged planets show similar properties, highlighting the challenges of studying substellar objects in the new physical regime of low-g.  For instance, the EGP HR 8799 b and the planetary mass companion  2M1207 b have L-dwarf like very red near-IR colors, but their luminosities and inferred temperatures (800-1000 K) are more like mid-T dwarfs \citep{2005A&A...438L..25C,2008Sci...322.1348M}.  This has been interpreted as a consequence of the gravity dependence of the L-T transition \citep{2006ApJ...651.1166M}.  Thick dust clouds, which cause the redward progression of the L dwarf sequence as temperature drops, persist to even lower temperatures at low-g \citep{2011ApJ...732..107S,2011ApJ...735L..39B, 2012ApJ...754..135M}.  In this framework extremely red and under-luminous HR 8799 b and 2M1207 b are objects which have yet to make the transition to the cloudless, bluer, T dwarf sequence, hence they are often thought of as extensions of the L dwarf sequence \citep{2010ApJ...723..850B,2011ApJ...733...65B, 2011ApJ...737...34M}.

The directly imaged EGP $\bPicb$, in contrast, is much hotter and its near-IR SED is much more typical when compared to field and low-g BDs \citep{2010Sci...329...57L,2010ApJ...722L..49Q, 2011AA...528L..15B, 2011ApJ...736L..33C, 2013arXiv1302.1160B, 2013ApJ...776...15C}.   $\bPicb$ is unique among the directly imaged EGPs in that we have a dynamical constraint on its mass from radial velocity (RV).  A complete orbit has not yet been observed, so RV monitoring constrains the mass depending on the semi-major  axis ($a$):  for $a < 8, 9, 10, 11, 12$ AU the upper mass limit is $M<10,12,15.5, 20, 25$ $M_{Jup}$  respectively \citep{2012A&A...542A..18L}.   The astrometry currently favors $8 \lesssim a \lesssim 9$ AU \citep{2012A&A...542A..41C}, hence $M\lesssim12$ $M_{Jup}$, though larger values are not ruled out.  We can expect to have a good dynamical understanding of $\bPicb$'s mass in the near future. 

$\bPicb$ is also  noteworthy in that we have relatively good constraints on the age of its primary star.  The age of the $\beta$ Pictoris moving group, of which $\bPicA$ is the eponymous member, has recently been revised upward to $21\pm4$ Myr \citep{2013arXiv1310.2613B} using the lithium depletion boundary technique.  Though somewhat larger than the earlier age estimate of $12^{+8}_{-4}$ Myr by \citet{2001ApJ...562L..87Z}, these two estimates are consistent at the $1\sigma$ level.  

A well determined age and a dynamical mass constraint make $\bPicb$ a valuable benchmark for understanding the formation and evolution of both low-mass BDs and giant planets.  Here we present the bluest observations of $\bPicb$ from the first light of the Magellan Adaptive Optics (MagAO) system, using its visible wavelength imager VisAO.  We also present detections with the Gemini Near Infrared Coronagraphic Imager (NICI).  In Section \ref{sec:magao} we describe MagAO and VisAO, and briefly discuss calibrations of this new high contrast imaging system.  In Section \ref{sec:obs} we present our observations and data reduction procedures.  We analyze the $0.9-2.4$ $\mu m$ SED of $\bPicb$ in Section \ref{sec:anal}, showing that this EGP looks like a typical early L dwarf.  We explore the ramifications of this for the physical properties ($L_{bol}$, mass, $T_{eff}$, and radius) of the planet.  Then in Section \ref{sec:discuss}, we compare these derived properties to field objects, and discuss the relationship of the measured characteristics of EGPs and BDs.  Finally, we summarize our conclusions in Section \ref{sec:conclude}.

\vspace{-.25in}
\section{The Magellan VisAO Camera \label{sec:magao}}

MagAO is a 585-actuator adaptive secondary mirror (ASM) and pyramid wavefront sensor (PWFS) adaptive optics (AO) system, installed at the 6.5 m Magellan Clay Telescope at Las Campanas Observatory (LCO), Chile.  The system is a near clone of the Large Binocular Telescope (LBT) AO systems \citep{2010ApOpt..49G.174E,2011SPIE.8149E...1E}.  MagAO has two science cameras: the Clio2 $1-5\microns$ camera \citep{2004SPIE.5492.1561F, 2006SPIE.6269E..27S} and the VisAO $0.5-1\microns$ camera.   Here we describe characterization and calibration of VisAO relevant to this report. For additional information about MagAO see \citet{2012SPIE.8447E..0XC}, \citet{2012SPIE.8447E..42M}, and \citet{2013arXiv1308.4844K}.  For additional information about the on-sky performance of VisAO see \citet{2013ApJ...774...94C}, \citet{2013ApJ...775L..13F}, and \citet{2013ApJ...774...45W}.  High contrast imaging of $\bPicb$ in the thermal-IR with Clio2 is presented in our companion paper \citep[in preparation, hereafter Paper II]{ktfave}.

\vspace{-.2in}
\subsection[The Ys Filter]{The $Y_S$ Filter}
\label{sec:Ys}
Since this was the first attempt to conduct high-contrast observations with VisAO, we used our longest wavelength bandpass to maximize Strehl ratio (SR) and flux from the thermally self-luminous planet.   We refer to this filter as ``Y-short'', or $Y_S$.  For a complete characterization of this filter see Appendix \ref{synphot}.  In brief, it is defined by a long-pass dichroic at $950\mu$m and the CCD QE cutoff at $\sim$$1.1\mu$m.  Including a representative atmosphere in the profile, the central wavelength of $Y_S$ is $0.985\mu$m and the width is $0.086\mu$m.

\vspace{-.2in}
\subsection{The VisAO Anti-blooming Occulting Mask}
\label{sec:mask}

The VisAO camera contains a partially transmissive occulting mask, used to prevent saturation of the CCD when observing bright stars.  The mask has a radius equivalent to $0.1''$, were it in the focal plane, but it is approximately 60mm out of focus in an f/52 beam.  This has two consequences of note here.  The first is that the mask has an apodized attenuation profile which extends to $\sim$$0.8''$ in radius, so at a separation of $\sim$$0.46''$ $\bPicb$ is under the mask.  The second, and more challenging, consequence is that the mask attenuation will depend on wavefront quality.  We did not fully characterize the mask using $\beta$ Pic itself, so we bootstrapped an attenuation profile from other measurements at different levels of wavefront correction.  We describe this process in detail in Appendix \ref{app:coron}.  We estimate that the mask transmission at the separation of $\bPicb$ was $0.60^{+0.05}_{-0.10}$.

\vspace{-.2in}
\subsection{Astrometric Calibration}

In \citet{2013ApJ...774...94C} we describe the calibration of platescale in several filters and North orientation of the VisAO camera.  Here we describe our calibrations in $Y_S$ and how we tied VisAO to the Clio2 astrometry.
The primary stars used for VisAO calibration were $\theta^1$ Ori $B1$ and $B2$ in the Trapezium.  These stars were observed repeatedly, and with a separation of $\sim$$0.94''$ $B2$ is well within the isoplanatic patch when guiding on $B1$.  \citet{2013ApJ...774...94C} recently showed that all four stars in  $\theta^1$ Ori $B$ are exhibiting orbital motion, so we measured their current astrometry using the wider FOV Clio2 camera.  A complete description of these measurements is provided in Paper II.  In brief, we used combinations of Trapezium stars other than $B1$ and $B2$ to measure the distortion, platescale, and orientation of Clio2.  This was done using the recent astrometry given in \citet{2012ApJ...749..180C} which used LBTAO/Pisces referenced to HST astrometry from \citet{2008AJ....136.2136R}.  We then compared measurements of $B1$ and $B2$ with Clio2 to those with VisAO.  This was done in Dec. 2012, in the $Y_S$ filter with the occulting mask out of the beam.  The extra glass added by the mask substrate changes the focus position slightly, decreasing the platescale.  We measured this change in May, 2013, also using $\theta^1$ Ori $B1$ and $B2$.  The ratio of platescales with and without the mask is 0.9972 $\pm$ 0.0003.  This was applied to the $Y_S$-open platescale to determine the $Y_S$-mask platescale.

We also determined the orientation of the CCD with respect to North, which we denote by $NORTH_{VisAO}$.    Our images are derotated counter-clockwise using the equation $DEROT_{VisAO} = ROTOFF + 90^o + NORTH_{VisAO}$, where $ROTOFF$ is the rotator offset, equal to rotator angle plus parallactic angle.      The astrometric calibration of VisAO is summarized in Table \ref{tab:visaoastro}.

\begin{table}[t]
\caption[VisAO Ys platescale and rotator calibration]{VisAO $Y_S$ platescale and rotator calibration.  Measurement uncertainty includes both Clio2 and VisAO scatter.  Astrometric uncertainty was propagated from the LBTAO/Pisces measurements of \citet{2012ApJ...749..180C}. \label{tab:visaoastro}}
\centering
\scriptsize
\begin{tabular}{lcccc}
\hline
\hline
                &    Value       &    Measurement     &     Astrometric      &     Total \\
                &                &     Uncertainty    &     Uncertainty      &   Uncertainty \\
\hline
$NORTH_{VisAO}$ &    -0.59 deg       &    0.04 deg        &     0.30 deg          &      0.30 deg\\
\hline
\multicolumn{5}{l}{$Y_S$+no mask:}\\
Platescale      &    7.896 mas/pix  &    0.004 mas/pix  &   0.019 mas/pix     &     0.019 mas/pix \\
\hline
\multicolumn{5}{l}{$Y_S$+mask:}\\
Platescale      &  7.874 mas/pix  &    0.005 mas/pix  &   0.019 mas/pix     &      0.020 mas/pix \\
\hline
\end{tabular}
\end{table}

By dithering $\theta^1$ Ori $B$ around the detector we diagnosed a slight focal plane tilt, which causes a small change in platescale, $<1$\%, from top to bottom predominantly in the y direction \citep{2013ApJ...774...94C}.  At the $\sim$$0.47''$ separation of $\bPicb$ this change in platescale amounts to less than 1 mas, so we neglect it.

\vspace{-.25in}
\section{Observations and Data Reduction \label{sec:obs}}

\subsection{VisAO}

We observed $\beta$ Pic with VisAO on the night of 2012 Dec 04 UT in the $Y_S$ filter using the 50/50 beamsplitter, sending half the $\lambda < 1$ $\microns$ light to the PWFS.  The image rotator was fixed to facilitate angular differential imaging \citep[ADI,][]{2004Sci...305.1442L,2006ApJ...641..556M}.  Conditions were photometric.   $V$ band seeing, as measured by a co-located differential image motion monitor (DIMM), varied from  $\sim0.45''$ $\sim0.75''$  during the 4.17 hours of elapsed time included in these observations.

\subsubsection{VisAO PSF and Strehl Ratio}

Towards the end of the observation we took off-mask calibration data to assess AO performance and calibrate our photometry.  We took 1061 0.283 sec frames off the occulting mask, at airmass 1.14. $V$ band seeing as measured by the DIMM was  $\sim0.65''$ during these measurements.   To avoid saturation this data was taken in our fastest full frame mode ($1024\times1024$ pixels at 3.51 frames-per-second) and in the lowest gain setting.  Even with these settings, we saturated the peak pixel in roughly a third of the exposures.  To compensate we selected frames with peak pixel between 8000 and 9000 ADU, where the detector is linear, such that we are using data between approximately the 75th and 25th percentiles.  

\begin{figure}[t]
\hspace{-55pt}
\includegraphics[height=4in,clip=true,angle=90]{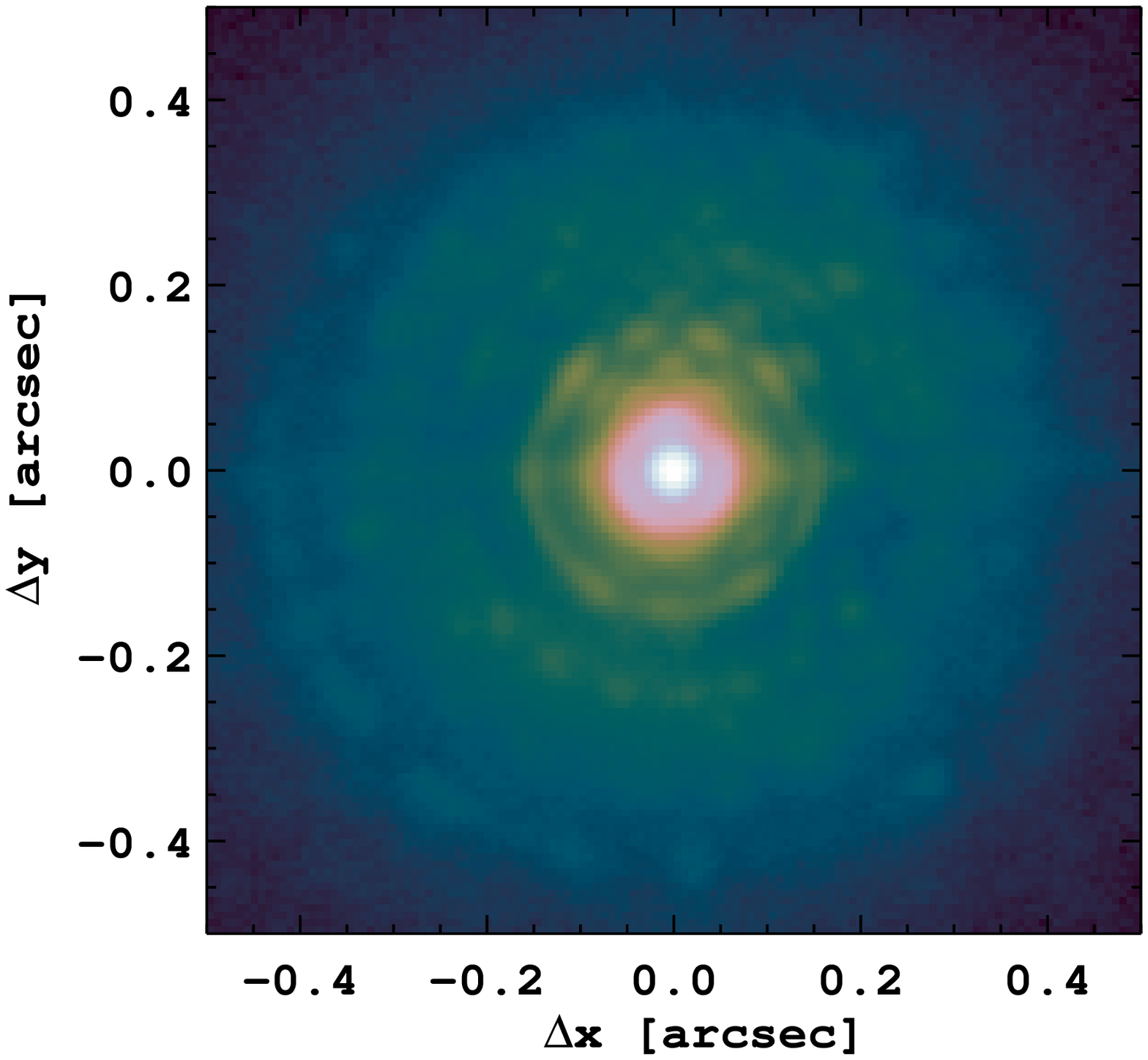}
\hspace{-30pt}
\includegraphics[height=4in,clip=true,angle=90]{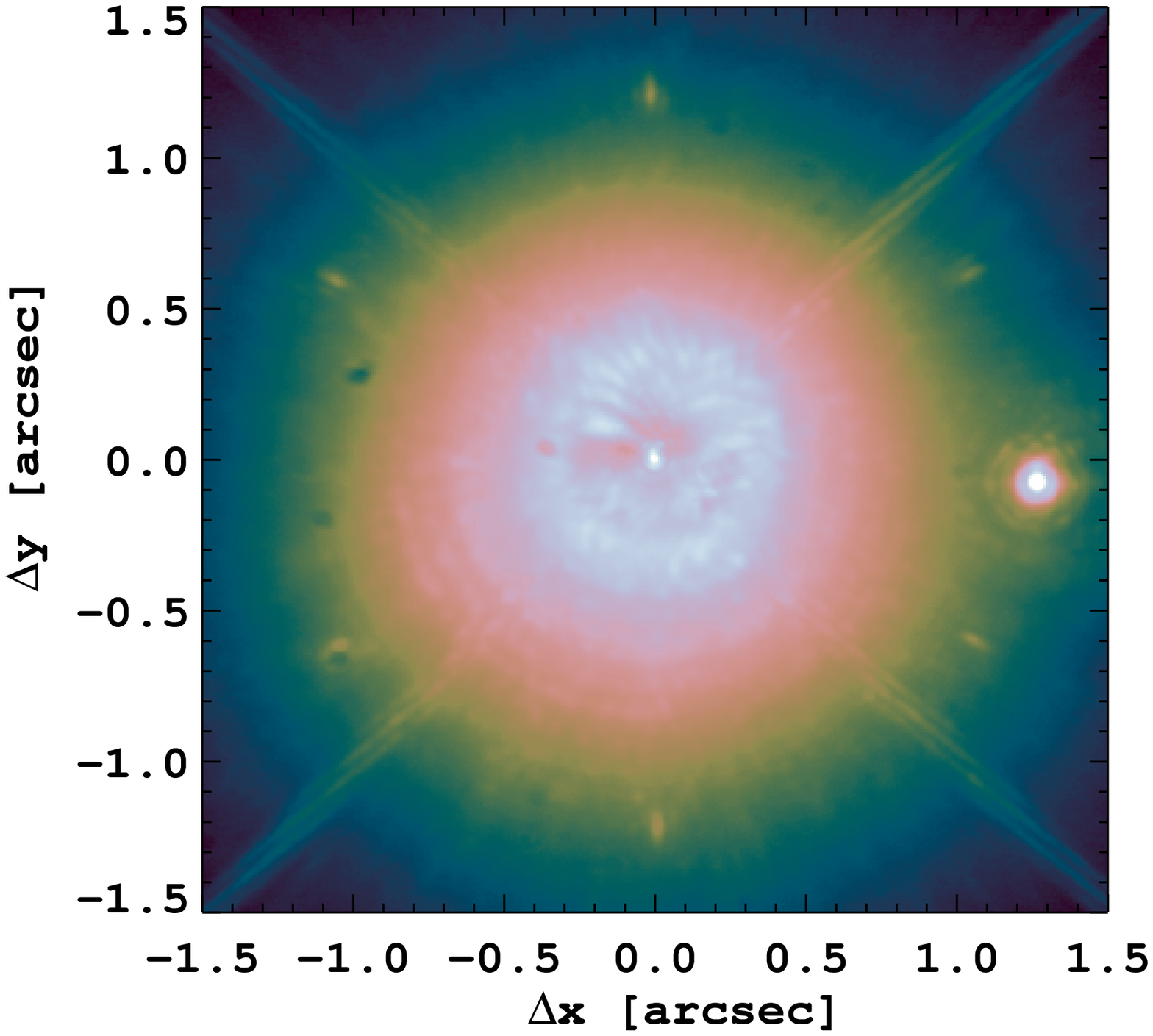}
\vspace{-.4in}
\caption{VisAO PSF measurements using $\beta$ Pictoris A, a $Y_s=3.561$ mag A6V star.  Left: the unsaturated (off the occulting mask) PSF made by shifting and adding 491 0.283 sec exposures.   Right: the on-mask PSF, formed by registering and combining 2.15 hours of 2.27 sec individual exposures. Images shown on  a log stretch.  The observed $Y_S$-band SR was $\sim$$32\%$, meaning the actual SR after correcting for the CCD's PRF was $\sim$$40\%$.  The bright signal $\sim$$1.3''$ to the right of the star is an in-focus beamsplitter ghost. 
\label{fig:bpicb_psf}}
\end{figure}

VisAO and Clio2 are operated simultaneously.  The dichroic entrance window of Clio2 transmits light with $\lambda \gtrsim 1.05\mu$m, while reflecting shorter wavelengths to the PWFS and VisAO camera. When working in the thermal IR with Clio2, we perform small pointing offsets (called ``nods'')  to facilitate background subtraction.  These nods occur at intervals ranging from $\sim2$ to $\sim5$ minutes depending on wavelength and star brightness.  The consequence for VisAO observations is occasionally short (5 to 15 second) periods of unusable data while the AO loop is paused during a nod. Pausing the loop causes wavefront error (WFE) to become much worse.  To automatically reject these periods during post-processing we apply a WFE cut using AO telemetry \citep{2012SPIE.8447E..42M}.  For all of these observations we used 130 nm RMS phase.  For the PSF measurement this rejected 84 frames.  We then registered and median combined the remaining 491 images, with the result shown in Figure \ref{fig:bpicb_psf}.   

The unsaturated un-occulted PSF core has a FWHM of 4.73 pixels (37 mas), compared to the 6.5 m diffraction limit at $Y_S$ of 3.87 pixels (31 mas).  We have identified two sources of broadening: $\sim$$7$ mas of residual jitter, most likely due to 60 Hz primary mirror cell fans; and the CCD's charge diffusion pixel response function (PRF).  For a near-Nyquist sampled CCD charge diffusion causes a blurring effect, which was well documented for the HST ACS and WFPC cameras.  See \citet{2003acs..rept....6K}, \citet{2006acs..rept....1A}, and the ACS handbook\footnote{\url{http://www.stsci.edu/hst/acs/documents/handbooks/cycle20/c05_imaging7.html}}.  We measured the PRF in the lab, and found that it broadens our PSF by $\sim$$0.4$ pixels, and lowers the peak such that SR due to PRF alone is 80\%.   SR measured using WFS telemetry and the unsaturated PSF was $32\pm2$\%.  Since this includes PRF, we can divide by 0.8 to estimate that the true $Y_S$ SR was $40\%$.

\subsubsection{High Contrast Observations}

The observations intended to detect $\bPicb$ were conducted in full frame mode (1024x1024 pixels, $8''\times8''$), in the camera's highest sensitivity gain setting, with an individual exposure time of 2.27 sec.  The star was behind the occulting mask, so no dithers were conducted.   The individual images were bias and dark subtracted, using shutter-closed darks taken at 15 min intervals throughout the observation.  We did not flat field.  The CCD-47 has very low pixel-to-pixel variation, and with an $8''\times8''$ FOV we expect only small illumination changes across an image.  Clio2 nods were rejected using WFE telemetry as described above.   We then examined each image by eye, rejecting those with apparent poor mask alignment or poor AO correction.   Finally, to reduce the number of images to process we median coadded the remaining 3399 dark-subtracted exposures until either 30 seconds had elapsed or 0.5 degrees of rotation had occurred.   This process resulted in 317 images corresponding to 2.5hrs of open-shutter data, with an elapsed clock-time of 4.17 hrs and 116 degrees of sky rotation from start to finish. Airmass was 1.12 at the beginning, reached 1.08 at transit, and was at a maximum of 1.29 at the end of the observations.

The  coadded images were first coarsely registered and centered using a beamsplitter ghost.  The mask is partially transmissive (ND$\approx 2.8$) in the core.  To more finely register the images we located the center of rotational symmetry of the attenuated star using cross correlation, with a tolerance of 0.05 pixels.  We median combined the registered images, forming our master PSF.  This is shown in Figure \ref{fig:bpicb_psf}.

We employed a reduction technique based on principal component analysis (PCA), using the Karhunen-Lo\'eve Image Processing (KLIP) algorithm of \citet{2012ApJ...755L..28S} (see also \citet{2012MNRAS.427..948A}).   We applied KLIP in search regions, as suggested by \citet{2012ApJ...755L..28S}, dividing the image in both radius and azimuth in ``optimization regions'' using the strategy developed by \citet{2007ApJ...660..770L} for the locally optimized combination of images (LOCI) algorithm.  In each optimization region we conducted the complete KLIP procedure.  Only a subsection of the optimization region, a ``subtraction region'', was kept.  We used the parameters specifying the regions given by \citet{2007ApJ...660..770L}, except that instead of having the same azimuthal width, our subtraction regions were 1/3 the width of the optimization regions in azimuth. This provided a noticeable ($\sim10\%$) improvement in signal-to-noise ratio (S/N) without significantly increasing the computational burden.  The final image is formed by combining the individually reduced subtraction regions.  Note that when we choose the number of KL modes, we apply this choice to all subtraction regions.

\begin{sidewaysfigure}
\hspace{-0.5in}
\includegraphics[height=3.5in,clip=false]{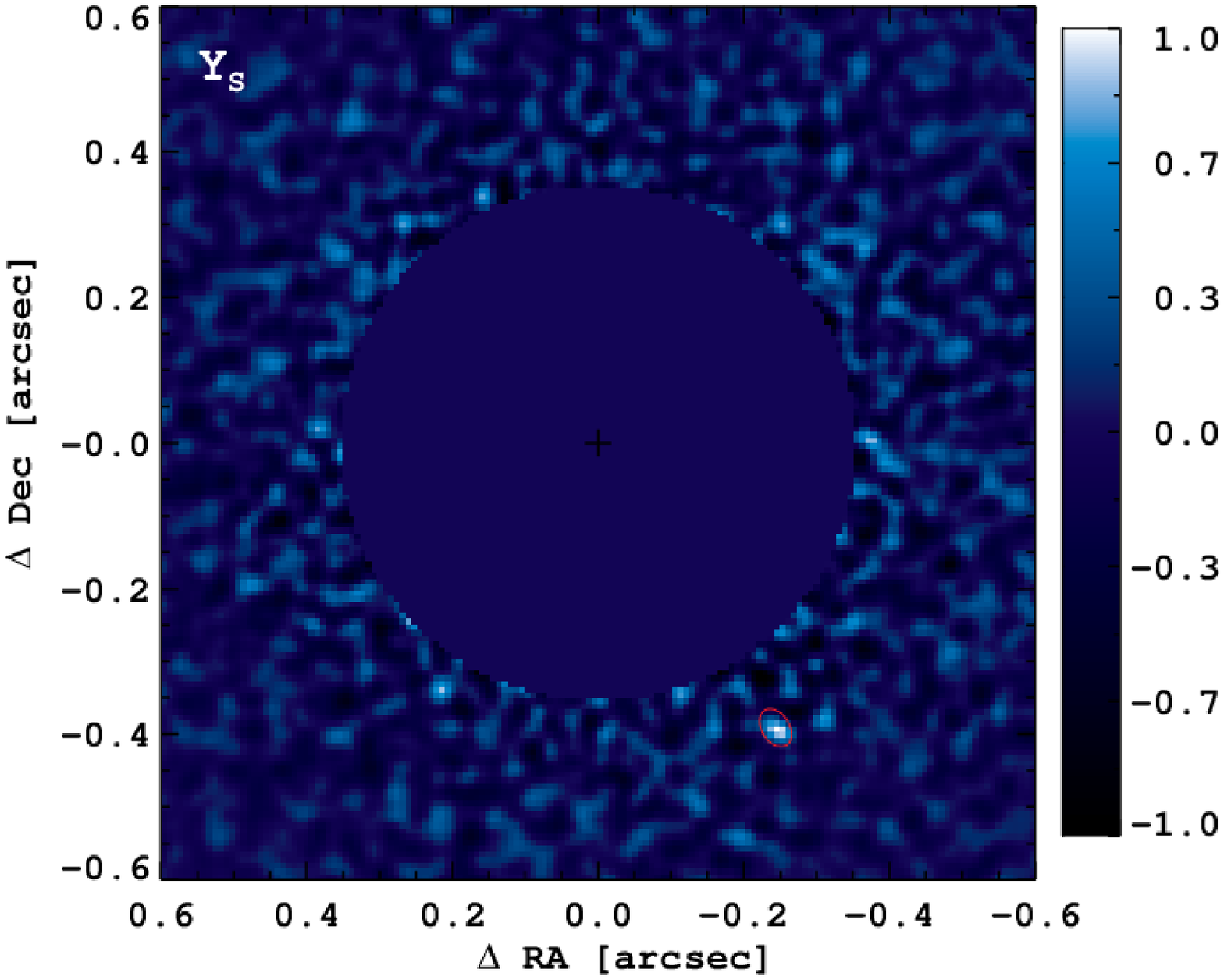}
\hspace{-0.1in}
\includegraphics[height=3.5in,clip=false]{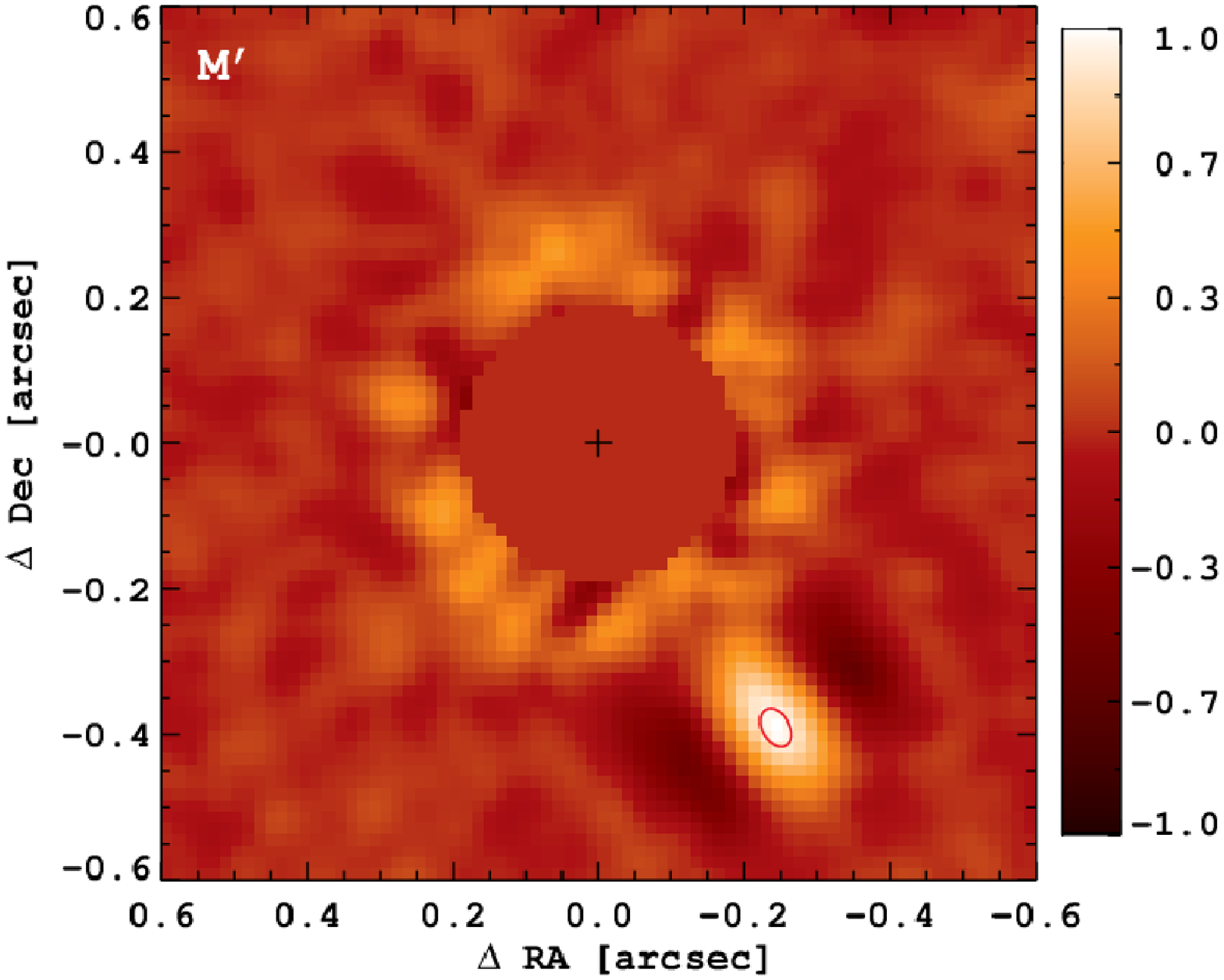}
\caption[beta Pic Detections]{Left: The MagAO+VisAO detection of $\bPicb$.   Right: The MagAO+Clio2 $M'$-band detection of $\bPicb$ (from Paper II), taken simultaneously with the VisAO image.  In each image the red ellipse denotes the $\pm 2\sigma$ uncertainty in the Clio2 position.  The color scales are relative to the peak of the planet. The inner regions of the images, which are dominated by residuals, have been masked out in post-processing.\label{fig:detections}}
\end{sidewaysfigure}

We compare the results of our KLIP reduction to the simultaneously obtained Clio2 $M'$ image, shown in Figure \ref{fig:detections} and described in Paper II.  The Clio2 position is shown in both images as an ellipse corresponding to the $2\sigma$ uncertainty.  Here we use the mean position and error using data taken with Clio2 in four filters on subsequent nights.  See Paper II for further details.

\begin{figure}
\includegraphics[width=2.8in,clip=true,angle=90]{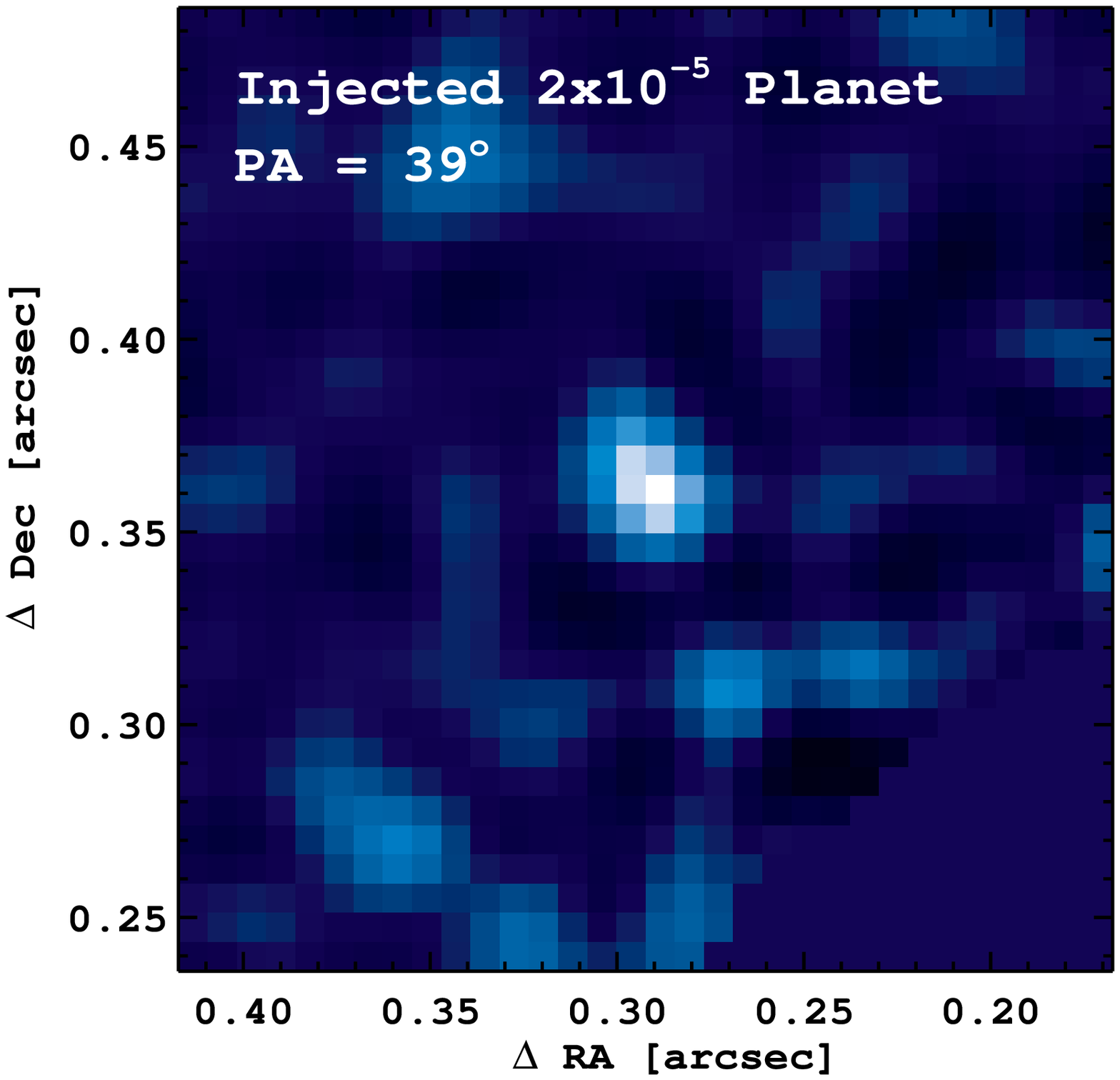}
\hspace{-1.in}
\includegraphics[width=2.8in,clip=true,angle=90]{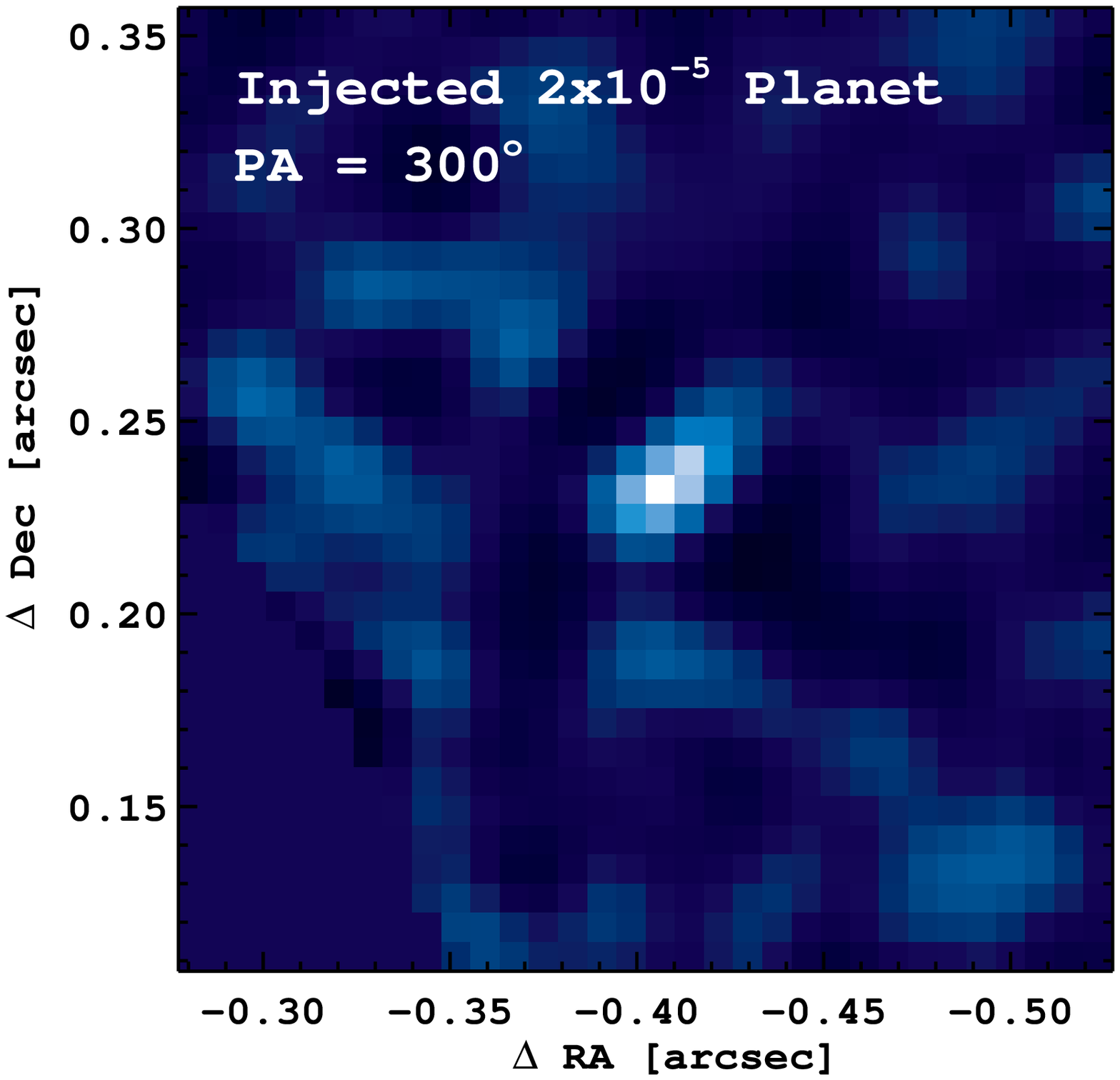}
\includegraphics[width=2.8in,clip=true,angle=90]{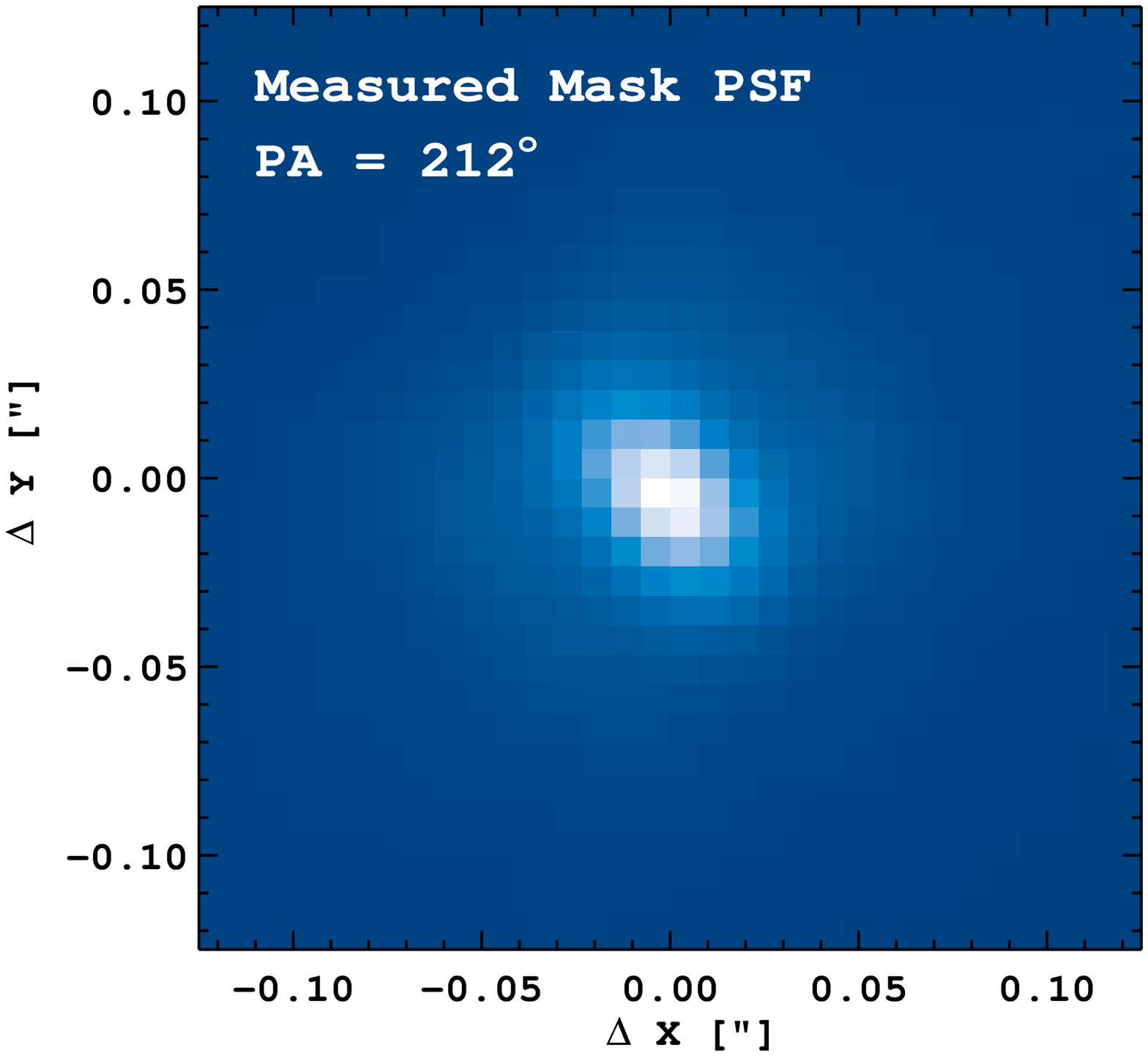}
\hspace{-1in}
\includegraphics[width=2.8in,clip=true,angle=90]{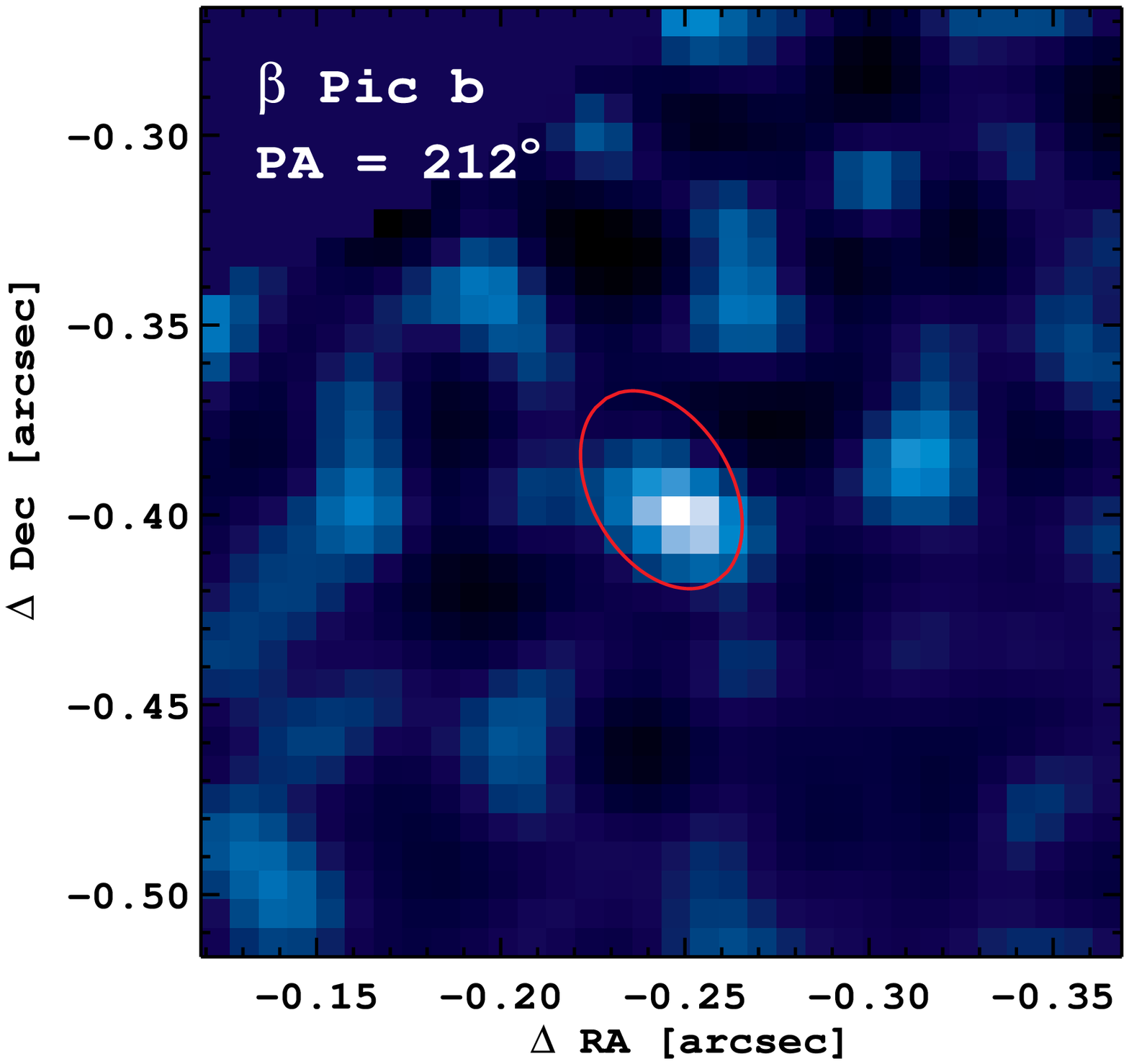}
\caption{ Comparison between our image of $\bPicb$, the occulting mask PSF, and two representative simulated planets which were injected into the data using the mask PSF.  The mask PSF was measured on-sky by scanning a star across the mask, and the image at the lower left corresponds to the separation of $\bPicb$.  At lower right we zoom in on the VisAO detection of $\bPicb$.  As in Figure \ref{fig:detections} the red ellipse corresponds to the $2\sigma$ Clio2 position uncertainty.  The recovered image of $\bPicb$ matches the PSF.  The top two images are of simulated planets, injected with contrast $2\times10^{-5}$, and recovered using the same data processing pipeline.  These show that our injected planets are correctly modeling the PSF.  Each image is plotted on the same spatial scale, and the color table is stretched relative to the peak of the object in each as in Figure \ref{fig:detections} .
\label{fig:zooms}}
\end{figure}

In Figure \ref{fig:zooms} we zoom in on the VisAO image of $\bPicb$ and compare it to the PSF at that position under the occulting mask.  This under-the-mask PSF was measured on-sky in closed-loop by scanning a star across the mask.  The mask radially elongates a point source, and the image produced by our KLIP analysis matches this expectation.  We also show two representative fake planets, injected using the same PSF (the details of this procedure are given below).  The signal recovered at the precisely known location of the planet closely matches the expected signal based on our PSF model.

Other tests conducted included a basic ADI-only reduction, which yields a lower significance detection (S/N $<$ 3).  We also tried many other search region geometries.  Though varying reduction parameters modulates various speckles throughout the image and changes algorithm throughput, the signal at the location of $\bPicb$ is always detected.  We also tested reducing half the data in alternating chunks (to preserve rotation), and even with ADI-only  we detected $\bPicb$ using only half the data (albeit with much lower S/N).  Along with our PSF comparisons, these results give confidence that our detection of $\bPicb$ is valid.  We quantify the significance of this detection next.

\subsubsection{Detection Significance}
\label{sec:snr}
\begin{sidewaysfigure}
\includegraphics[width=5in,clip=true]{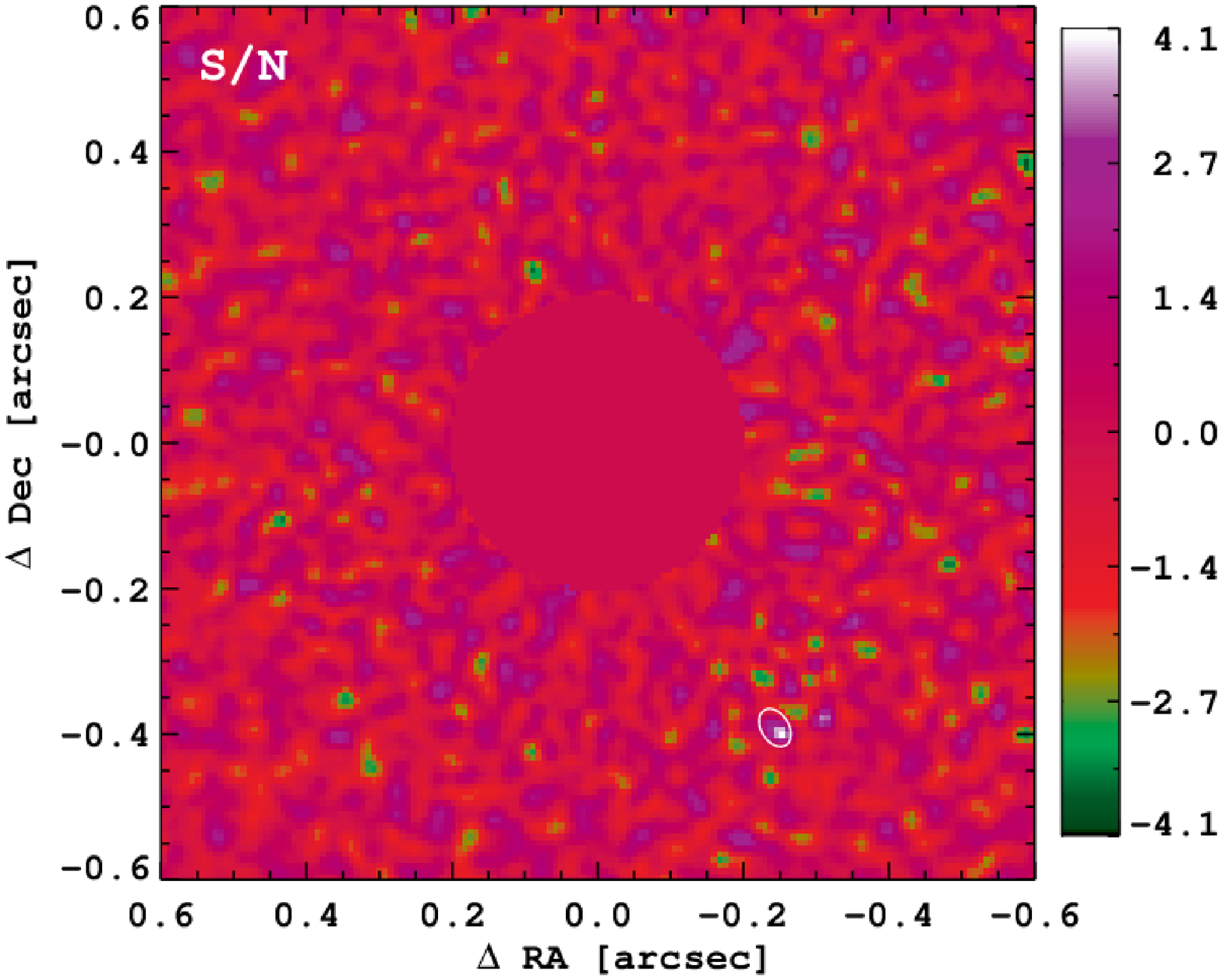}
\includegraphics[width=4in,clip=true,angle=90]{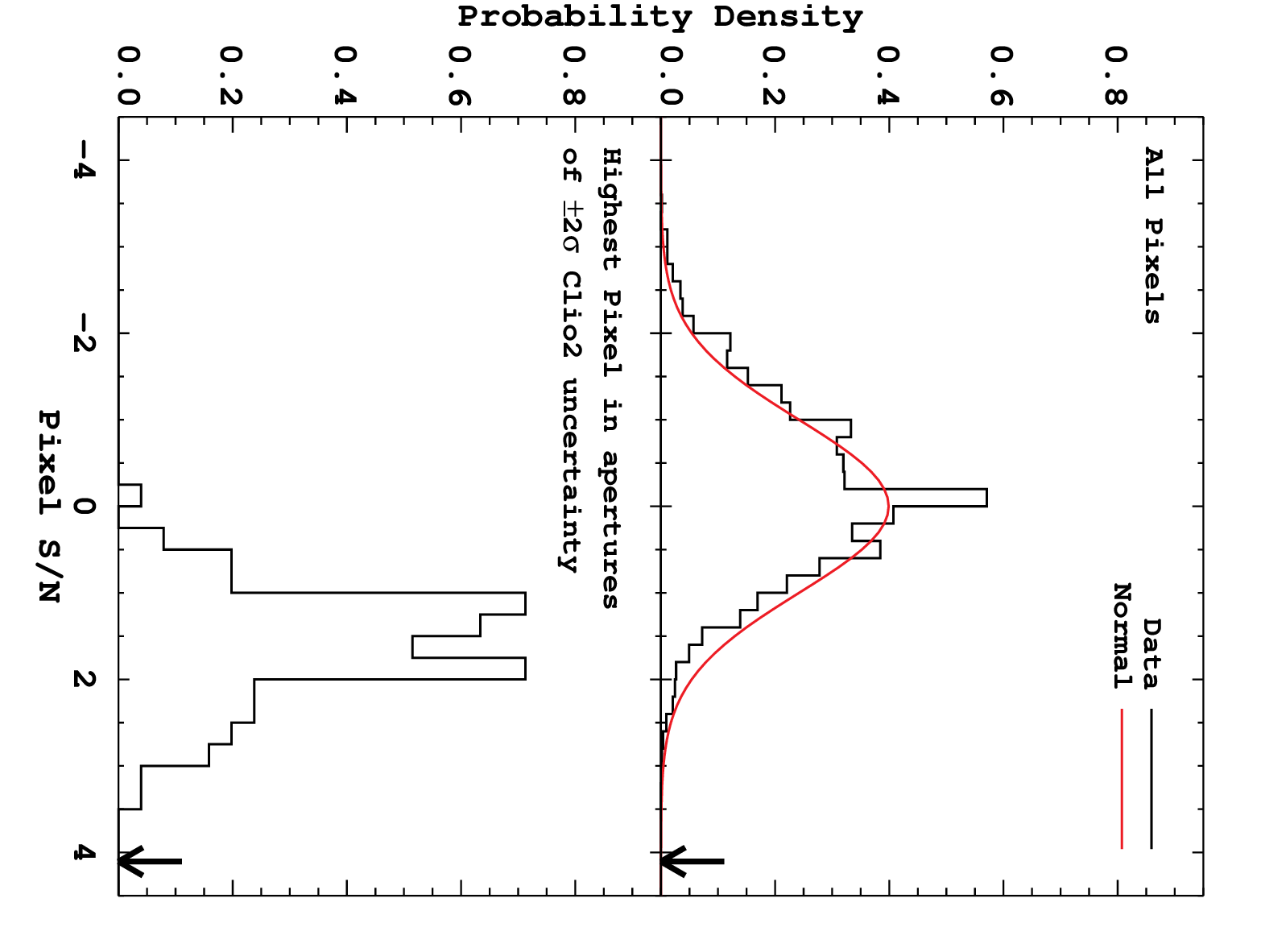}
\caption[beta Pic b S/N]{Left: S/N map.  The white circle shows the $2\sigma$ uncertainty in the Clio2 position.  We detect $\bPicb$ with VisAO at S/N = 4.1 at the location of the Clio2 detection.   Right: Pixel S/N histograms.  The top panel shows all pixels within the annulus of width $\pm2\sigma$ (Clio2 uncertainty) around the detection.  The red curve is the Standard Normal distribution (not fit to the data), which we show for comparison.  The arrow shows the detection of $\bPicb$.  The bottom panel shows the distribution of the maximum S/N in apertures with radius of the twice the Clio2 position uncertainty ($\pm2\sigma$).   We estimate a conservative upper limit on false alarm probability of $FAP = 1.0\%$.      \label{fig:snr_fap}}
\end{sidewaysfigure}

To assess the statistical significance of the VisAO detection we next calculated a S/N map. For this analysis we used 150 KL modes (we discuss choosing the number of modes in Section \ref{sec:fake}).  The final image was Gaussian smoothed with a kernel of width 3 pixels.   Each pixel was divided by the standard deviation calculated in an annulus  1 pixel in width.  Pixels within 1 FWHM radius of the location of the planet were excluded from the standard deviation calculation.  The S/N map is shown at left in Figure \ref{fig:snr_fap}.  Within the $2\sigma$ uncertainty of the Clio2 detection $\bPicb$ is detected by VisAO with S/N$ = 4.1$.  Other choices of smoothing kernels, different numbers of modes, and different techniques for calculating the noise can change this value by $\pm$$\sim$$25\%$ (see Figure \ref{fig:contrastvsnmodes}).  Regardless, this is the maximum S/N pixel at or near the separation of $\bPicb$, so these choices have minimal impact on the following analysis.

As a starting point we first calculated the histogram of all pixels within the annulus of width $\pm2\sigma$ (Clio2 uncertainty) centered on the location of the planet, excluding the planet location.  We restrict ourselves to this annulus to ensure the statistics are representative.   The histogram is shown in Figure \ref{fig:snr_fap}, where we also overplot a Gaussian distribution with $\sigma=1$ for comparison.  There are 2628 pixels included in the histogram, all with S/N$ < 4.1$.  We note that these pixels will tend to be correlated across the PSF width and by the smoothing kernel, so we do not estimate a false alarm probability (FAP) from the histogram of individual pixels.  However, this does show that the S/N = 4.1 peak at the location of $\beta$ Pic b is not expected from the distribution of pixels in the image.

In truth, we would have considered any signal close to the location determined by Clio2 as a VisAO detection.  We next analyze the 101 unique apertures with a radius of $2\sigma$ (Clio2 radial uncertainty) which fit in the same annulus, choosing the highest S/N pixel in each.  These apertures have a diameter of $\sim7$ VisAO pixels, larger than the 4.7 pixel FWHM, so they are uncorrelated with respect to both the PSF and the smoothing kernel. The histogram for these trials is shown at lower right in Figure \ref{fig:snr_fap}.  The probability of having the aperture with the highest S/N pixel occur at the Clio2 position by chance is $1/102 = 1.0\%$ (adding one aperture at the location of the planet).  This then sets an empirical upper limit on FAP.  

\subsubsection{Photometry with Simulated Planets}
\label{sec:fake}

We calibrated our photometry by injecting simulated planets.    We used our on-sky under-the-mask PSF measurement, shown in Figure \ref{fig:zooms} (see also Appendix \ref{app:coron}) to simulate a planet, which we scaled based on the under-the-mask image of the star in each frame, using a 3 pixel radius photometric aperture.  Note that this accounts for Strehl variation and airmass effects.  The mask transmission was also applied, taking into account the changing position of $\bPicA$ under the mask over the course of the observation due to flexure and re-alignment. This caused the transmission at the location of $\bPicb$ to change by roughly $\pm3\%$.  With this procedure we injected planets with contrasts of 1, 2, and $3\times10^{-5}$ at 21 locations $13.5^o$ apart in the $270^o$ opposite the location of $\bPicb$, at the same 59 pixel ($\sim$$0.47''$) separation.  This was done prior to registration and then the complete reduction was carried out, meaning an entirely new set of KL modes was calculated in each search region.

We conducted aperture photometry on these simulated planets, and on $\bPicb$.   Using a reduction with no injected planets but otherwise having the same parameters, we estimated the noise in our photometry by sampling 59 locations spaced by $2$ FWHM, at the same separation, avoiding the known location of $\bPicb$.  Using a simple grid search strategy, we tested various aperture sizes and Gaussian smoothing kernels.  Using the mean photometry of the simulated planets we found the aperture radius and smoothing width which maximized S/N.  These values vary depending on the number of modes and contrast. We used the values determined for the $2\times10^{-5}$ planets for photometry on the actual $\bPicb$.

\begin{figure}[t]
\includegraphics[width=2.4in,clip=true,angle=90]{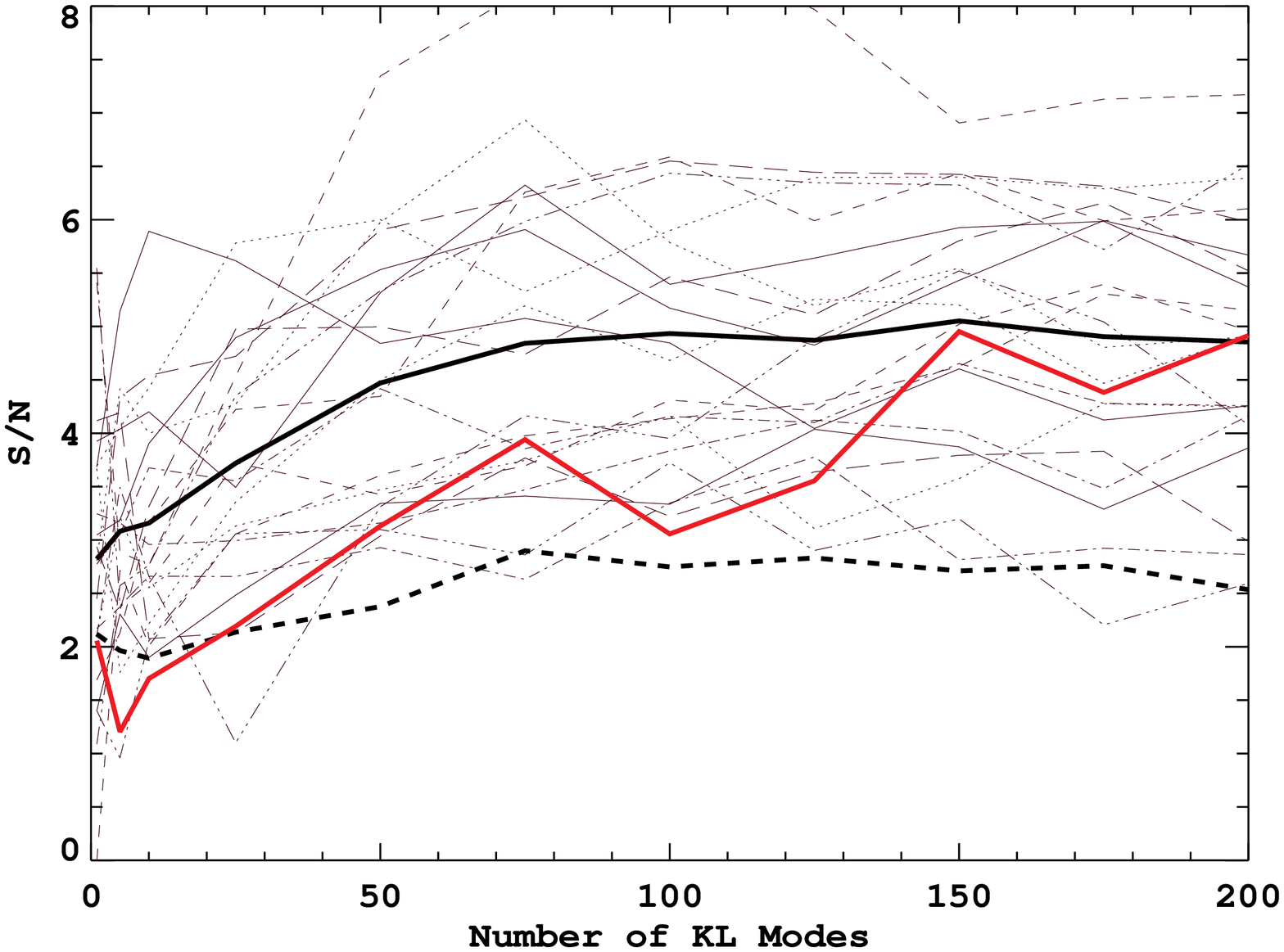}
\includegraphics[width=2.4in,clip=true,angle=90]{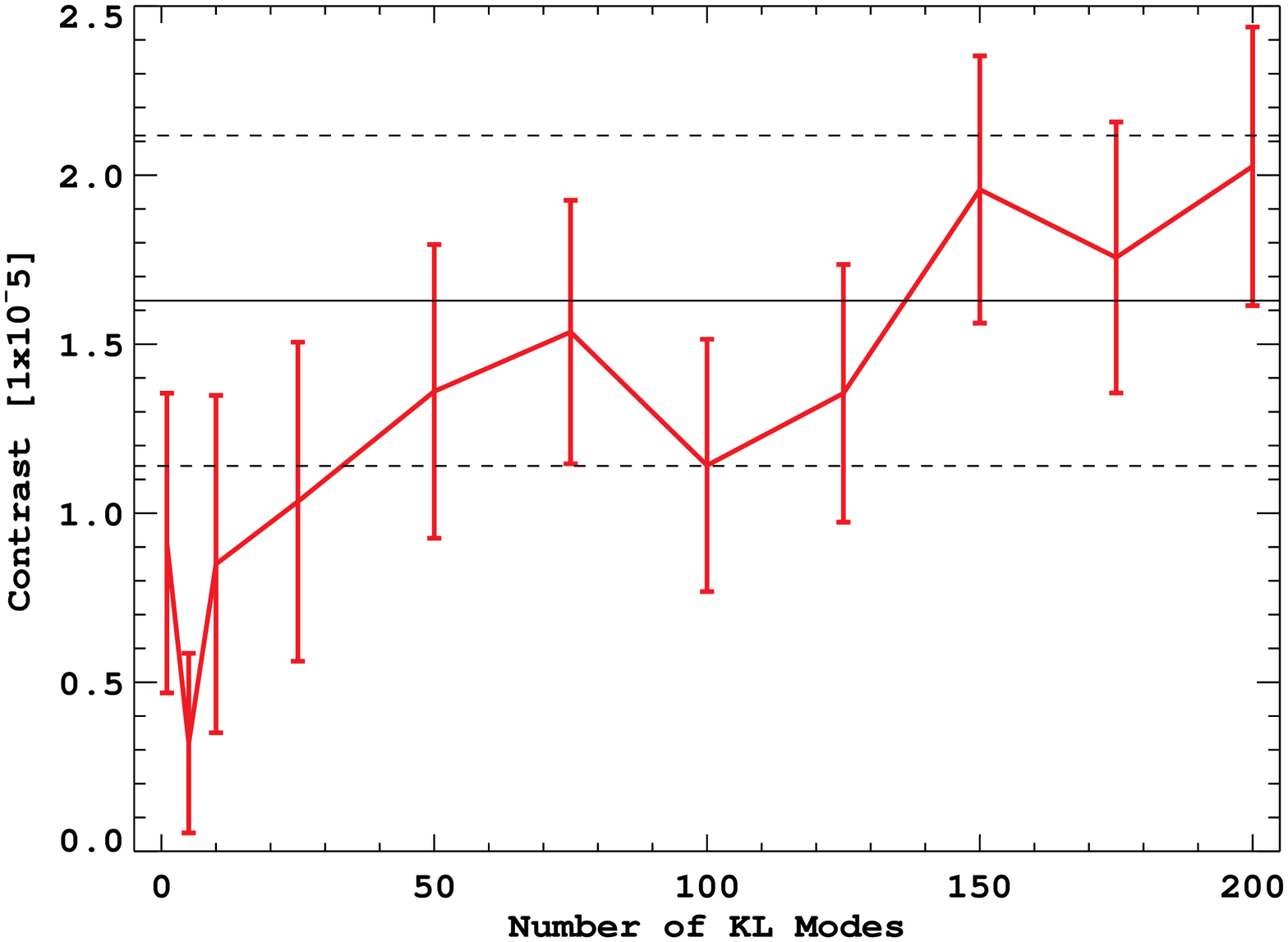}
\vspace{-.3in}
\caption[S/N vs.\ Number of KL Modes]{ Left: S/N vs.\ number of KL modes.  The thick black curve is the mean of 21 fake planets injected with contrast $2\times10^{-5}$.  The thick dashed black curve is the same for $1\times10^{-5}$ fake planets.  The red curve is for $\bPicb$.  Also shown are the results for each of the 21 $2\times10^{-5}$ injected planets, showing that the $\bPicb$ result is typical.  Right: The resulting contrast measurements.  Given that the mean S/N of the fake planets has flattened at 75 modes, we have no clear way to choose how many modes to use.  As a result, we average the six contrast measurements from 75 to 200 modes.  We adopt an uncertainty of $30\%$.  Though this is larger than implied by our detection $S/N=4.1$, it accounts for the additional uncertainty from choosing reduction parameters.  The resulting raw contrast measurement is $(1.63 \pm 0.49)\times 10^{-5}$, indicated by the horizontal lines.\label{fig:contrastvsnmodes}}
\end{figure}

In Figure \ref{fig:contrastvsnmodes} we show S/N vs.\ number of KL modes for the mean of the simulated $1\times10^{-5}$ and $2\times10^{-5}$ planets.  In each case, the mean S/N has become nearly constant once 75 modes are included in the reduction.    We also show the individual results for each of the 21 injected $2\times10^{-5}$ planets, and the result for $\bPicb$.  The S/N vs.\ number of modes curve for $\bPicb$ varies significantly up to 200 modes, but we see that this appears rather typical compared to the individual injected planets.  Combined with the comparisons shown in Figure \ref{fig:zooms}, it appears that our injected fake planets model the true signal well.

In the right-hand panel of Figure \ref{fig:contrastvsnmodes} we show the contrast of $\bPicb$.  This was calibrated using the injected planets.  We linearly interpolated between the mean values of the injected planet photometry, finding the contrast which corresponds to the photometry of $\bPicb$. The problem we face, illustrated in Figure \ref{fig:contrastvsnmodes}, is that there is no clear choice for number of KL modes.  Increasing the number of modes from 75 to 200 does not improve the mean S/N of the injected planets, but it does have a large ($\sim$$30\%$) impact on the measured contrast of $\bPicb$.  Rather than choose a single number of modes, we instead average the contrast measurements from 75 to 200 modes.  This gives a contrast of $(1.63 \pm 0.49)\times 10^{-5}$.  We have adopted an uncertainty of $\pm30\%$.  Though this is larger than the $\sim$$25\%$ expected from the S/N found in Section \ref{sec:snr}, it accounts for the additional uncertainty in choosing the number of KL modes.  We added the uncertainty in mask transmission ($+5\%,-10\%$) in quadrature, so our total uncertainty in contrast is $+32\%, -30\%$.  In magnitudes we have $\Delta Y_S = 11.97^{+0.34}_{-0.33}$.

\subsubsection{VisAO Astrometry}

We also used the simulated planets to calibrate our astrometric precision, finding that Gaussian centroiding on the injected planets gave unbiased results.  We measured the positions of the $1\times10^{-5}$ and $2\times10^{-5}$ planets by Gaussian centroiding, and used these to estimate the statistical uncertainties.  We found $\sigma_{sep} = 0.82$ pixels and  $\sigma_{PA} = 0.63$ degrees for $\bPicb$.  In tests using binary stars under the occulting mask, we found a $\sim$$1$ pixel scatter in recovered positions (at the separation of $\bPicb$), which we attribute to uncertainty in centroiding under the mask.  Finally, we include the astrometric calibration uncertainty (see Table \ref{tab:visaoastro}).  We also measured the position of $\bPicb$ using Gaussian centroiding.  Our astrometry for $\bPicb$ is presented in Table \ref{tab:visaoastrom}.

\begin{table}
\scriptsize
\caption{VisAO Astrometry of $\beta$ Pictoris b
\label{tab:visaoastrom}}
\centering
\begin{tabular}{lccccc}
\hline
\hline
Date       & Filter & Separation      &    Separation\tablenotemark{1}     &   PA & Notes\\
           &        &  ($''$)         &       (AU)        &  degrees &\\
\hline
\hline
\multicolumn{4}{l}{VisAO}\\

\hline
2012-12-04 & $Y_S$  & $0.470\pm0.010$ &    $9.14\pm0.20$ & $211.95\pm1.19$ & \\ 
\hline
\multicolumn{4}{l}{Clio2}\\
\hline
2012-12-01/02/04/07 & (mean)  & $0.461\pm0.014$ &  $8.96\pm0.27$ & $211.9\pm 1.2$& Mean of [3.1], [3.3], $L'$, and $M'$ from Paper II\tablenotemark{2}\\

\multicolumn{4}{l}{\tablenotetext{1}{Using a distance of $19.44\pm0.05$ pc from \citet{2007A&A...474..653V}.}}\\
\multicolumn{4}{l}{\tablenotetext{2}{Clio2 observations, provided here for comparison.}}
\end{tabular}
\vspace{-.35in}
\end{table}

\vspace{-.2in}
\subsection{NICI}

We present observations of $\bPicb$ taken during the course of the Gemini Near-Infrared Coronagraphic Imager \citep[NICI,][]{2008SPIE.7015E..49C} campaign \citep{2010SPIE.7736E..53L, 2013ApJ...773..179W, 2013arXiv1309.1462B, 2013ApJ...776....4N}.  We observed $\beta$ Pic on 25 Dec 2010 UT, in $K_S$, and again on 20 Oct 2011 UT in $CH_{4S,1\%}$ and $K_{cont}$.  The $K_S$ data was independently analyzed by \citet{2013A&A...551L..14B}, but with an extrapolated calibration of mask transmission and hence large photometric uncertainties. Here we provide photometry based on a direct measurement of the focal plane mask transmission.  We reduced the NICI data using the well tested tools developed for the NICI campaign (see references above), with some differences as noted next.  The VisAO reduction techniques described above were developed independently.

The images were reduced as described in \citet{2013ApJ...779...80W}, but with the addition of smart frame selection for PSF building similar to \citet{2006ApJ...641..556M} and \citet{2007ApJ...660..770L}. The standard ADI reduction procedure median combines all the science images in the pupil-aligned orientations to make a source-less PSF and thereby subtract the star from each individual image.  In our method, we median-combine only the frames which are similar to the image which is to undergo PSF subtraction. This similarity is measured in the difference of the target image and the candidate reference image using the RMS of pixel values at radii between $0.3''$ and $0.6''$ separation. Only the best 20 images are used to make the reference PSF.  We differ from \citet{2007ApJ...660..770L} and \citet{2006ApJ...641..556M} in that we do not reject frames because they have too little rotation relative to the target image. Instead we require that the range of reference frames selected have total rotation $> 2$ times the NICI FWHM. These parameters were optimized by comparing the S/N maps resulting from the reductions, and we found that this algorithm performed better than either basic ADI or LOCI for data taken with NICI.  

\begin{sidewaysfigure}
\hspace{-0.5in}
\includegraphics[width=2.75in,angle=90]{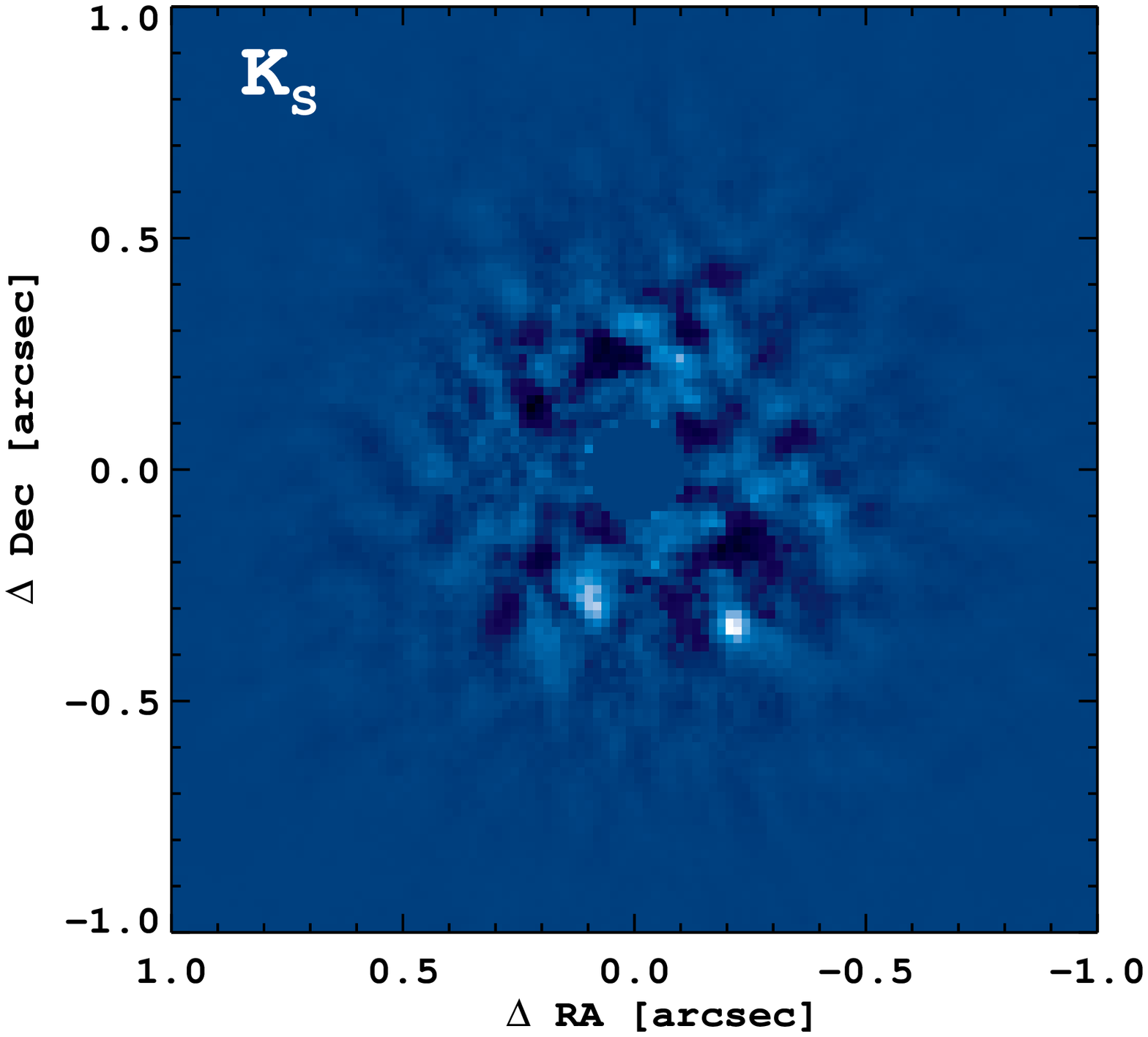}
\hspace{-1in}
\includegraphics[width=2.75in,angle=90]{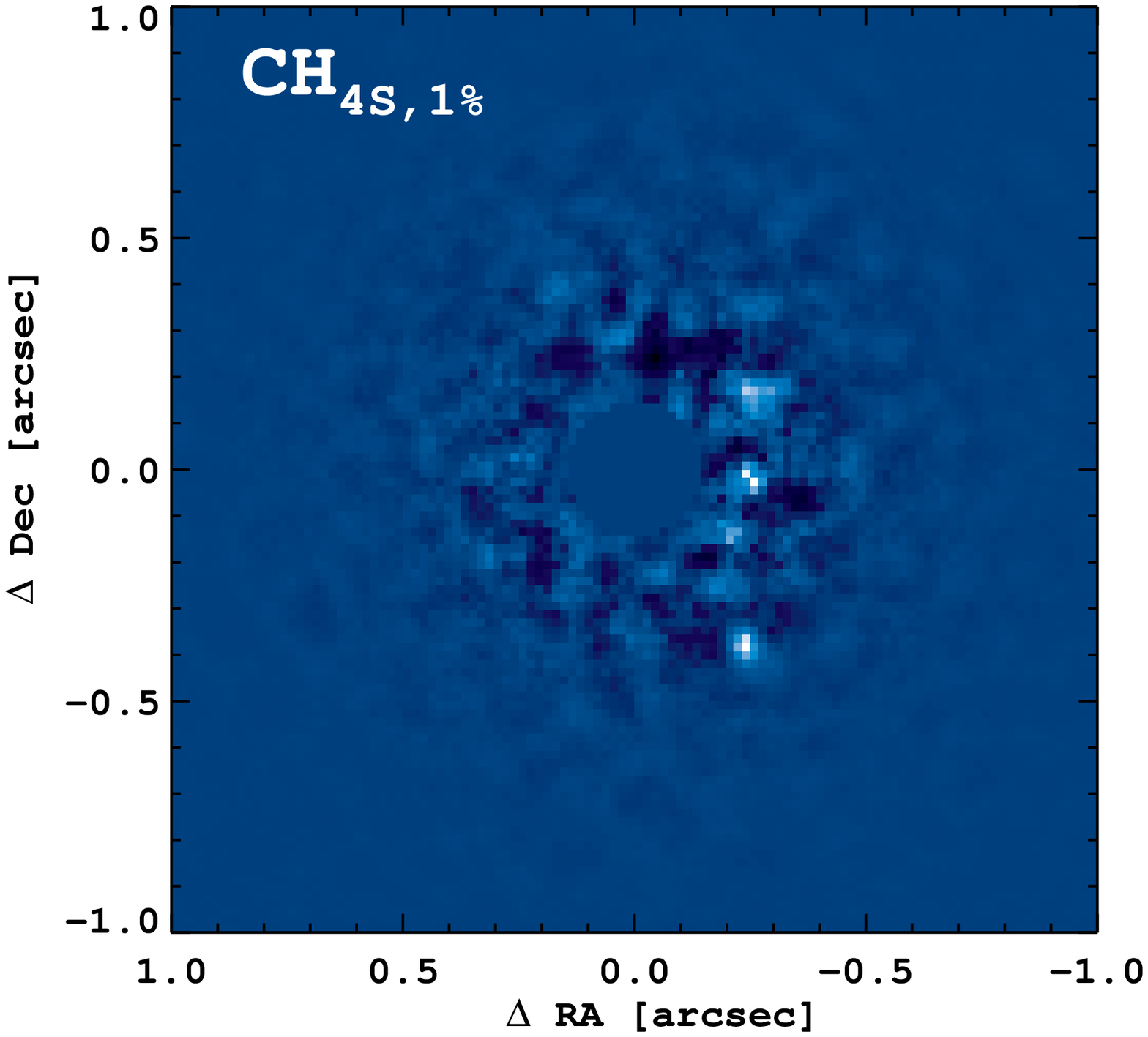}
\hspace{-1in}
\includegraphics[width=2.75in,angle=90]{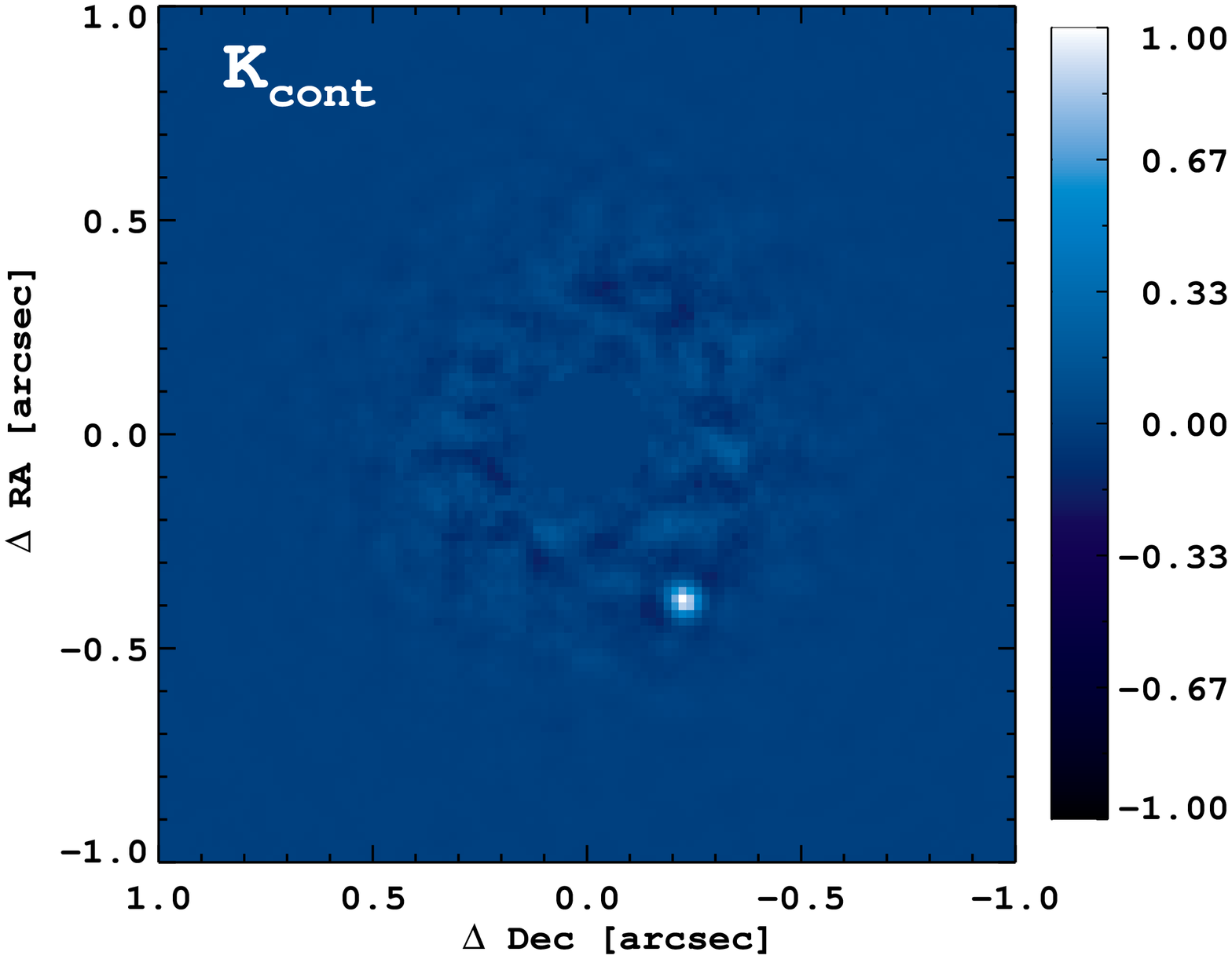}
\caption{NICI images of $\bPicb$.  The color scale is given at right, relative to the peak of the planet in each image.
\label{fig:nici}}
\end{sidewaysfigure}

The NICI detections of $\bPicb$ are presented in Figure \ref{fig:nici}.  To estimate the S/N of these detections, each pixel was divided by the standard deviation of the pixels in an annulus of width 5 pixels centered on the same separation.  No smoothing was applied.  A robust standard deviation algorithm was used, so the much higher pixels at the location of the planet were not included in the estimated noise.  The peak S/N in  $CH_{4S,1\%}$ was 10.4, in $K_S$ it was 6.8, and in the $K_{cont}$ filter it was 27.6.
  
NICI photometry was calibrated by injecting simulated planets.  The simulated planet signals were scaled from the star PSF, and were injected into the data at the same separation as the real planet.  The star was nonlinear in the $CH_{4S,1\%}$ and $K_{cont}$ science exposures, so we instead used short acquisition exposures appropriately scaled.  The simulated planets were injected into the data at twenty position angles opposite the planet, one at a time, and a complete reduction was carried out.  The stellar PSF halo was occasionally saturated near the radius of the planet in the $K_S$ observations, and such frames were removed from the reduction process.

We used aperture photometry with an aperture radius of 3 pixels.  ADI self-subtraction was corrected using the difference between the injected and recovered flux of the simulated planets.  We estimated the uncertainty in the measured flux as the standard deviation of the recovered fluxes.  For $CH_{4S,1\%}$ and $K_{cont}$ we estimated an additional uncertainty of $8\%$ due to variability of the star peak.

These observations were made using a $0.22''$ radius coronagraphic focal plane mask. The $0.22''$ mask opacity falls to zero at $0.4''$ from the center of the mask, so we have to correct for the mask opacity at the location of $\bPicb$. As described in \citet{2011ApJ...729..139W} for the $0.32''$ mask, we measured the $0.22''$ mask opacities using a pair  of 2MASS stars, both off the mask, and then one under the mask at different separations from the center. The mask opacities in the $K_S$, $K_{cont}$ and $CH_{4S,1\%}$ bands were $5.03\pm0.03$, $5.05\pm0.03$ and $5.47\pm0.09$ mags, respectively.  Typically, more than 5 measurements were taken at each position of the target under the mask, so that our opacity measurement uncertainties should be well-estimated \citep[see][]{2011ApJ...729..139W}.  The sampling of separations are sparse (in steps of $0.1''$) and any discontinuities could introduce systematics into our estimates, which are based on fitting a smooth curve through the opacities as a function of separation. However, we have no evidence that such discontinuities exist, and the uncertainties are estimated with this assumption.

Contrasts were calculated as $-2.5 \log (\mbox{star counts}/\mbox{planet counts}) + \mbox{mask opacity} $.  The uncertainties described above were added in quadrature.  We measured contrasts of  $\Delta CH_{4S,1\%} = 9.65 \pm 0.14$ mag, $\Delta K_S = 8.92 \pm 0.13$ mag, and $\Delta K_{cont} = 8.23 \pm 0.14 $ mag.

Astrometry from NICI, as well as an analysis of all available astrometry, is presented in Nielsen et al. (2014, submitted).

\begin{wrapfigure}[16]{r}{4in}
\vspace{-.3in}
\centering
\includegraphics[width=2.75in,angle=90]{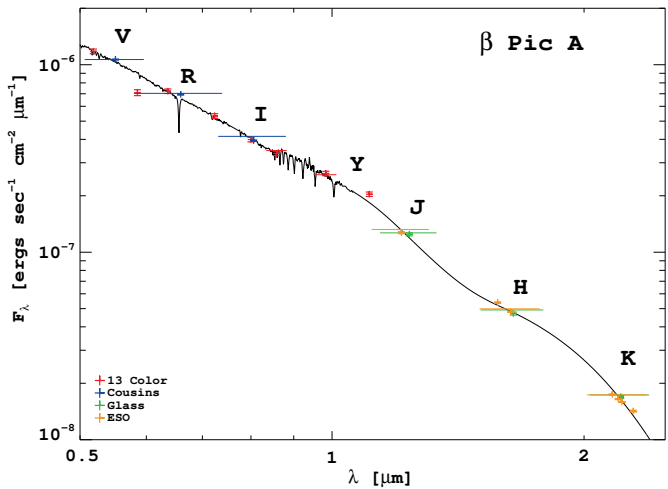}
\vspace{-.15in}
\caption{The SED of $\beta$ Pic A.  The synthetic spectrum was normalized to the $V$ photometry only.  We assume that the lack of reddening also holds for the planet.  Also note the lack of significant variability implied by these different measurements.
\label{fig:bpicA_SED}}
\end{wrapfigure}

\vspace{-.2in}
\subsection[The SED of beta Pic A]{The SED of $\beta$ Pic A}

We used archival photometry of $\beta$ Pictoris A to find its brightness in the filters used here, and also to investigate whether there is any reddening, which could affect our $\lambda<1\mu$m photometry.  We used $V,R_C,I_C$ photometry from \citet{1980SAAOC...1..234C} and \citet{1980SAAOC...1..166C}.  Especially useful was photometry in the 13-color system, which spans $0.33\mu$m to $1.10\mu$m in somewhat narrow passbands, from \citet{1969CoLPL...8....1M} and \citet{1975RMxAA...1..299J}.    In the IR we used Johnson-Glass $J,H$, and $K$ from \citet{1974MNSSA..33...53G}, and ESO broadband $J,H,K$ and narrowband $H_0$, $Br_\gamma$, $K_0$, and $CO$ photometry from  \citet{1996A&AS..119..547V}.  Finally, following \citet{2013arXiv1302.1160B} we synthesize an A6V spectrum by averaging the A5V and A7V spectra in the Pickles Spectral Atlas \citep{1998PASP..110..863P}.  We then normalized this spectrum to Cousins $V$.  We compare this spectrum to the photometry in Figure \ref{fig:bpicA_SED}.  Based on the good agreement we conclude that reddening is not significant for $\beta$ Pic A, and assume that this will also be true for the planet (though there could be circumplanetary reddening that this analysis would miss).   This exercise also demonstrates that variability of the primary star is not a concern when using it as a photometric reference for AO observations.

The central wavelength of the `99' filter of \citet{1969CoLPL...8....1M} is nearly identical to VisAO $Y_S$, so we adopt their measurement of $Y_S=3.561\pm0.035$ mag for $\bPicA$.  Using the interpolated A6V spectrum we find that $CH_{4S,1\%} - H_{ESO} = 0.024 \pm 0.05$, so we have $CH_{4S,1\%} = 3.526 \pm 0.05$ mag for $\bPicA$ \citep{2013arXiv1302.1160B}.  We find $K_{S,NICI} - K_{ESO} = -0.026 \pm 0.05$  so we use $K_{S,NICI} = 3.468\pm0.05$ mag.  We find $K_{cont}-K_{ESO}=0.068\pm0.05$ mag for an A6V, so for $\bPicA$ we have $K_{cont} = 3.563 \pm 0.05$ mag.  We combine these with our contrast measurements, adding uncertainties in quadrature. We present our new photometry of $\bPicb$ in Table \ref{tab:bpicparams}, along with prior measurements in $J$, $H$, and $K$.  Measurements at longer wavelengths are considered in Paper II.

\begin{table}
\scriptsize
\caption{$YJHK$ Photometry of $\beta$ Pictoris b \label{tab:bpicparams}}
\vspace{-.25in}
\begin{center}
\begin{tabular}{cccccl}
\hline
\hline
Filter &  Instrument & Date        & Apparent       &      Absolute      &  Notes \\
       &             &             & Magnitude      &     Magnitude      &   \\
\hline
\vspace{3pt}\\
$Y_S$  &    VisAO    &  2012/12/04 & $15.53^{+0.34}_{-0.33}$   &   $14.09^{+0.34}_{-0.33}$     &  [1]\\
\\
$J$           &  NACO        &  2011/12/16 & $14.0\pm0.3$ & $12.6\pm0.3$ & [2] \\
           &     ---           &     ---         & $14.11\pm0.21$ & $12.68\pm0.21$ & [3] \\
\\
$CH_{4S,1\%}$ & NICI &  2011/10/20 & $13.18\pm0.15$ & $11.74\pm0.15$ & [1]\\
\\
$H$    &  NACO       & 2012/01/11  & $13.5\pm0.2$ & $12.0\pm0.2$  & [2]\\ 
       &   ---       &   ---       & $13.32\pm0.14$ & $11.89\pm0.14$  & [3]\\
       &  NICI       & 2013/01/09  & $13.25\pm0.18$ & $11.82\pm0.18$  & [3]\\      
\\
$K_S$ &  NACO        & 2010/04/10  & $12.6\pm0.1$   & $11.2\pm0.1$  & [4]\\ 
      &  NICI        & 2010/12/25  & $12.39\pm0.14$ & $10.94\pm0.14$ & [1]\\
      &  NICI        & 2013/01/09  & $12.47\pm0.13$ & $11.04\pm0.13$  & [3]\\
\\
$K_{cont}$ & NICI    & 2011/10/20  & $11.79\pm0.15$ & $10.34\pm0.15$ & [1]\\
\\
\hline
\multicolumn{6}{l}{Notes: [1] this work, [2] \citet{2013arXiv1308.3859B}, [3] \citet{2013ApJ...776...15C}, [4] \citet{2011AA...528L..15B}}\\
\end{tabular}
\end{center}
\end{table}

\vspace{-.25in}
\section{Analysis \label{sec:anal}}

We now analyze the combined $Y_S,J, H$, and $K_S$ photometry of $\bPicb$.  Our analysis is based on a library of over 500 field BDs, low-g BDs, and low-mass companions, which we describe in Appendix \ref{synphot}.  

\vspace{-.2in}
\subsection{Color-Color and Color-Magnitude Comparisons}

\begin{figure}
\includegraphics[width=3in,clip=true,angle=90]{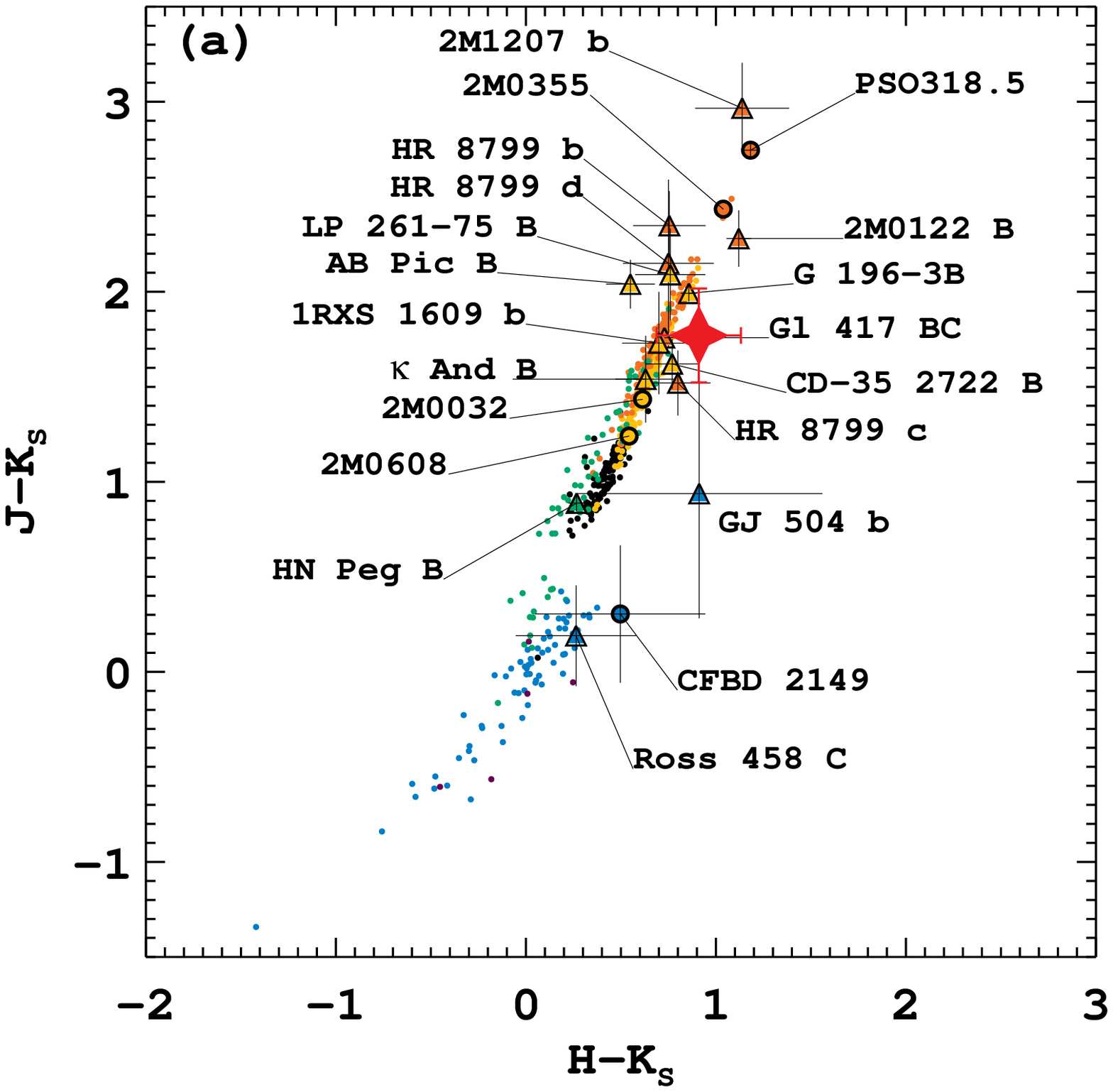}
\hspace{-1.in}
\includegraphics[width=3in,clip=true,angle=90]{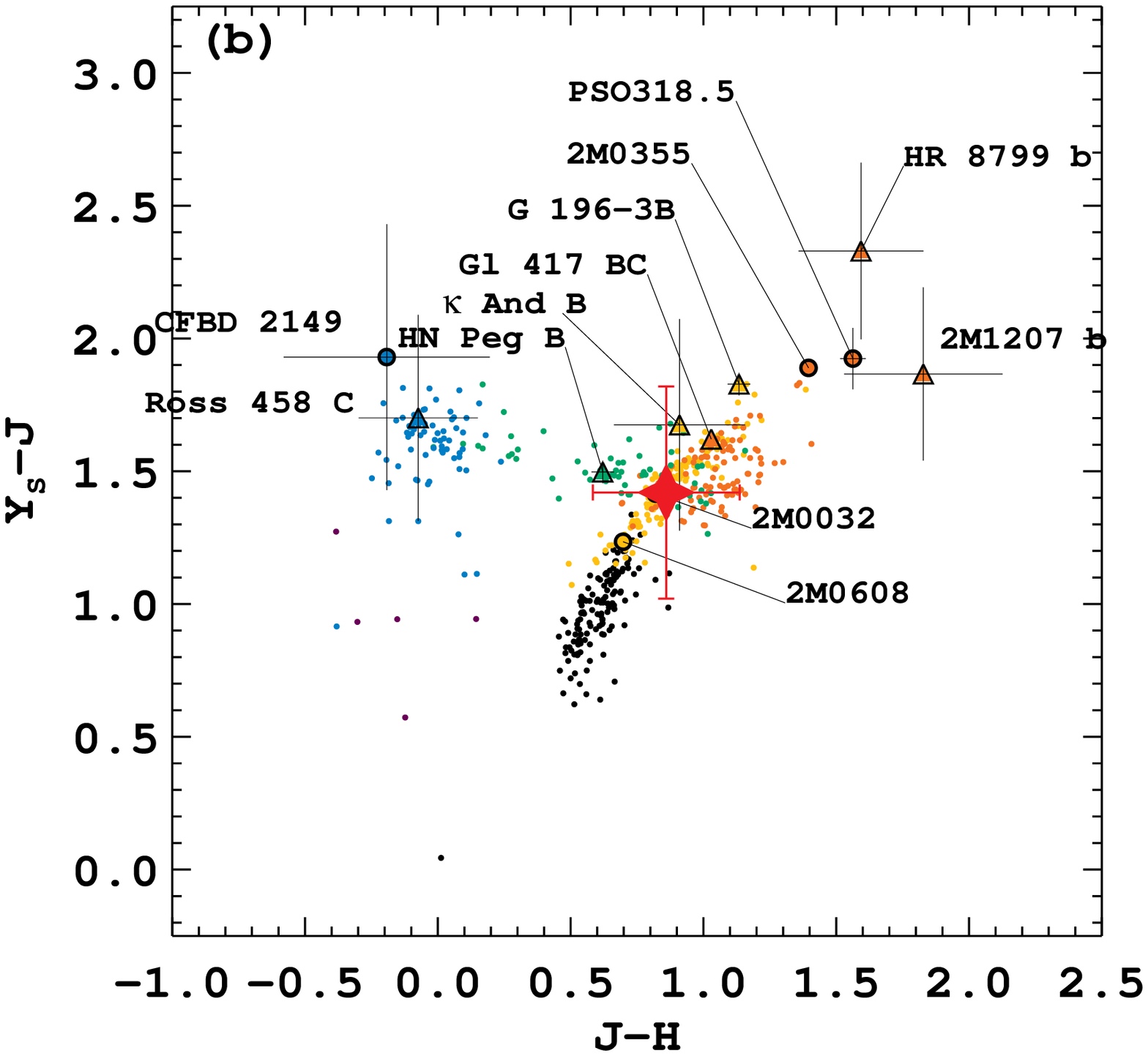}
\includegraphics[width=3in,clip=true,angle=90]{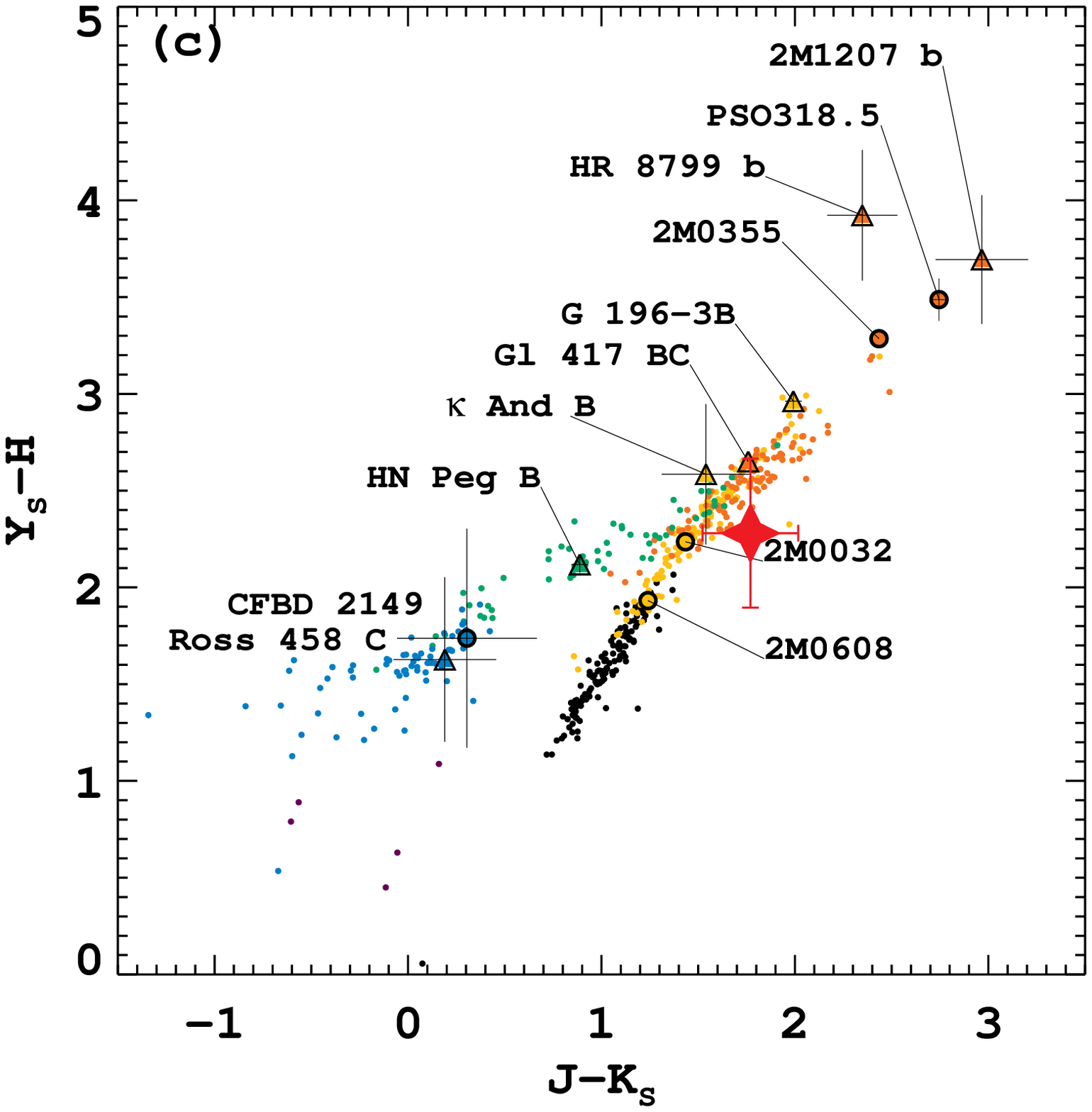}
\hspace{-1in}
\includegraphics[width=3in,clip=true,angle=90]{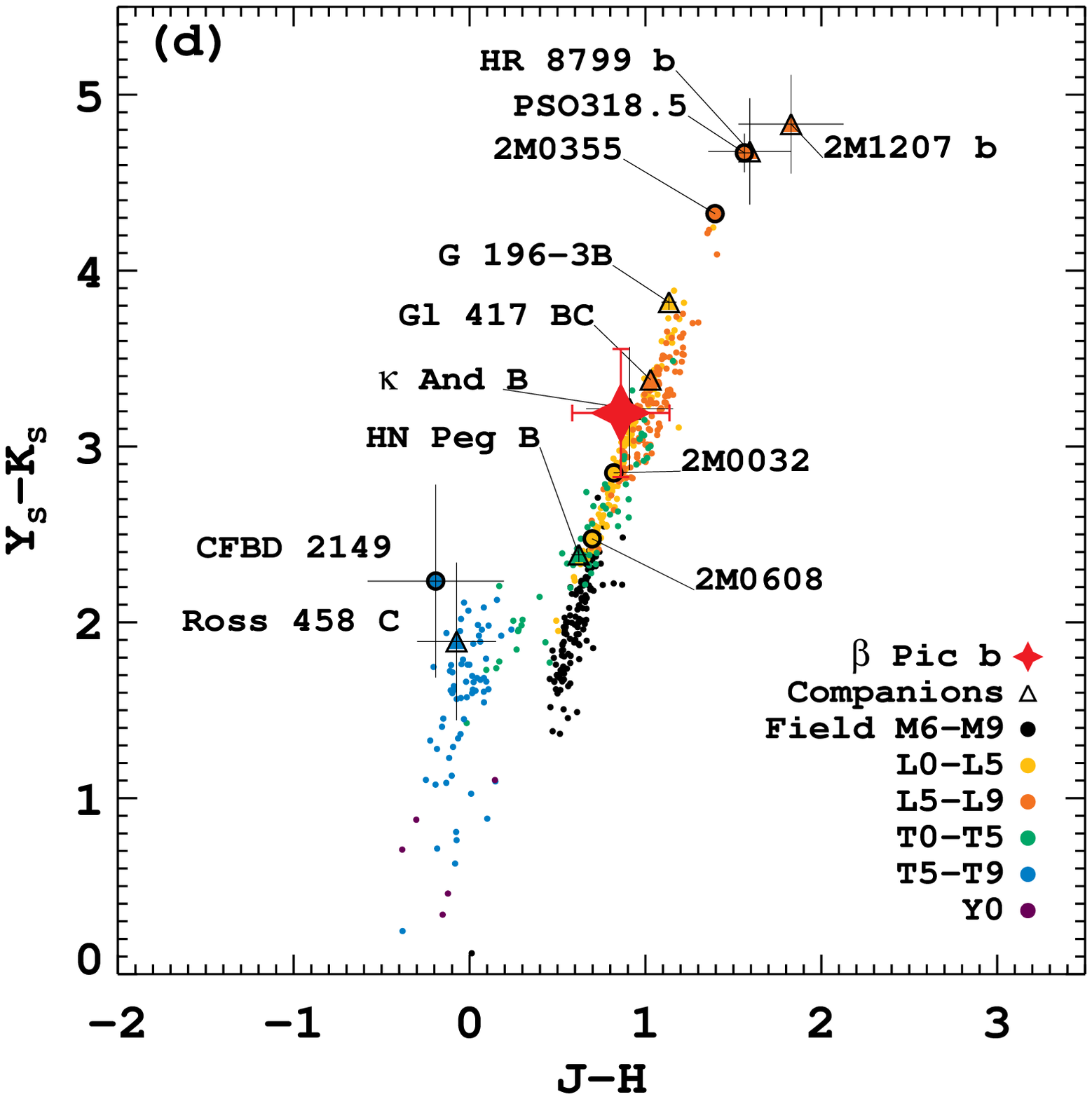}
\caption{(a): $J-K_S$ vs $H-K_S$ colors of $\bPicb$  compared to field BDs, low-g BDs, and companions.  (b),(c), and (d):  Various permutations in color-color space including $Y_S$ photometry.   We see that $\bPicb$ has colors very similar to L0-L5 field dwarfs, though within the $1\sigma$ uncertainties it is consistent with an early T dwarf.  These diagrams emphasize the diversity of low-mass, low-temperature companions.  In (d) we see that $\bPicb$ is separated from HR 8799 b and 2M1207 b by 2 magnitudes in $Y_S-K_S$ color, and 2M1207 b is separated from the coolest Y dwarfs by nearly 5 magnitudes. 
\label{fig:ccdiagrams}}
\end{figure}

\begin{figure}[p!]
\begin{center}
\includegraphics[width=3.in,angle=90]{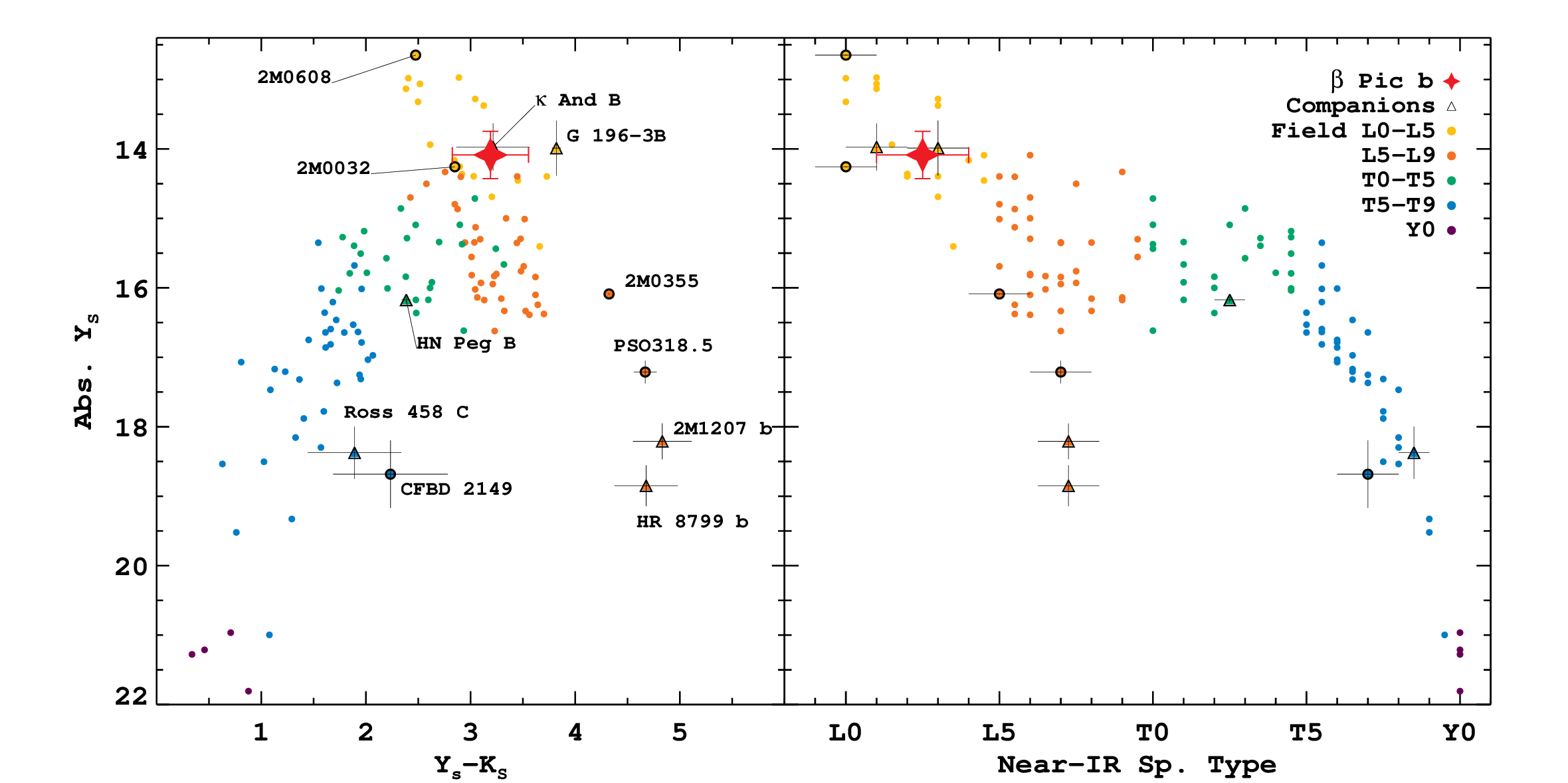}
\includegraphics[width=3.in,angle=90]{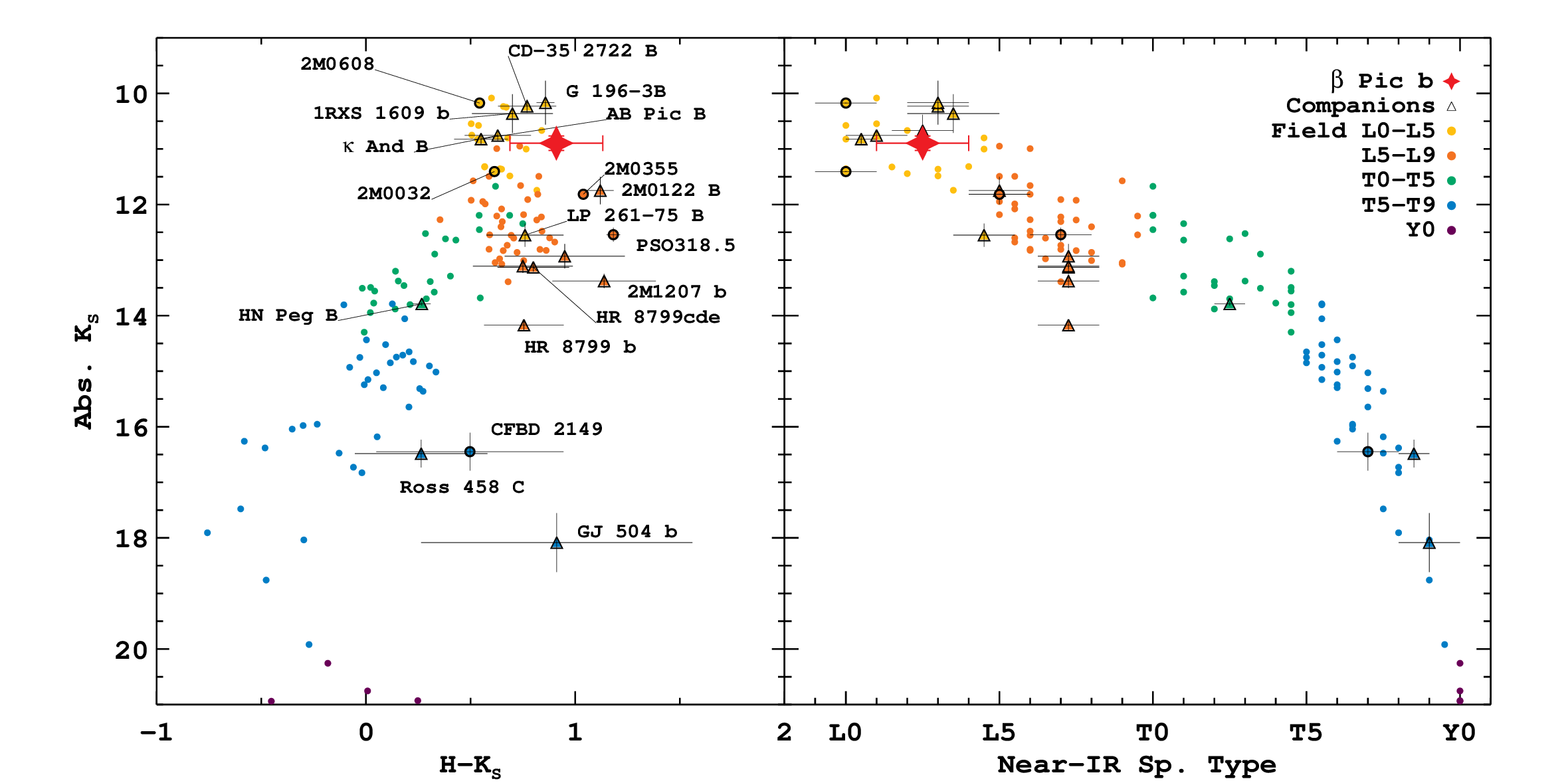}
\end{center}
\caption{Color-magnitude diagrams.  In the left hand panel of each plot we show absolute magnitude vs.\ color.   $\bPicb$ falls on or near the early L dwarf locus. Though it is somewhat red in $H-K_S$ it is $1\sigma$ consistent with the field.  In the right hand panels we show the same data organized by SpT.  We plot $\bPicb$ with a SpT of L$2.5\pm1.5$ based on our analysis of its colors and absolute magnitudes.  The HR 8799 EGPs and 2M1207 b are plotted with SpTs of L7.25 based on their locations in the color-magnitude plots.
\label{fig:cmds}}
\end{figure}

In Figure \ref{fig:ccdiagrams} we compare $\bPicb$ to field BDs and other low-mass companions in color-color plots.  We show $J-K_S$ vs $H-K_S$ colors, where nearly all of the companions have photometry (HR 8799 e is the notable exception in $J$).  $Y$ band is relatively unexplored territory for low-mass companions,  so we present three different permutations of $Y_S$ colors.  The field objects are plotted in the NICI system using synthetic photometry as described in Appendix \ref{synphot}.  We found that conversions between photometric systems are typically $<0.1$ mags, especially for spectral type L (see also \citet{2004PASP..116....9S}), so in Figure \ref{fig:ccdiagrams} the companion objects are plotted in the $J$, $H$, and $K_S$ photometric system in which they were observed.    We used available spectra to estimate $Y_S$ photometry for the companions shown.  For $\bPicb$ we plot our new $K_S$ point from NICI, and the NICI $H$ point from \citet{2013ApJ...776...15C}.  In all cases $\bPicb$ has colors consistent with early to mid L-dwarfs.  It is also consistent with an early T, a consequence of the blueward progression at the L/T transition.    

The color degeneracy of the L/T transition is broken by absolute magnitude.  In Figure \ref{fig:cmds} we show $Y_S$ and $K_S$ color-magnitude diagrams.  These plots firmly place $\bPicb$ in the early L-dwarfs.  As noted by \citet{2013arXiv1302.1160B} and \citet{2013ApJ...776...15C} it is somewhat redder than the L dwarfs in $K_S$ vs $H-K_S$.  This is consistent with other low surface-gravity L dwarfs such as AB Pic B, 2M0122B, and PSO318.5, though within $1\sigma$ $\bPicb$ is consistent with the field.

\vspace{-.2in}
\subsection{Spectral Fitting}

\begin{wrapfigure}[17]{r}{4in}
\vspace{-.25in}
\centering
\includegraphics[width=2.75in,angle=90]{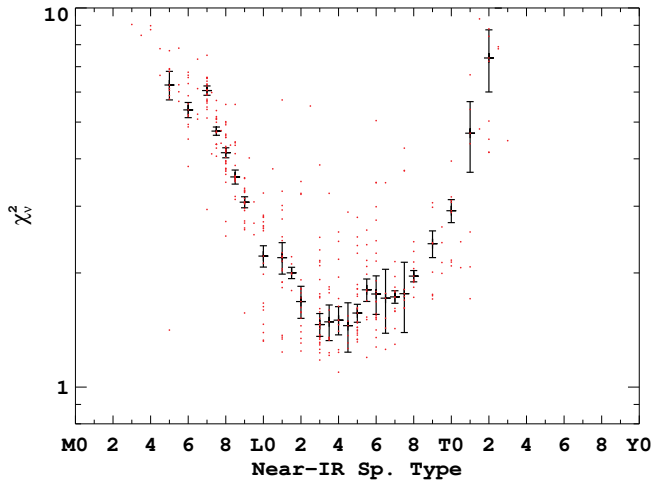}
\vspace{-.2in}
\caption{Reduced $\chi^2$ from fitting $\bPicb$ photometry to field BD spectra.  The red points are individual objects, and the black points indicate the median and standard deviation in each SpT. The apparent minimum in the early to mid L dwarfs gives a range of L2-L5 for the spectral type of $\bPicb$. 
\label{fig:spectype_fit}}
\end{wrapfigure}

We next fit each of 499 BD spectra collected from various sources to the $Y_S$ through $K_{cont}$ photometry of $\bPicb$ (treating each measurement independently).  We did this by computing the flux of each spectrum in each of the bandpasses, then finding the single scaling factor which minimized $\chi^2$.  Only spectra with a complete $Y$ band measurement were used, which tends to select for high S/N.  We adopted an error of 0.05 mag for our synthetic photometry in each bandpass based on the findings of \citet{2013arXiv1310.0457L}.  This was added in quadrature to the $\bPicb$ measurement errors.  The results are presented in Figure \ref{fig:spectype_fit}, where we show $\chi^2_{\nu}$ vs.\ SpT for each of the field objects, as well as the median in each SpT.  The error bars indicate the standard deviation in each SpT.  There is a minimum in the early to mid L dwarfs, giving a range of L2 to L5.

\begin{wrapfigure}[26]{r}{4in}
\vspace{-.35in}
\centering
\includegraphics[width=5in,angle=90]{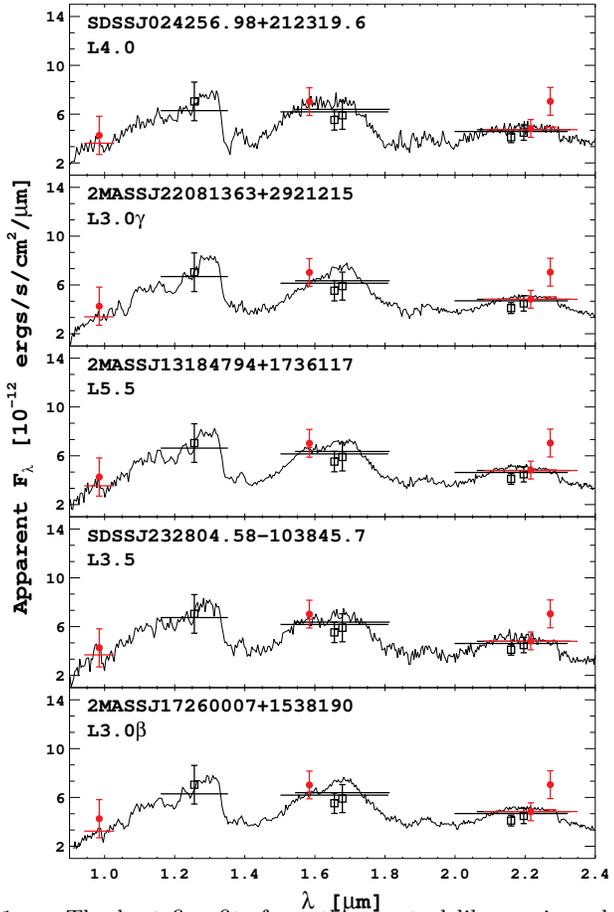}
\vspace{-.25in}
\caption{The best five fits from the spectral library, in order of $\chi^2_{\nu}$ from top to bottom.  The best single match is an L4, but all of these have similar values of $\chi^2_{\nu} \sim 1$.  The red circles are our new measurements, and the black open squares are prior measurements.   
\label{fig:spectfit_best5}}
\end{wrapfigure}

In Figure \ref{fig:spectfit_best5} we show the 5 best fitting spectra, which range from L3 to L5.5.   Our new $Y_S$ photometry is well fit by these spectra, and our new $CH_{4S,1\%}$ and $K_S$ point are consistent with the prior measurements.  The $K_{cont}$ point is only consistent with prior measurements at the $2\sigma$ level, though.  A possible culprit is the noted non-linearity of the star, which may have biased the result.  However, the same procedure was applied to the  $CH_{4S,1\%}$ measurement from the same night. Based on the many measurements of $\bPicA$ we discount the primary as a source of variability at this level. BDs are known to be variable \citep[e.g.][]{2012ApJ...760L..31B} but typically with low amplitude for early L dwarfs \citep{2008A&A...487..277G}.  We do not have enough evidence to speculate other than to note variability in $\bPicb$ as a possible explanation.

Though it established that the $Y$ through $K$ photometry of $\bPicb$ does not appear to be atypical compared to the field, this exercise ultimately does not yield a well-determined SpT.

\subsection{Near-IR SpT}

\begin{figure}
\includegraphics[width=3.6in]{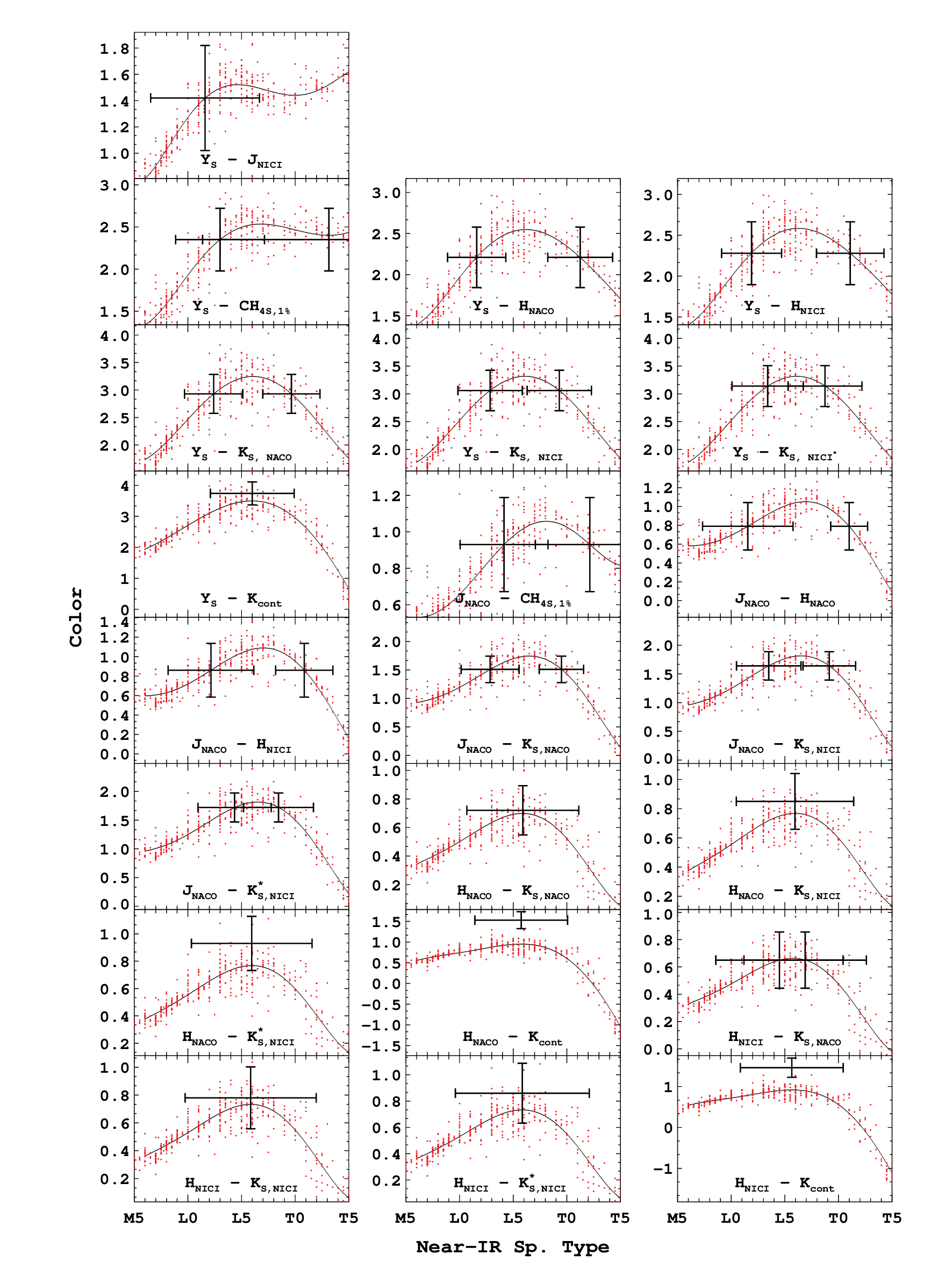}
\includegraphics[width=2.4in]{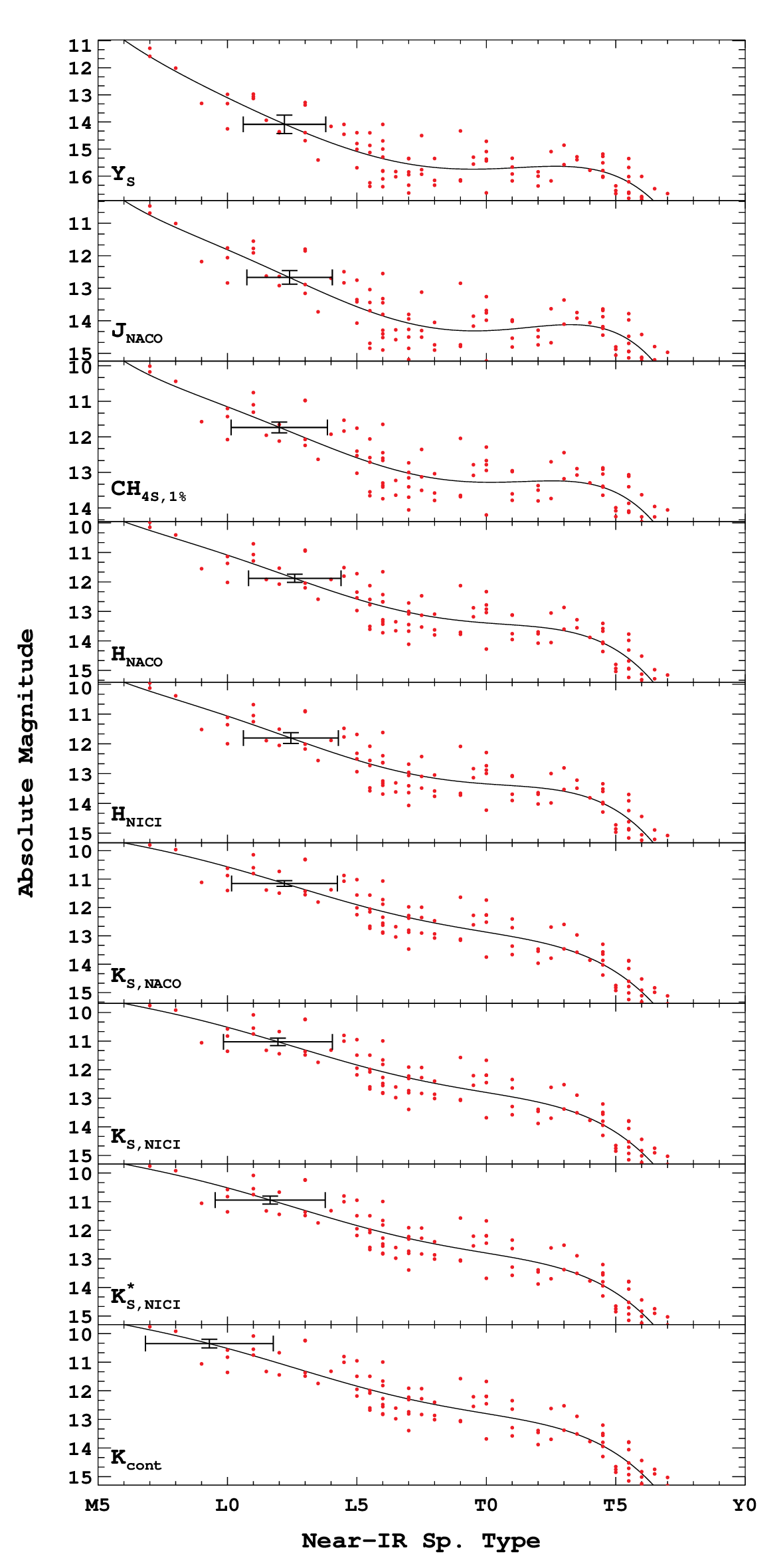}
\vspace{-0.in}
\caption{Left: A color by color comparison of $\bPicb$ to the field BDs, with each possible combination of the available photometry.  The L/T transition causes the resultant SpT estimate to be dual-valued.  As we saw in Figures \ref{fig:ccdiagrams} and \ref{fig:cmds} the planet is modestly redder than the field in the redder colors, especially $H-K$.  Right: A filter by filter comparison of the absolute magnitude of $\bPicb$ to field BDs.  For the early to mid L dwarfs the sequence is single valued in SpT.  In terms of its absolute magnitudes $\bPicb$ is quite clearly an early L.  
\label{fig:sptype_col}}
\end{figure}

Having established that the photometry of $\bPicb$ places it in the early to mid L dwarfs, we next attempt to estimate its SpT.  Spectral typing is usually done by comparing the spectral morphology of the object to spectral standards.   Since there is not yet a measurement of the $Y$ through $K$ spectrum of $\bPicb$, we instead compare its broadband photometry to field objects.  We first plot color vs.\ SpT for the 499 spectra in our library, shown in Figure \ref{fig:sptype_col}.  For each color  we fit a polynomial to the sequence, and then find the root(s) of this polynomial corresponding to the photometry of $\bPicb$.  A consequence of the L/T transition is that the BD SpT sequence is dual valued in each permutation of the $YJHK$ colors, hence the double values in most of the panels of Figure \ref{fig:sptype_col} (and the broad minimum in Figure \ref{fig:spectype_fit}).  The $H-K_S$ color measurements do not intersect the polynomial, though they are consistent with field dwarfs.  For these $H-K_S$ measurements we assign the SpT of the turnover in color vs.\ SpT. 

The bluer colors of $\bPicb$ (e.g. $Y_S-J$, $J-H$) indicate an SpT of early-L or mid-T. The redder colors --- especially $H-K_S$ --- favor mid-L. To break the degeneracy from colors we turn to absolute magnitudes.  In similar fashion, we fit polynomials to the objects in our library with parallaxes.  These fits, shown at right in Figure \ref{fig:sptype_col} are single valued, and all favor an SpT of early-L.  This paints a picture of $\bPicb$ as having the luminosity of an early L dwarf but being somewhat redder than typical for the field, much like many well studied low-gravity field BDs and companions.

We next compared all of these SpT determinations.  It has been shown that low-g BDs do not follow the field BD sequence in near-IR absolute magnitudes \citep{2013AJ....145....2F,2013AN....334...85L}, so we must be cautious about using the absolute magnitudes of $\bPicb$ to directly estimate SpT.  However, according to \citep{2013AN....334...85L} in the SpT range of roughly L0-L3 the absolute J-band magnitudes of low-g objects do match the field.  We also note that for spectral types later than L3, the absolute J-band magnitudes tend to be fainter than the field for low-g objects, which does not appear to be true for $\bPicb$.  The absolute magnitudes of $\bPicb$ all correspond to the early-L dwarf SpTs, and the weighted mean using just the absolute magnitudes is L$2.1\pm0.8$.  Turning to the colors next, we reject the T classifications based on the absolute photometry.  We form a best estimate of SpT by averaging all of the L values, finding L$3.6\pm2.2$ using only the colors --- as expected from the range we derived from using Figure \ref{fig:spectype_fit}.

Based on the consistency and the findings of \citet{2013AN....334...85L}, we give somewhat more weight to the type estimate derived from absolute magnitudes than the color derived estimate.  We also take into account that low-g objects tend to be red compared to the field for their spectral types. Based on all of these considerations we adopt L$2.5\pm1.5$ as the near-IR spectral type of $\bPicb$.  This is consistent with \citet{2010ApJ...722L..49Q}, who found an approximate spectral type of L4, with a range of L0-L7, using $L'$ and narrow-band 4.05 $\mu$m color. Our estimate is also consistent with the spectral type of L2$\gamma \pm2$ estimated by \citet{2013arXiv1302.1160B}, and the range of L2-L5 given by \cite{2013ApJ...776...15C}.

\afterpage{\clearpage}

\subsection{Physical Properties}

Armed with our SpT estimate, we next estimated the bolometric luminosity of $\bPicb$.  We first convert to the MKO photometric system, using the relationships we derive in Appendix \ref{synphot}.  We converted each measurement independently, then calculated the uncertainty-weighted mean for each bandpass.  The results are presented in Table \ref{tab:mkomags}.  From there, we determined the BC for each of $J,H,K$ using the polynomials given in \citet{2010ApJ...722..311L}.  We then took the uncertainty-weighted mean value of the bolometric magnitude and its uncertainty, also given in Table \ref{tab:mkomags}.    The resulting uncertainty in $M_{bol}$ is a relatively small 0.11 mag.  There are likely additional systematic errors, for example from using the field relationships on a low-g object.

\begin{table}
\small
\caption{MKO photometry, bolometric corrections, and bolometric magnitudes for $\bPicb$. \label{tab:mkomags}}
\begin{center}
\begin{tabular}{lccc}
\hline
\hline
Band         &       Abs. Mag.       &         BC      &      $M_{bol}$    \\
\hline
$Y_{MKO}$    &   $13.78 \pm 0.34$    &       ---         &        ---        \\
$J_{MKO}$    &   $12.61 \pm 0.21$    & $1.89 \pm 0.23$   &   $14.50 \pm 0.31$\\
$H_{MKO}$    &   $11.87 \pm 0.11$    & $2.67 \pm 0.10$   &   $14.54 \pm 0.14$\\
$K_{MKO}$    &   $11.00 \pm 0.07$    & $3.33 \pm 0.08$   &   $14.33 \pm 0.11$\\
Mean:        &       ---             &       ---         &   $14.40 \pm 0.11$\\
\hline
\end{tabular}
\end{center}
\end{table}

Converting to luminosity gives us $\log (L_{bol}/L_\sun ) = -3.86 \pm 0.04$ dex.  This is very close to the value of $-3.87\pm0.08$ dex by \citet{2013arXiv1302.1160B}.  They  used a somewhat different method, determining only the K band based $M_{bol}$ by using the mean BC of two similar field dwarfs. They also reported $-3.83\pm0.24$ dex based on model atmosphere fitting, including longer wavelengths.  Our value is modestly fainter than the estimate of \cite{2013ApJ...776...15C}, $-3.80\pm0.02$ dex, which was also inferred from model fitting including longer wavelength measurements.    

\begin{wrapfigure}[16]{r}{4in}
\vspace{-.35in}
\centering
\includegraphics[width=2.25in,clip=true,angle=90]{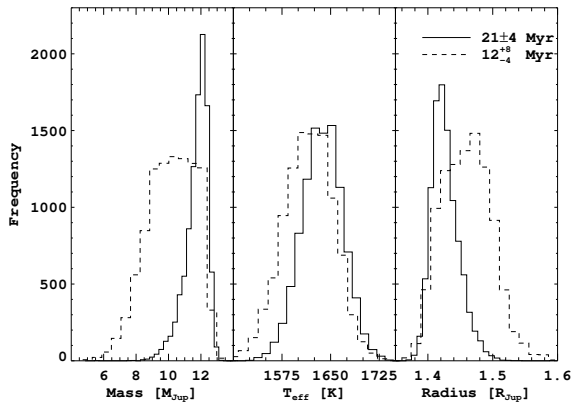}
\vspace{-0.05in}
\caption{Results of Monte Carlo experiments using our luminosity estimate, $\log(L_{bol}/L_{\sun})=-3.86\pm0.04$, a $\beta$ Pic moving group age of $21\pm4$ Myr (solid line) and $12^{+8}_{-4}$ Myr (dashed line), and hot start evolutionary models  \citep{2000ApJ...542..464C, 2003A&A...402..701B}.  Note the negative skewness in mass. We can place an upper limit at $M\approx13$ $M_{Jup}$.  
\label{fig:hotstart}}
\end{wrapfigure}

We next calculated estimates of mass, temperature, and radius with Monte Carlo trials based on the hot start evolutionary models  \citep{2000ApJ...542..464C, 2003A&A...402..701B}.  We adopt the lithium depletion boundary age of $21\pm4$ Myr for the $\beta$ Pic moving group from \citet{2013arXiv1310.2613B}.  We treated our estimate for $\log(L_{bol})$ and this age estimate as Gaussian distributions, and drew $10^4$ trial values.  For each random pair of $\log(L_{bol})$ and age we calculated the hot start model prediction of mass, temperature, and radius by linear interpolation on the model grid.  The resulting distributions are shown in Figure \ref{fig:hotstart}, and summarized in Table \ref{tab:physpars}.  The mass distribution has mean and standard deviation $11.9 \pm 0.7$ $M_{Jup}$,  and median  12.0 $M_{Jup}$.  The mass distribution has negative skewness, and there appears to be an upper limit at $M\approx13$ $M_{Jup}$.   For temperature we find $T_{eff} = 1643 \pm 32$ K.  For radius we find $R = 1.43 \pm 0.02$ $R_{Jup}$ with a small positive skewness.  These very small uncertainties are the formal errors from our Monte Carlo analysis of the model grid, and do not include any additional terms such as the uncertainties in the models themselves.

Repeating the Monte Carlo experiment with the prior age estimate of $12^{+8}_{-4}$ Myr from \citet{2001ApJ...562L..87Z}, we found $M=10.5\pm1.5$ $M_{Jup}$, $T_{eff} = 1623\pm38$ K, and $R=1.47\pm0.04$ $R_{Jup}$.  To account for the asymmetric uncertainty we used the skew-normal distribution for age, choosing its parameters such that the mode was 12 Myr, the left-half-width was 4 Myr, and the right-half-width was 8 Myr.  The resulting 12 Myr distributions are also shown in \ref{fig:hotstart} and Table \ref{tab:physpars}.  There is again an upper limit on mass at roughly $13$ $M_{Jup}$.  We surmise that this is due to our luminosity estimate being too faint to allow deuterium burning at these ages according to the models, hence the mass is strongly constrained to be below $13$ $M_{Jup}$ regardless of age.  Note that we consider additional models, using the complete 1-5$\mu$m photometry of $\bPicb$ in Paper II.

\citet{2013arXiv1302.1160B} arrive at an estimate $T_{eff} = 1700\pm100$ K from fitting PHOENIX based atmosphere models coupled with various cloud models \citep{2000ApJ...542..464C, 2001ApJ...556..357A, 2003A&A...402..701B}  and found $R = 1.4\pm 0.2$ $R_{Jup}$.  \cite{2013ApJ...776...15C} estimated a range of $1575-1650$ K using the various cloud + atmosphere models of \citet{2006ApJ...640.1063B}, \citet{2011ApJ...737...34M}, and \citet{2011ApJ...729..128C}, and estimated $R=1.65\pm0.06$ $R_{Jup}$. Both of these efforts assumed the previous age estimate of $12^{+8}_{-4}$ Myr.

\begin{table}
\footnotesize
\caption{Estimated Physical Properties of $\bPicb$. \label{tab:physpars}}
\centering
\begin{tabular}{cccccc}
\hline
\hline
 Near-IR                    & log($L_{bol}/L_{\sun}$)         &  Age           &     Mass     &  $T_{eff}$  &      Radius\\
Sp.\ Type                    &                                 & [Myr]          &  [$M_{Jup}$] &     [K]     &     $[R_{Jup}]$\\ 
\hline
\multirow{2}{*}{L2.5$\pm$1.5} & \multirow{2}{*}{-3.86$\pm$0.04} & $21\pm4$       & 11.9$\pm$0.7 & 1643$\pm$32 & 1.43$\pm$0.02  \\                    
                            &                                 & $12^{+8}_{-4}$ & 10.5$\pm$1.5 & 1623$\pm$38 & 1.47$\pm$0.04  \\
\hline
\end{tabular}
\end{table}

\vspace{-.25in}
\section{Discussion \label{sec:discuss}}

It has recently become clear that the SpT to $T_{eff}$ relationship for low-gravity objects is different from the relationship for the field \citep[e.g.][]{2013ApJ...774...55B,2013arXiv1310.0457L}.  Like other low-g objects, $\bPicb$ appears to be cooler than field BDs with the same SpT.  The SpT to $T_{eff}$ relationship of \citet{2009ApJ...702..154S}  gives an estimate  of $T_{eff} = 1904$ K for an L2.5 field brown dwarf, which is $\sim$$250$ K hotter than we calculated assuming hot start evolution for this young object.

\begin{wrapfigure}[19]{r}{4in}
\centering
\vspace{-0.4in}
\includegraphics[width=3.in,clip=true,angle=90]{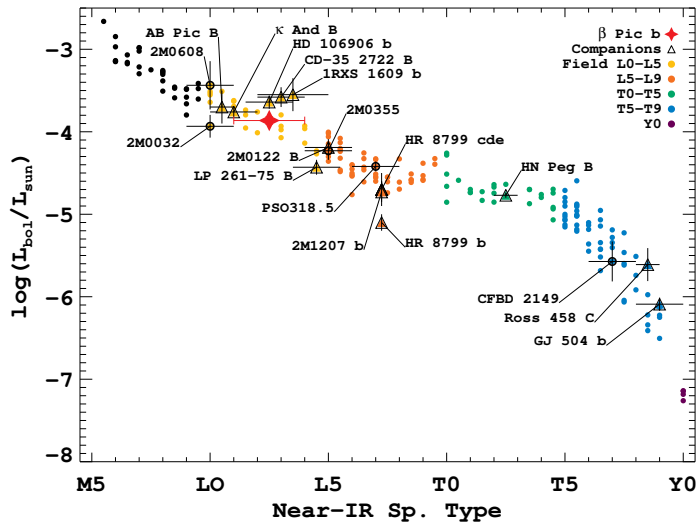}
\vspace{-0.4in}
\caption{Bolometric luminosity vs.\ SpT.  We expect a temperature sequence to produce a luminosity sequence under the assumption of nearly constant radius.  The young low-gravity BDs and EGPs also appear to fall on the same luminosity sequence as the field BDs despite their lower effective temperatures. The HR 8799 planets and 2M1207 b, objects which don't fit in the standard L spectral types, are plotted as L7.25. 
\label{fig:lumseq}}
\end{wrapfigure}

It was noted by \citet{2013arXiv1310.0457L} that low-g objects tend to have $L_{bol}$ consistent with the field for their SpT, despite their lower temperatures and red near-IR colors.  In Figure \ref{fig:lumseq} we show log$(L_{bol}/L_{\sun})$ vs.\ SpT for field brown dwarfs and low-g brown dwarfs and companions.  The field BDs are plotted using BCs from \citet{2010ApJ...722..311L}, and companions are plotted using published values of $L_{bol}$ (see Table \ref{tab:compobjs} for references).  Attempts have been made to use spectral indices and morphology to assign SpTs or SpT ranges to 2M1207 b and HR 8799 b \citep{2010A&A...517A..76P, 2010ApJ...723..850B, 2011ApJ...733...65B, 2013ApJ...772...79A}.  As evident in Figures \ref{fig:ccdiagrams} and \ref{fig:cmds}, however, these objects do not fit within the field brown dwarf sequence in either colors or absolute magnitudes.  Many authors have noted that 2M1207 b and the HR 8799 EGPs appear to be extensions of the L dwarf sequence  \citep{2008Sci...322.1348M, 2010ApJ...723..850B,2011ApJ...733...65B, 2011ApJ...737...34M}, and in Figure \ref{fig:lumseq} we have plotted them with a spectral type of L7.25 which appears to be a likely place to expand the L dwarf sequence to include these objects.  

Interestingly, $\bPicb$'s luminosity is consistent with the field for a spectral type of L2.5.  Like the other directly-imaged planets, $\bPicb$ is younger than field brown dwarfs of similar luminosity.  Its young age corresponds to a lower mass and larger radius, hence it has lower surface gravity.  Given its comparable luminosity, these properties imply a lower effective temperature.  However, in the case of $\bPicb$, these physical differences do not appear to have a large effect on its overall appearance in near-IR color-color and color-magnitude diagrams.  Though it is somewhat red in $H-K$, $\bPicb$'s colors and luminosity are otherwise consistent with field early-L brown dwarfs,  implying that it shares basic atmospheric properties with such objects.  We contrast this with other directly-imaged planets: the four HR 8799 planets and 2M1207 b are systematically fainter and redder than the field brown dwarf sequence \citep{2005A&A...438L..25C, 2008Sci...322.1348M, 2010Natur.468.1080M}.  

The difference appears to result from (1) the persistence of clouds at luminosities where field brown dwarfs have transitioned from cloudy to cloud-free \citep{2011ApJ...729..128C,2011ApJ...737...34M,2011ApJ...732..107S}, and (2) the absence of methane absorption at luminosities where field brown dwarfs are beginning to show strong methane absorption \citep{2011ApJ...733...65B,2011ApJ...735L..39B,2013arXiv1311.2085S}.  \citet{2012ApJ...754..135M} present a phenomenological explanation for why clouds may exist at higher altitudes in low-gravity objects.  The methane transition may occur on a different timescale, also as a result of a youth-based mechanism \citep{2013arXiv1311.2085S}.  The fact that $\bPicb$ has more ``normal'' near-IR colors separates it from the HR 8799 EGPs, 2M1207 b, and the unusually red low-gravity L dwarfs \citep{2013AJ....145....2F, 2013ApJ...774...55B, 2013arXiv1310.0457L}.

\vspace{-.25in}
\section{Conclusions \label{sec:conclude}}

We have presented the first high-contrast far-red optical observations of an EGP with MagAO's VisAO CCD camera, detecting $\bPicb$ in $Y_S$ at a contrast of $(1.63\pm0.49)\times10^{-5}$, at a separation of $0.470\pm0.010''$.  The VisAO detection has S/N = 4.1, and a conservative upper-limit on false alarm probability of 1.0\%.  We also present observations of $\bPicb$ in the near-IR made with the NICI instrument at the Gemini-South Telescope.  Combining our VisAO $Y_S$ and NICI $CH_{4S,1\%}$, $K_S$, and $K_{cont}$ photometry with previous measurements in $J,H$ and $K$, we estimated that $\bPicb$ has a spectral type of L$2.5\pm1.5$.  In color-color and color-magnitude plots, $\bPicb$ fits very well with other early-L dwarfs, perhaps being slightly redder in $H-K$.  

We used our spectral type estimate to evaluate the physical properties of $\bPicb$.  Using field brown dwarf bolometric corrections, we estimate $\log(L/L_\sun) = -3.86\pm0.04$ dex. This is consistent with previous estimates.  Using hot-start evolutionary models at an age of $21\pm4$ Myr,  our $L_{bol}$ measurement yields a mass estimate of $M=11.9 \pm 0.7$ $M_{Jup}$, with an upper limit at $M\approx13$ $M_{Jup}$ due to the model treatment of deuterium burning.  For temperature we find $T_{eff} = 1643 \pm 32$ K.  For radius we find $R = 1.43 \pm 0.02$ $R_{Jup}$. All of these results are consistent with those of prior studies.

If we instead used the field BD sequence to estimate temperature, we would find a $T_{eff}$ $\sim$$250$K hotter than expected from the evolutionary models.  The population of low surface gravity ultracool dwarfs and directly-imaged EGPs likewise have low effective temperatures compared to field brown dwarfs of similar spectral type. However, these objects tend to be very red in near-IR colors, and so don't follow the field brown dwarf sequence in color-magnitude diagrams.  In contrast to other directly-imaged young EGPs (such as HR 8799 b and 2M 1207 b), $\bPicb$ looks much more like a typical early L dwarf in the near-IR, both in terms of its colors and luminosity, despite its inferred low gravity and cooler temperature.
  
\vspace{-.25in}
\section {Acknowledgements}

We thank the anonymous referee for many helpful comments and suggestions which greatly improved this manuscript.  J.R.M. is grateful for the generous support of the Phoenix ARCS Foundation. J.R.M and K.M.M. were supported under contract with the California Institute of Technology (Caltech) funded by NASA through the Sagan Fellowship Program. L.M.C.'s research was supported by NSF AAG and NASA Origins of Solar Systems grants. A.J.S. was supported by the NASA Origins of Solar Systems Program, grant NNX13AJ17G.  The MagAO ASM was developed with support from the NSF MRI program.  The MagAO PWFS was developed with help from the NSF TSIP program and the Magellan partners. The Active Optics guider  was developed by Carnegie Observatories with custom optics from the MagAO team. The VisAO camera and commissioning were supported by the NSF ATI program. We thank the LCO and Magellan staffs for their outstanding assistance throughout our commissioning runs. We also thank the teams at the Steward Observatory Mirror Lab/CAAO (University of Arizona), Microgate (Italy), and ADS (Italy) for their contributions to the ASM.  

The NICI campaign was supported in part by NSF grants AST-0713881 and AST-0709484 awarded to M. Liu, NASA Origins grant NNX11 AC31G awarded to M. Liu, and NSF grant AAG-1109114 awarded to L. Close.  The Gemini Observatory is operated by the Association of Universities for Research in Astronomy, Inc., under a cooperative agreement with the NSF on behalf of the Gemini partnership: the National Science Foundation (United States), the Science and Technology Facilities Council (United Kingdom), the National Research Council (Canada), CONICYT (Chile), the Australian Research Council (Australia), CNPq (Brazil), and CONICET (Argentina). 

This work made use of the General Catalogue of Photometric Data\footnote{{\url http://obswww.unige.ch/gcpd/}} \citep{1997A&AS..124..349M}. This research has benefited from the SpeX Prism Spectral Libraries, maintained by Adam Burgasser\footnote{{\url http://pono.ucsd.edu/~adam/browndwarfs/spexprism}}.  We thank Davy Kirkpatrick for providing WISE spectra, Jackie Faherty for providing the 2M0355 spectrum, Travis Barman for the 2M1207b model, and Sasha Hinkley and Laurent Pueyo for the $\kappa$ And B spectrum.  We made use of the Database of Ultracool Parallaxes maintained by Trent Dupuy\footnote{{\url https://www.cfa.harvard.edu/\textapprox tdupuy/plx/Database\_of\_Ultracool\_Parallaxes.html}}. This research has made use of the SIMBAD database, operated at CDS, Strasbourg, France, and NASA's Astrophysical Data System. 

\vspace{-.25in}
\bibliographystyle{apj}
\footnotesize
\bibliography{bpicb}

\normalsize

\appendix
\section{Synthetic Photometry and Conversions} 
\label{synphot}
In this appendix we provide details of our synthetic photometry.   The primary purpose of this is to verify the methodology used for our analysis, but we also determine transformations between various filter systems used in brown dwarf and exoplanet imaging which may be useful to others.  

\subsection{Filters}

\begin{wraptable}[12]{r}{4in}
\vspace{-15pt}
\scriptsize
\caption{Atmospheres. \label{tab:atmospheres}}
\centering
\begin{tabular}{lcccc}
\hline
\hline
System          &  Model                 &  Airmass     &  PWV (mm)  &  Notes      \\
\hline
\hline
VisAO           &  ATRAN, Cerro Pachon   &  1.0         &  2.3       & 1,2 \\
2MASS           &  PLEXUS                &  1.0         &  5.0       & 3\\
MKO             &  ATRAN, Mauna Kea      &  1.0         &  1.6       & 1\\
UKIDSS          &                        &  1.3         &  1.0       & 4\\
NACO            &  Paranal-like          &  1.0         &  2.3       & 5\\
NICI            &  ATRAN, Cerro Pachon   &  1.0         &  2.3       & 1\\
\hline
\multicolumn{5}{l}{Notes:}\\
\multicolumn{5}{l}{[1] \citet{1992NASATM}}\\ 
\multicolumn{5}{l}{[2] C. Manqui elevation is 2380 m, C. Pachon is 2700 m.}\\
\multicolumn{5}{l}{[3] \citet{2003AJ....126.1090C}}\\
\multicolumn{5}{l}{[4] \citet{2006MNRAS.367..454H}}\\
\multicolumn{5}{l}{[5] 0.4-6.0 $\mu$m atmospheric transmission for Paranal from ESO.}\\
\end{tabular}
\end{wraptable}

We obtained a transmission profile and atmospheric transmission profile appropriate for each site.  Table \ref{tab:atmospheres} summarizes the atmosphere assumptions and models used.  We converted to photon-weighted ``relative spectral response'' (RSR) curves, using the following equation \citep{2000PASP..112..961B}
\[
T(\lambda) = \frac{1}{hc}\lambda T_0(\lambda),
\]
where $T_0$ is the raw energy-weighted profile.  We calculate the central wavelength as
\[
\lambda_0 = \frac{\int_0^\infty \lambda T(\lambda)d\lambda }{ \int_0^\infty T(\lambda)d\lambda}.
\]
We calculate the effective width $\Delta \lambda$, such that
\[
F_{\lambda}(\lambda_0)\Delta\lambda = \int_0^\infty F_{\lambda}(\lambda)T(\lambda)d\lambda.
\]
To calculate the magnitude of some object with a spectrum given by $F_{\lambda,\: obj}$ we used
\[
m = -2.5\log\left[ \frac{\int_0^\infty R(\lambda)F_{\lambda,\: obj}d\lambda}{\int_0^\infty R(\lambda)F_{\lambda,\: vega}d\lambda}\right]
\]
using the Vega spectrum of \citet{2007ASPC..364..315B}, which has an uncertainty of 1.5\%.  These calculations are summarized in Table \ref{tab:synphotchar}.  We next describe details particular to the different photometric systems.

\textbf{The $Y$ Band: } The $Y$ band was first defined in \cite{2002PASP..114..708H}.  We follow \citet{2012ApJ...758...57L} and assume that the UKIDSS $Y$ filter defines the MKO system passband, as the largest number of published observations in this passband are from there (cf. \citet{2013MNRAS.433..457B}).  This is a slightly narrow version of the filter.  The UKIDSS $Y$ RSR curve is provided in \citet{2006MNRAS.367..454H}, which is already photo-normalized and includes an atmosphere appropriate for Mauna Kea.

We also consider the unfortunately named $Z$ filter used at Subaru/IRCS and Keck/NIRC2, which is actually in the $Y$ window rather than in the traditionally optical $Z/z$ band.  To add to the confusion the filter has been labeled with a lower-case $z$, but the scanned filter curves and Alan Tokunaga's website\footnote{\url{ http://www.ifa.hawaii.edu/~tokunaga/MKO-NIR_filter_set.html}} indicate that it was meant to be capital $Z$.  In any case, it is a narrow version of the $Y$ passband.  Here we follow \citet{2012ApJ...758...57L} and refer to it as $z_{1.1}$ to emphasize its location in the $Y$ window, and that it is not related to the optical bandpasses of the same name.  We used the same atmospheric assumptions as for the MKO system (see below).

\begin{wrapfigure}[18]{r}{3.5in}
\vspace{-.35in}
\centering
\includegraphics[width=2.5in,angle=90]{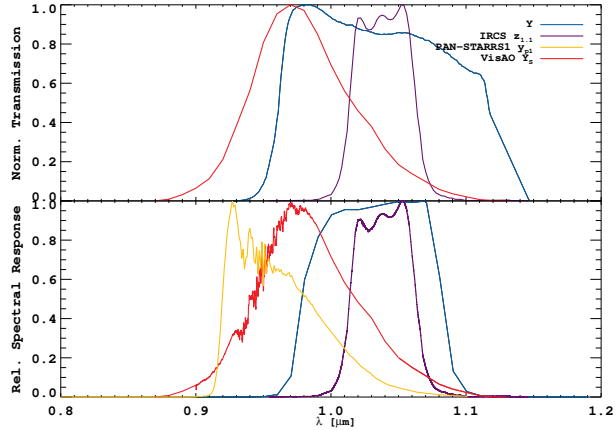}
\vspace{-.2in}
\caption{Comparison of the VisAO $Y_S$ bandpass with other $Y$ band filters.  Top: raw filter profiles.  The $Y$ bandpass is from \citet{2002PASP..114..708H}. The ``$z_{1.1}$'' bandpass, used at Subaru/IRCS and Keck/NIRC2 (where it is called either ``z'' or ``Z'').  Bottom: the filters after photon-weighting and including models for atmospheric transmission, here we show the UKIDSS ``Y'' RSR curve from \citet{2006MNRAS.367..454H}, and the $y_{p1}$ filter of PAN-STARRS \citep{2012ApJ...750...99T}. 
\label{fig:Yfilters}}
\end{wrapfigure}

\textbf{VisAO $\mathbf{Y_S}$:} The VisAO Y-short ($Y_S$) filter is defined by a Melles-Griot long wavepass filter (LPF-950), which passes $\lambda \gtrsim 0.95\microns$, and the quantum efficiency (QE) limit of our near-IR coated EEV CCD47-20 detector.  We convolved the transmission curve with the QE for our EEV CCD47-20 (both provided by the respective manufacturers), and included the effects of 3 Al reflections.  We also include the Clio2 entrance window dichroic which reflects visible light to the WFS and VisAO, cutting on at 1.05$\mu$m.  We next multiply the profile by a model of atmospheric transmission, using the 2.3 mm of precipitable water vapor (PWV),  airmass (AM) 1.0 ATRAN  model for Cerro Pachon provided by Gemini Observatory\footnote{{\url http://www.gemini.edu/?q=node/10789}} \citep{1992NASATM}.  Cerro Pachon, $\sim2700\meters$, is slightly higher than the Magellan site at Cerro Manqui, $\sim2380\meters$ (D. Osip, private communication), so this will slightly underestimate atmospheric absorption.  We finally determine the photon-weighted RSR.   We refer to this filter as ``Y-short'', or $Y_S$.  This is  similar to the $y_{p1}$ bandpass of the PAN-STARRS optical survey \citep{2012ApJ...750...99T}.  $Y_S$ is $\sim 22$nm redder than $y_{p1}$, and we prefer to emphasize that we are working in the $Y$ atmospheric window.  $Y$ band filter profiles are compared in  Figure \ref{fig:Yfilters}.  

The $Y_S$ filter is slightly affected by telluric water vapor.  Using the ATRAN models  we assessed the impact of changes in both AM and PWV.  The mean transmission changes by $\pm 3\%$ over the ranges $1.0 \le \mbox{ AM } \le 1.5$ and $2.3 \le \mbox{ PWV } \le 10.0$ mm.    This change in transmission has little impact on differential photometry so long as PWV does not change significantly between measurements.  AM has almost no effect on $\lambda_0$ but changes in PWV do change it by 2 to 4 nm as expected given the $H_2O$ absorption band at $\sim0.94$ $\microns$.  This is relatively small and since we have no contemporaneous PWV measurements for the observations reported here we ignore this effect.

\textbf{The 2MASS System: } The 2MASS $J$, $H$, and $K_S$ transmission and RSR profiles were collected from the 2MASS website\footnote{\url{http://www.ipac.caltech.edu/2mass/releases/allsky/doc/sec6_4a.html}}.  The RSR profiles are from \citet{2003AJ....126.1090C}.  They used an atmosphere based on the PLEXUS model for AM 1.0.  This model does not use a parameterization corresponding directly to PWV, but according to the website it is equivalent to ~5.0 mm of PWV.

\begin{wraptable}[18]{r}{3.5in}
\vspace{-12pt}
\caption[Synthetic Photometric System]{Synthetic Photometric System Characteristics\label{tab:synphotchar}}
\centering
\scriptsize
\begin{tabular}{lccc}
\hline
\hline
Filter & $\lambda_0$  & $\Delta \lambda$  & 0 mag $F_{\lambda}$ \\
       & ($\mu$m)     &(     $\mu$m)      & ($10^{-6}$ ergs/s/cm$^2$/$\mu$m) \\
\hline

\multicolumn{4}{l}{$Y$ Band}\\
\hline
PAN-STARRS1 $y_{p1}$ & 0.9633 & 0.0615 & 7.17\\
VisAO $Y_S$ & 0.9847 & 0.0855 & 6.75\\
MKO $Y$ & 1.032 & 0.101 & 5.82\\
IRCS $z_{1.1}$ & 1.039 & 0.049 & 5.73\\
\hline
\multicolumn{4}{l}{$J$ Band}\\
\hline
2MASS $J$ & 1.241 & 0.163 & 3.14\\
MKO $J$ & 1.249 & 0.145 & 3.03\\
NACO $J$ & 1.256 & 0.192 & 3.02\\
\hline
\multicolumn{4}{l}{$H$ Band}\\
\hline
NICI $CH_{4S,1\%}$ & 1.584 & 0.0167 & 1.28\\
MKO $H$ & 1.634 & 0.277 & 1.19\\
2MASS $H$ & 1.651 & 0.251 & 1.14\\
NACO $H$ & 1.656 & 0.308 & 1.14\\
NICI $H$ & 1.658 & 0.27 & 1.15\\
\hline
\multicolumn{4}{l}{$K$ Band}\\
\hline
MKO $K_S$ & 2.156 & 0.272 & 0.438\\
NACO $K_S$ & 2.16 & 0.323 & 0.438\\
2MASS $K_S$ & 2.166 & 0.262 & 0.431\\
NICI $K_S$ & 2.176 & 0.268 & 0.424\\
MKO $K$ & 2.206 & 0.293 & 0.403\\
NICI $K_{cont}$ & 2.272 & 0.0375 & 0.356\\

\hline
\end{tabular}
\end{wraptable}

\textbf{The MKO System: }We used the Mauna Kea filter profiles provided by the IRTF/NSFCam website\footnote{\url{http://irtfweb.ifa.hawaii.edu/~nsfcam/filters.html}} for the MKO $J$, $H$, $K_S$, and $K$ passbands which correspond to the 1998 production run of these filters.  We again used the ATRAN model atmosphere from Gemini Observatory, now for Mauna Kea with 1.6 mm precipitable water vapor (PWV) at AM 1.0.

\textbf{The NACO System:}  We obtained transmission profiles for NACO from the instrument website\footnote{\url{http://www.eso.org/sci/facilities/paranal/instruments/naco/inst/filters.html}}.  We used the ``Paranal-like'' atmosphere provided by ESO\footnote{\url{http://www.eso.org/sci/facilities/eelt/science/drm/tech_data/data/atm_abs/}}, which is for AM 1.0 and 2.3 mm of PWV.  The NACO filters are close to the 2MASS system, but there are subtle differences, which are somewhat more pronounced once the atmosphere appropriate for each site is included.

\textbf{The NICI System: } Profiles for the NICI filters were obtained from the instrument websites for NICI\footnote{\url{http://www.gemini.edu/sciops/instruments/nici/}} and NIRI\footnote{\url{http://www.gemini.edu/sciops/instruments/niri/}}.  The Cerro Pachon ATRAN atmosphere was used, with AM 1.0 and 2.3 mm of PWV.  The NICI $J$, $H$, and $K_S$ bandpasses are intended to be in the MKO system and so should be insensitive to atmospheric conditions, but there are subtle differences between the filter profiles.

\subsection{Photometric Conversions}

To quantify the differences between these systems and to accurately compare results for objects measured in the different systems, we used the library of brown dwarf spectra we compiled from various sources (described in Section \ref{sec:library}).  We calculated the magnitudes in each of the various filters and then fit a 4th or 5th order polynomial to the results.  Our notation is
\begin{equation}
m_1-m_2 = c_0 + c_1 \mbox{SpT}+ c_2 \mbox{SpT}^2+ c_3 \mbox{SpT}^3+ c_4 \mbox{SpT}^4 + c_5 \mbox{SpT}^5
\label{eqn:photconvcoeff}
\end{equation}
where SpT is the near-IR spectral type given by
\begin{eqnarray*}
\mbox{SpT} &=& 0 ... 9, \mbox{ for } \mbox{M0} ... \mbox{M9}\\
\mbox{SpT} &=& 10 ... 19, \mbox{ for } \mbox{L0} ... \mbox{L9}\\
\mbox{SpT} &=& 20 ... 29, \mbox{ for } \mbox{T0} ... \mbox{T9}
\end{eqnarray*}
We provide the coefficients determined in this manner for a variety of transformations in Table \ref{tab:photconvcoeff}.  

\begin{figure}

\centering
\includegraphics[height=2.5in,angle=90]{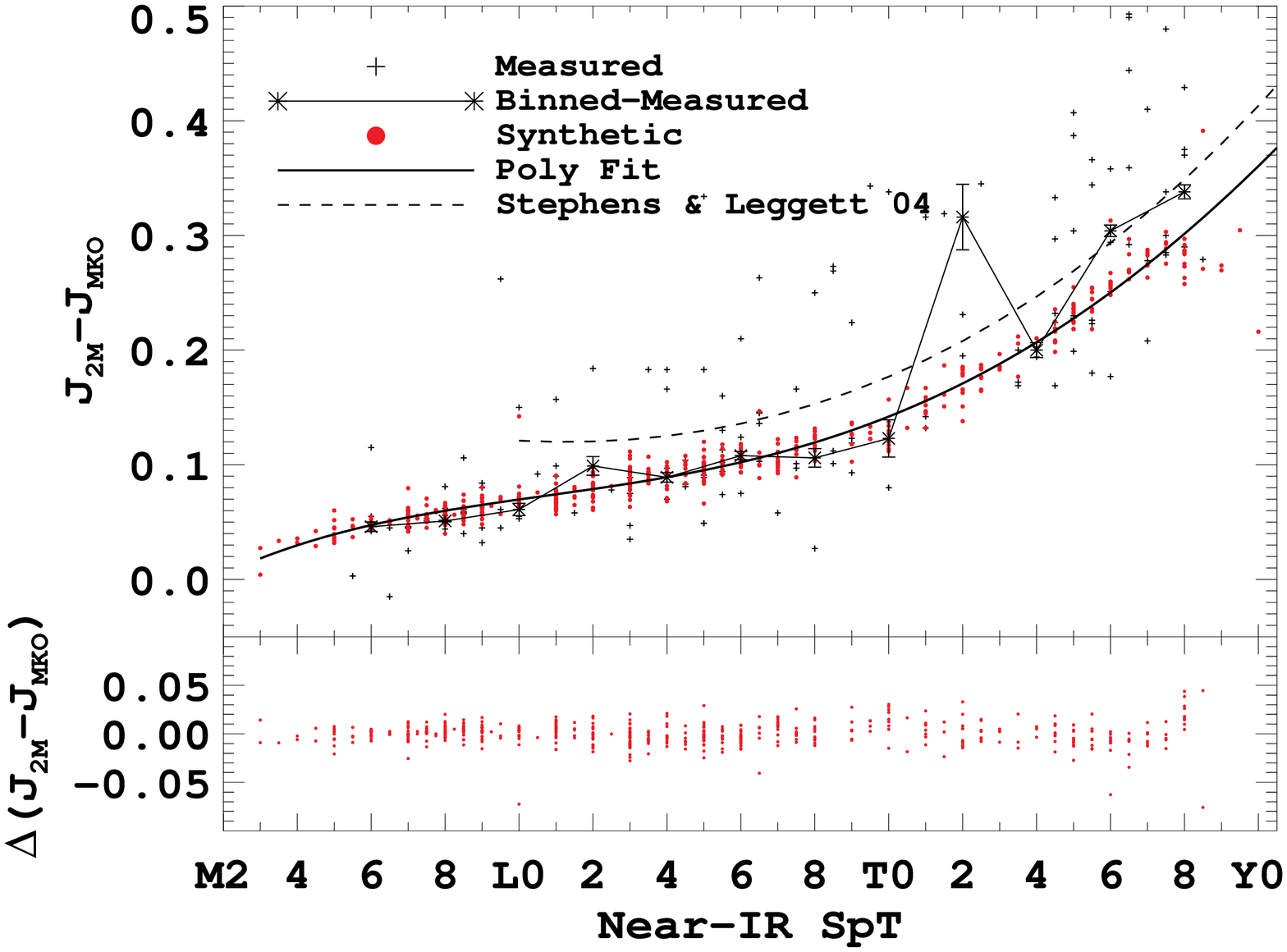}
\includegraphics[height=2.5in,angle=90]{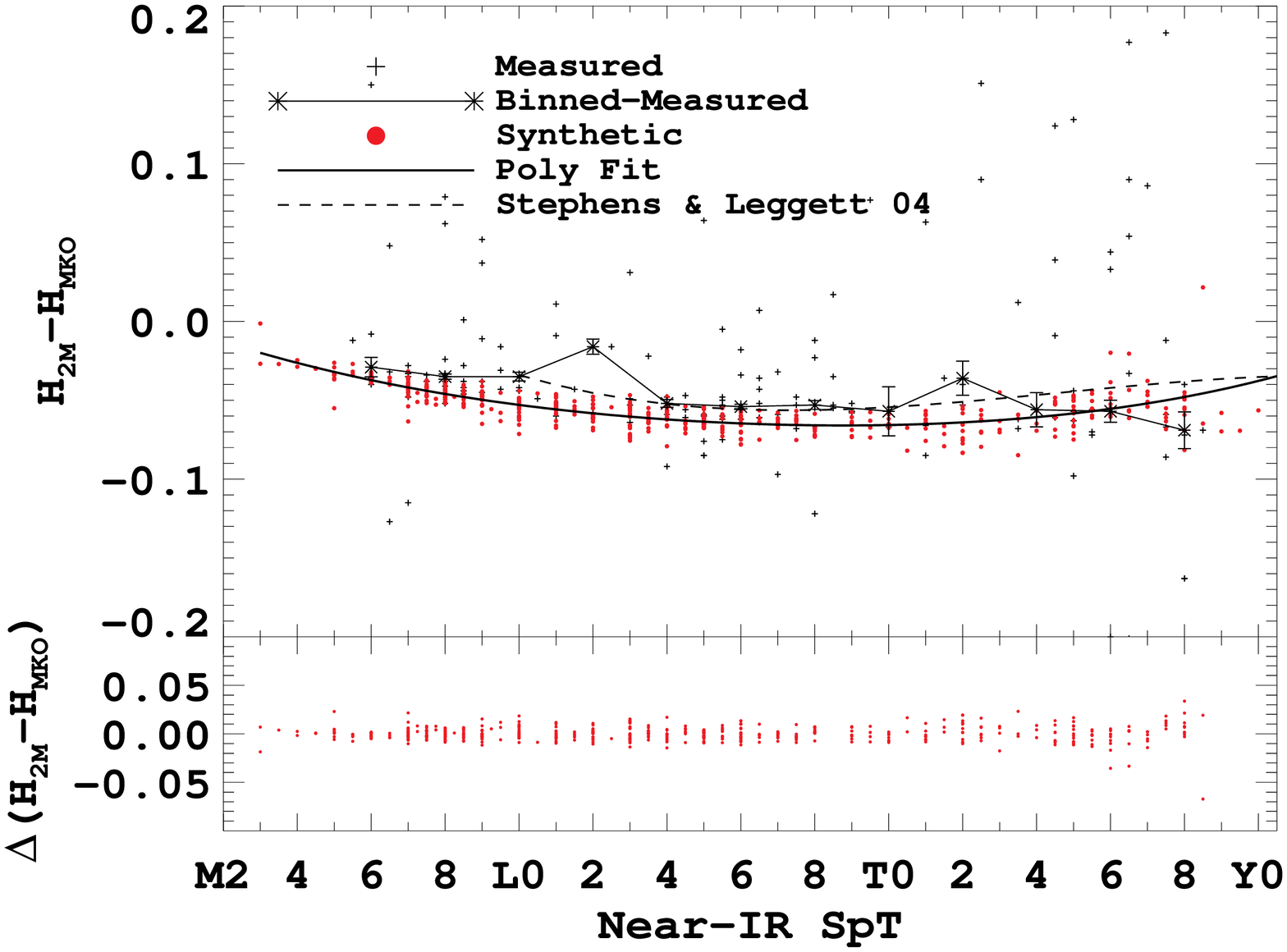}
\includegraphics[height=2.5in,angle=90]{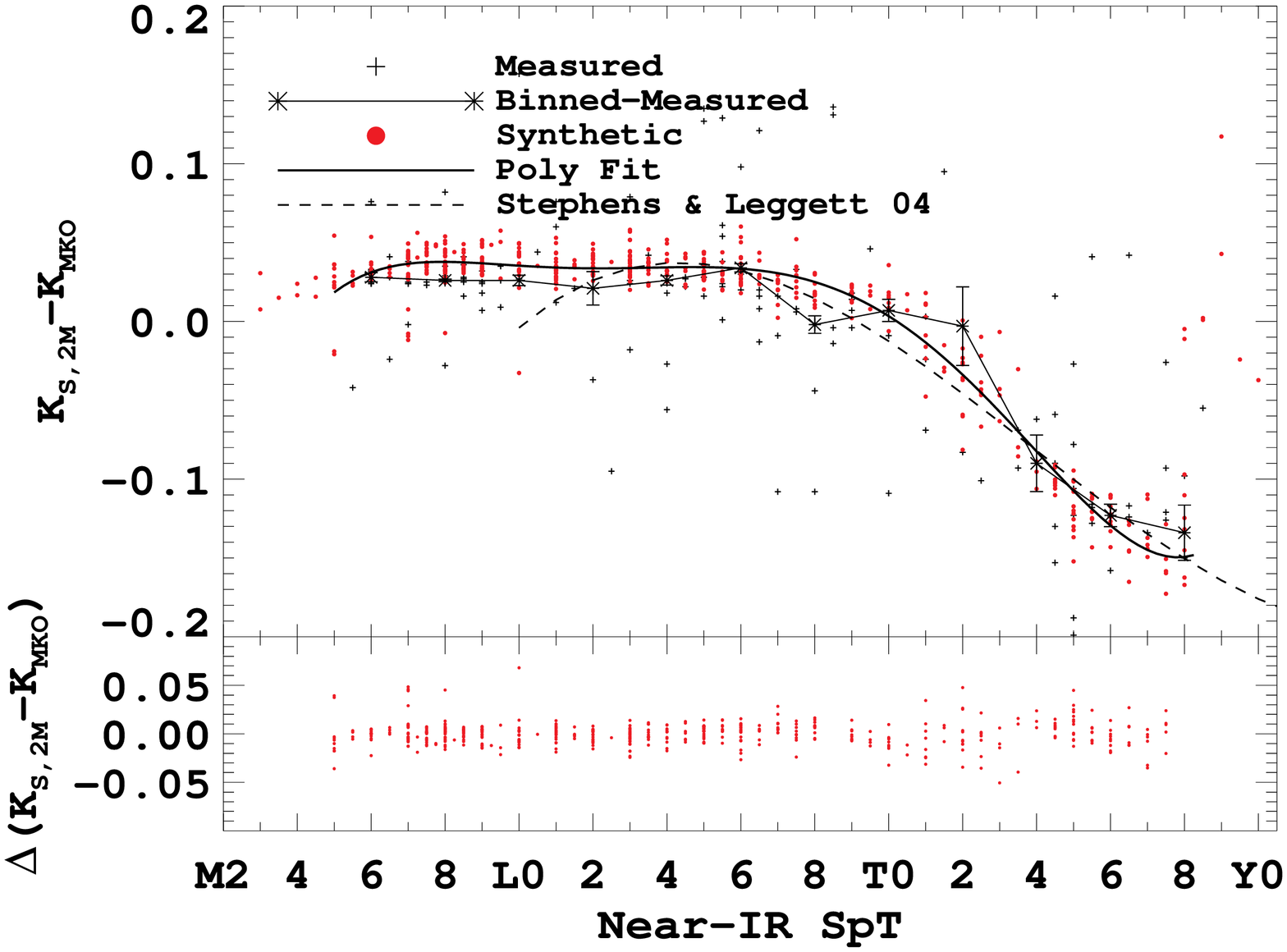}
\caption[2MASS to MKO Conversions]{Synthetic photometry (red points) in the 2MASS and MKO systems and  measurements made in the 2 systems (crosses, from \citet{2012ApJS..201...19D}).  We also plot the binned median of the measurements.  Our polynomial fit is shown as the solid black line.  For comparison we show the fit determined by \citet{2004PASP..116....9S}, who also used synthetic photometry.  Our fit to the synthetic photometry appears to be a better match to the actual measurements in J.  In the H and K bands both fits appear to be acceptable, with \citet{2004PASP..116....9S} being somewhat better for M and L dwarfs in H.   \label{fig:2MtoMKO}}
\end{figure}

As a check on our calculations consider the conversions from 2MASS to MKO.  There are many objects with measurements in both systems, which allows us to directly compare our synthetic photometry to actual measurements.  We here use the compilation of \citet{2012ApJS..201...19D}.  This also allows a comparison to the previous work of \citet{2004PASP..116....9S}, who employed similar methodology to ours but with fewer objects.  The results are shown graphically in Figure \ref{fig:2MtoMKO}.  In all three bands our synthetic photometry and fit appear to be a good match to the measurements.  In $J$ our results appear to be an improvement over \citet{2004PASP..116....9S}, and in $H$ and $K$ either fit appears to be reasonable.  These results give confidence that our synthetic photometry reproduces the variations in these systems reasonably well.

\begin{sidewaystable}
\small
\caption{Photometric conversion coefficients. \label{tab:photconvcoeff}}
\begin{center}
\begin{tabular}{lccccccc}
\hline
\hline
Filters                 &  $c_0$    &  $c_1$    &  $c_2$     &  $c_3$       &  $c_4$       &  $c_5$       & $\sigma$ (mag)   \\
\hline
\hline

$Y_{S}-Y_{MKO}        $ & 0.00904   & 0.0247    &  0.00411   &$-0.000633$  & $2.89\times10^{-5}$       & $-3.96\times10^-7$ & 0.038 \\
$z_{1.1} - Y_{MKO}$     & 0.00424   & $-0.00911$ &0.000823  & $-2.85\times10^{-5}$ & $1.23\times10^{-7}$ & 0.0 & 0.011\\
$J_{2M}-J_{MKO}$       & $-0.0308$   & 0.0208    & $-0.00165$ & $6.4\times10^{-5}$ & $-5.83\times10^{-7}$ & 0.0  &  0.012\\
$J_{NACO}-J_{MKO}$    & $-0.0895$    &  0.0343   & $-0.00331$  & 0.000145 & $-1.87\times10^{-6}$ & 0.0 &  0.015 \\
$H_{2M}-H_{MKO} $     & 0.00358      & $-0.00903$ & 0.000435 & $-1.14\times10^{-5}$ & $1.82\times10^{-7}$  & 0.0 & 0.008\\
$H_{NACO}-H_{MKO} $     & $-0.0521$   & 0.0152   & $-0.00182$ & $8.61\times10^{-5}$ & $-1.25\times10^{-6}$ & 0.0 & 0.011\\
$H_{NICI}-H_{MKO}     $ & $-0.0138$ & $-0.000532$ & $-0.000303$ & $1.92\times10^{-5} $ & $-2.69\times10^{-7}$ & 0.0 &    0.008\\
$H_{NACO}-H_{NICI}$    &  $-0.0383$ & 0.0157  & $-0.00151$  & $6.69\times10^{-5} $ & $-9.76\times10^{-7}$ & 0.0 &   0.007\\
$H_{NICI}-CH_{4S,1\%}$ & $-1.28  $  & 0.695   & $-0.152 $  & 0.0164  & $-9.31\times10^{-4}$ & 0.0 &   0.037\\
$K_{S,2M}-K_{MKO}$    & $-0.324$   & 0.149 & $-0.0233$  & 0.00171 & $-5.91\times10^{-5} $ & $7.54\times10^{-7}$ &    0.014\\
$K_{S,2M}-K_{S,MKO}$   & 0.0272     & $-0.0155$ & 0.00227  & $-1.71\times10^{-4}$  & $6.12\times10^{-6}$ & $-7.8\times10^{-8}$ & 0.006\\
$K_{S,NACO}-K_{S,MKO}$ &   0.0428   & $-0.0209$ & 0.00422  & $-3.39\times10^{-4}$ & $1.22\times10^{-5} $ & $-1.59\times10^{-7}$ & 0.007\\
$K_{S,NICI}-K_{MKO}$ & $-0.154$ & 0.0828  & $-0.0142$ & 0.00112 & $-4.06\times10^{-5}$ & $5.35\times10^{-7}$ & 0.011\\
$K_{S,NACO}-K_{S,NICI}$ & 0.00554  & 0.000497 & 0.00128 & $-1.24\times10^{-4}$ & $4.59\times10^{-6}$& $-6\times10^{-8}$ &0.007\\
$K_{S,NICI}-K_{cont}$  & 1.07 & $-0.524$ & 0.108 & $-0.00999$ & 0.000428 & $-6.96\times10^{-6}$  &  0.058\\
\hline

\end{tabular}
\end{center}
\end{sidewaystable}

\subsection{Comparison Objects}
\label{sec:library}
\textbf{$\mathbf{Y_S-K_S}$ Spectral Library:} We analyzed the $Y_SJHK_S$ photometry of $\beta$ Pic b by comparison with a library consisting of 499 brown dwarf spectra. Of these: 441 are from the SpeX Prism Spectral libraries maintained by Adam Burgasser\footnote{\url{http://pono.ucsd.edu/~adam/browndwarfs/spexprism/}} (from various sources), 23 are WISE brown dwarfs from \citet{2011ApJS..197...19K}, and 35 are young field BDs from \citet{2013ApJ...772...79A}.  We correlated 115 of these with parallax measurements, either listed in the SpeX Library (from various sources) or from \citet{2012ApJS..201...19D}. We conducted synthetic photometry on these spectra as described above.  For the objects with parallaxes we normalized the spectra to available photometry.  In most cases there is a near-IR SpT assigned which we use here.  In the few cases where there is no near-IR SpT we use the optical SpT.

\begin{wrapfigure}[19]{r}{3.4in}
\vspace{-0.5in}
\centering
\includegraphics[width=2.5in,clip=true,angle=90]{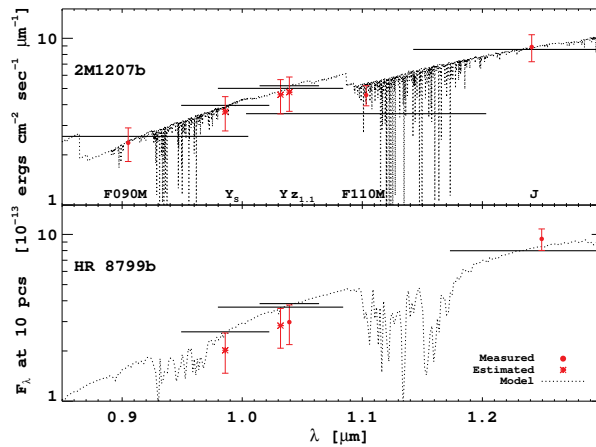}
\vspace{-.2in}
\caption[2M1207b and HR 8799b]{Optical and near-IR photometry and models of 2M 1207 b and HR 8799 b.    To estimate photometry in the Y atmospheric window, we used the models to calculate colors and then extrapolated or interpolated from the measured photometry.  Measured photometry is indicated by filled circles, and our estimated photometry is indicated by asterisks.  The 2M 1207 b model is from \citet{2011ApJ...735L..39B}, and the HR 8799 b model is from \citet{2011ApJ...737...34M}.  See also Table \ref{tab:yphotb}. \label{fig:2mhr}}
\end{wrapfigure}

We also compiled very late-T and early-Y dwarf photometry from \citet{2012ApJ...758...57L} and \citet{2013ApJ...763..130L}.  Here we lack spectra, so instead we use the above transformations from the MKO system determined using our compiled spectra.  

\textbf{Y Band Photometry of EGPs:} The other young planetary mass companions with $Y$ band photometry are HR 8799 b and 2M 1207 b.  \citet{2011ApJ...729..128C} detected HR 8799 b at $1.04 \microns$ in the $z_{1.1}$ filter with Subaru/IRCS.    \citet{2013ApJ...768...24O} also have the ability to work in the $Y$ band with Project 1640, and reported low S/N detections of HR 8799 b and HR8799 c at $1.05 \microns$.  2M 1207 b was observed with the Hubble Space Telescope by \citet{2006ApJ...652..724S}.  To compare our results  we must convert these measurements into our $Y_S$ bandpass.  The SEDs of these two objects do not correspond to those of field brown dwarfs of comparable temperature, so we turn to published best-fit model spectra:  for 2M 1207 b we use the model of \citet{2011ApJ...735L..39B} and for HR 8799 b we use the best-fit model from \citet{2011ApJ...737...34M}.  We use the models to calculate color between bandpasses.  We illustrate this in Figure \ref{fig:2mhr} and summarize the results in Table \ref{tab:yphotb}.

\begin{table}
\footnotesize
\caption{$Y$-band photometry of 2M 1207 b and HR 8799 b \label{tab:yphotb}}
\begin{center}
\begin{tabular}{lcccc}
\hline
\hline
Filter   &  $\lambda_0$ & \multicolumn{2}{c}{Abs. Magnitude}  &  References \\
         &  $(\microns)$  & Measured  &  Estimated  &     \\
\hline
\multicolumn{5}{l}{2M1207b} \\
\hline
F090M         &  0.905  &  18.86$\pm$0.25 &   ---            & [1] \\
$Y_S$         &  0.986  &    ---          &  18.2$\pm$0.26   & [2]\\
$Y$             &  1.032  &    ---          &  17.79$\pm$0.26  & [2] \\
$z_{1.1}$     &  1.039  &    ---          &  17.73$\pm$0.26  & [2] \\
F110M         &  1.102  &  17.01$\pm$0.16 &   ---            & [1] \\
$J_{2MASS}$   &  1.241  &  16.40$\pm$0.21 &   ---            & [3]\\
\hline
\multicolumn{5}{l}{HR 8799b} \\
\hline
$Y_S$         &  0.986  &    ---          &  18.84$\pm$0.29   & [2]\\
$Y$             &  1.032  &   ---           &  18.31$\pm$0.29   & [2]\\
$z_{1.1}$     &  1.039  &  18.24$\pm$0.29 &   ---             & [4]\\
$J_{MKO}$     &  1.249  &  16.30$\pm$0.16 &   ---             & [5] \\

\hline
\multicolumn{5}{l}{Notes: [1] \citet{2006ApJ...652..724S}, [2] this work,}\\ 
\multicolumn{5}{l}{[3] \citet{2007ApJ...657.1064M}, [4] \citet{2011ApJ...729..128C},}\\
\multicolumn{5}{l}{[5] \citet{2008Sci...322.1348M}}\\
\end{tabular}
\end{center}
\end{table}

\textbf{Other Comparison Objects: } We also collected a number of low-temperature, low-mass objects from the literature to compare to $\bPicb$, which are listed in Table \ref{tab:compobjs}.  Most of these are companions.  Of particular interest are the low-surface gravity objects.  This includes the four HR 8799 planets and the planetary mass BD companion 2M1207b.  Two other companions with photometric properties similar to these faint red companions are AB Pic b \citep{2005A&A...438L..29C} and 2MASS 0122-2439B \citep{2013ApJ...774...55B}.  

Both \citet{2013arXiv1302.1160B} and \citet{2013ApJ...776...15C} noted that the near-IR SED of $\bPicb$ resembled that of an early to mid-L dwarf. For comparison then, we use companions such as $\kappa$ And B  \citep[L$1\pm1$,][]{2013arXiv1308.3859B, 2013arXiv1309.3372H}) and CD-35 2722 B \citep[L4$\pm$1,][]{2011ApJ...729..139W}). We used the Project 1640 spectrum of $\kappa$ And B from \citet{2013arXiv1309.3372H} to estimate its Y band photometry.  We also highlight several field dwarfs. The L0$\gamma$ object 2MASS 0608-2753 is believed to be a member of the $\beta$ Pic moving group \citep{2010ApJ...715L.165R}, giving us a potentially co-eval object to compare with $\bPicb$.  We include two field objects which appear to have low surface gravity, dusty photospheres, and are young moving group members: 2MASS 0355+1133 \citep{2013AJ....145....2F,2013arXiv1310.0457L} and PSO318.5 \citep{2013arXiv1310.0457L}. PSO318.5 is also a possible member of the $\beta$ Pic moving group.  2MASS 0032-4405, an L0$\gamma$, was discussed as a possible proxy for both $\bPicb$ and $\kappa$ And B by \citet{2013arXiv1302.1160B}.

\begin{sidewaystable}
\footnotesize
\caption{Comparison Objects \label{tab:compobjs}}
\begin{center}
\begin{tabular}{lcccccccl}
\hline
\hline
Name &  Near-IR       &  Distance  &  Group   &  Age      &   log(L$_{bol}$/L$\sun$)   & T$_{eff}$   &   Mass      &   Ref. \\
     &  Sp. Type      &    (pc)    &          &   (MYr)   &                            &   (K)        & (M$_{Jup}$) &    \\
\hline
\multicolumn{9}{l}{Companions}\\
\hline
  HR 8799 b & late-L/early-T & 39.4$\pm$1 & Col & 30-60 & $-5.1\pm0.1$ & 850 & 7 &1,2,3,4\\
 HR 8799 c & --- & 39.4$\pm$1 & Col & 30-60 & $-4.7\pm0.1$ & $1100\pm100$ & 10 &1,2,5\\
 HR 8799 d & --- & 39.4$\pm$1 & Col & 30-60 & $-4.7\pm0.1$ & 900 & 10 &1,2\\
 HR 8799 e & --- & 39.4$\pm$1 & Col & 30-60 & $-4.7\pm0.2$ & 1000 & 7-10 &6,7\\
 2MASS 1207-3932 b & M8.5-L4, L3 & 52.4$\pm$1.1 & TWA & 8 & $-4.73\pm0.12$ & 1000 & 5 &8,9,10,11,12\\
 1RXS 1609 b & L4$^{+1}_{-2}$ & 145$\pm$20 & USco & 5-11 & $-3.55\pm0.2$ & 1800$\pm$200 & 6-15 &13,14\\
 HD 106906 b & L2.5$\pm1$ & $92\pm6$ & LCC & $13\pm2$ & $-3.64\pm0.08$* & $1800\pm100$ & $11\pm2$ &15\\
 GJ 504 b & late-T/Y & $17.56\pm0.08$ & --- & $105^{+30}_{-20}$ & $-6.09\pm0.08$ & & $4.0^{+4.5}_{-1.0}$ &16\\
 AB Pic B & L0 VL-G & $46.1\pm1.4$ & Tuc-Hor & 30 & $-3.7\pm0.2$ & $2000^{+100}_{-300}$ & 13-14 &8,17,18,19,20\\
 $\kappa$ And B & L$1\pm1$ & 51.6$\pm$0.5 & --- & $220\pm100$ & $-3.76\pm0.06$ & $2040\pm60$ & $50^{+16}_{-13}$ &21,22,23\\
 CD-35 2722 B & L3$\pm$1 INT-G & 22.7$\pm$1.0 & AB Dor & 125$\pm$25 & $-3.51\pm0.12$* & 1700-1900 & 31$\pm$8 &8,24,25\\
 G 196-3 B & L3 VL-G & 14.9$\pm$2.7$^\dagger$ & --- & $\approx$100 & $-4.15\pm0.16$ & 1800$\pm$200 & $25^{+15}_{-10}$ &8,26,27,28\\
 2MASS 0122-2439 B & L4-L6 & 36$\pm$4$^\dagger$ & AB Dor (?) & 120$\pm$10 & $-4.19\pm0.1$ & 1300-1500 & 12-14 $/$ 23-27 &29\\
 LP 261-75 B & L4.5$\pm$ 1.0 & $32.95^{+2.80}_{-2.40}$ & --- & 100-200 & $-4.43\pm0.09$ & $\sim1400$ & 13 $/$ 22 &29,30\\
 Gl 417 BC & L$4.5\pm1$ $+$ L$6\pm1$ & 21.93$\pm$0.21 & --- & 80-300 & $-3.80\pm0.04$* & 1600-1800 & 35$\pm$15 &8,31,32,17,33\\
 HN Peg B & T2.5 $\pm$ 0.5 & 17.99$\pm$0.14 & --- & $237\pm33$ & $-4.77\pm0.03$ & 1115 & 28 &17,34,35,33,36\\
 Ross 458 C & T8.5p$\pm$0.5 & $11.69\pm0.21$ & --- & $\le 1000$ & $-5.61\pm0.20$ & $695\pm60$ & 5-20 &17,37,38\\
  \hline
 \multicolumn{9}{l}{Field Objects}\\
 \hline
 2MASS 0608-2753 & M8.5$\gamma$, L0 VL-G & 30$\pm$10 & $\beta$ Pic & $21\pm4$ & $-3.43\pm0.29$* & 2529 & 15-35 &39,40,8\\
 2MASS 0032-4405 & L0$\gamma$, VL-G & 26$\pm$3.3 & Field & --- & $-3.93\pm0.14$* & --- & --- &8,41\\
 2MASS 0355+1133 & L5$\gamma$, VL-G & $9.1\pm0.1$ & AB Dor & $125\pm25$ & $-4.23\pm0.11$ & $1420^{+80}_{-130}$ & $24^{+3}_{-6}$ &
42,43,44,45\\
 PSO318.5-22 & L7$\pm1$ VL-G & $24.6\pm1.4$ & $\beta$ Pic & $21\pm4$ & $-4.42\pm0.06$ & $1160^{+30}_{-40}$ & $6.5^{+1.3}_{-1.0}$ &43
\\
 CFBDSIR2149-0403 & T7 & 40$\pm$3$^\dagger$ & AB Dor & 125$\pm$25 & $-5.57\pm0.24$* & 650-750 & 4-7 &46\\

\hline
\multicolumn{9}{l}{$^\dagger$photometric distance. Trigonometric parallax otherwise.}\\
\multicolumn{9}{l}{*quantity estimated from field relations, either in this work or in references}\\
\multicolumn{9}{l}{\parbox[t]{22cm}{ References:
[1] \citet{2008Sci...322.1348M}, [2] \citet{2011ApJ...737...34M}, [3] \citet{2010ApJ...723..850B}, [4] \citet{2011ApJ...733...65B}, [5] \citet{2013Sci...339.1398K}, [6] \citet{2010Natur.468.1080M}, [7] \citet{2012ApJ...753...14S}, [8] \citet{2013ApJ...772...79A}, [9] \citet{2005A&A...438L..25C}, [10] \citet{2008A&A...477L...1D}, [11] \citet{2011ApJ...735L..39B}, [12] \citet{2010A&A...517A..76P}, [13] \citet{2010ApJ...719..497L}, [14] \citet{2012ApJ...746..154P}, [15] \citet{bailey2013}, [16] \citet{2013ApJ...774...11K}, [17] \citet{2007A&A...474..653V}, [18] \citet{2005A&A...438L..29C}, [19] \citet{2010A&A...512A..52B}, [20] \citet{2003ApJ...599..342S}, [21] \citet{2013arXiv1308.3859B}, [22] \citet{2013arXiv1309.3372H}, [23] \citet{2013ApJ...763L..32C}, [24] \citet{2011ApJ...729..139W}, [25] \citet{2012ApJ...758...56S}, [26] \citet{1998Sci...282.1309R}, [27] \citet{2009ApJ...699..649S}, [28] \citet{2010ApJ...715.1408Z}, [29] \citet{2013ApJ...774...55B}, [30] \citet{2006PASP..118..671R}, [31] \citet{2001AJ....121.3235K}, [32] \citet{2010ApJ...710.1142B}, [33] \citet{2012ApJS..201...19D}, [34] \citet{2007ApJ...654..570L}, [35] \citet{2008ApJ...682.1256L}, [36] \citet{2007ApJ...669.1167B}, [37] \citet{2011MNRAS.414.3590B}, [38] \citet{2010MNRAS.405.1140G}, [39] \citet{2010ApJ...715L.165R}, [40] \citet{2010ApJS..186...63R}, [41] \citet{2012ApJ...752...56F}, [42] \citet{2013AJ....145....2F}, [43] \citet{2013arXiv1310.0457L}, [44] \citet{2013AN....334...85L}, [45] \citet{2013ApJ...766....6B}, [46] \citet{2012A&A...548A..26D}

}}\\
\end{tabular}
\end{center}
\end{sidewaystable}

\afterpage{\clearpage}

\clearpage
\section{VisAO Occulting Mask Transmission} 
\label{app:coron}

\begin{figure}
\centering
\includegraphics[width=2.in,angle=90]{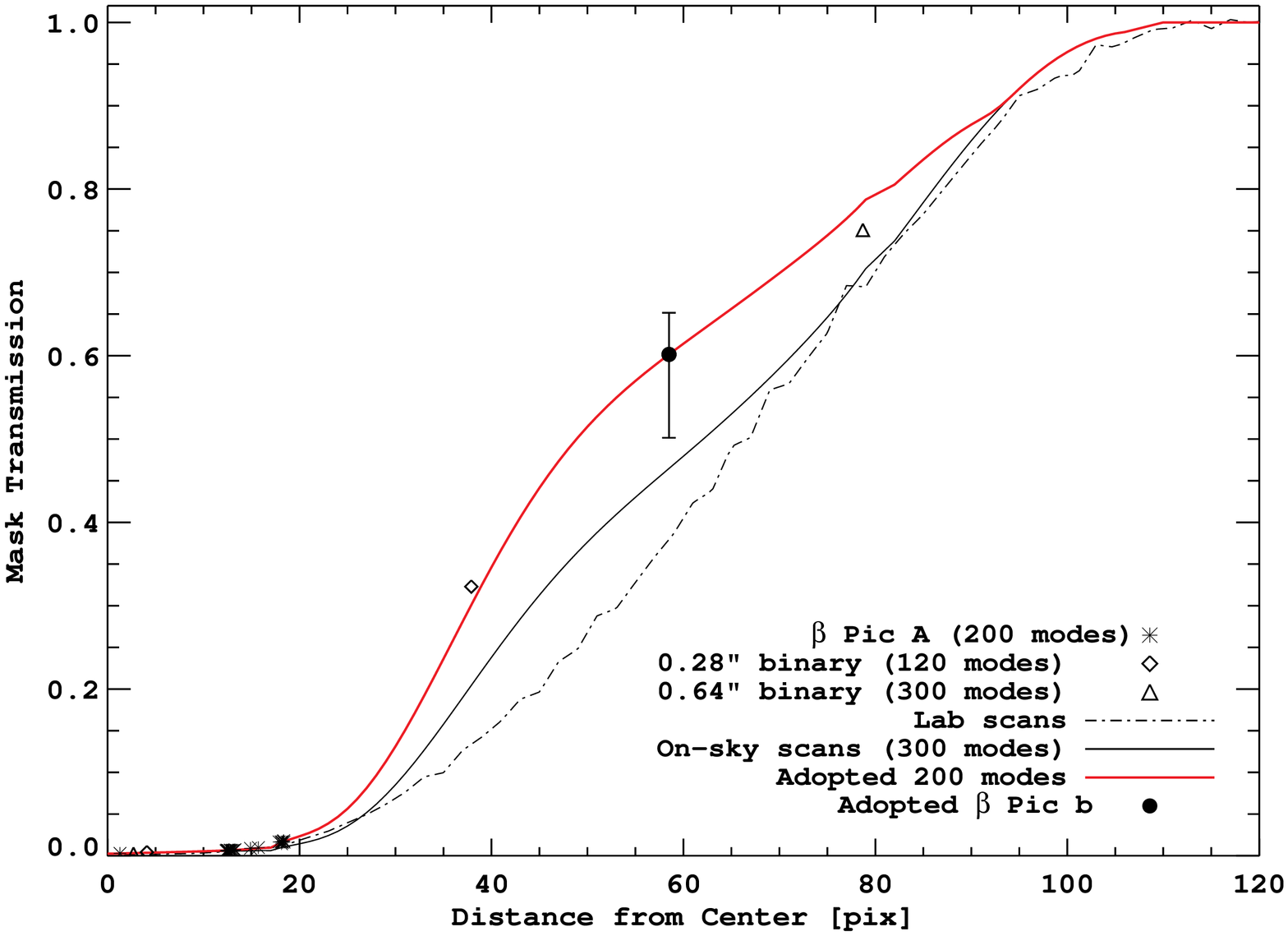}
\includegraphics[width=2.in,angle=90]{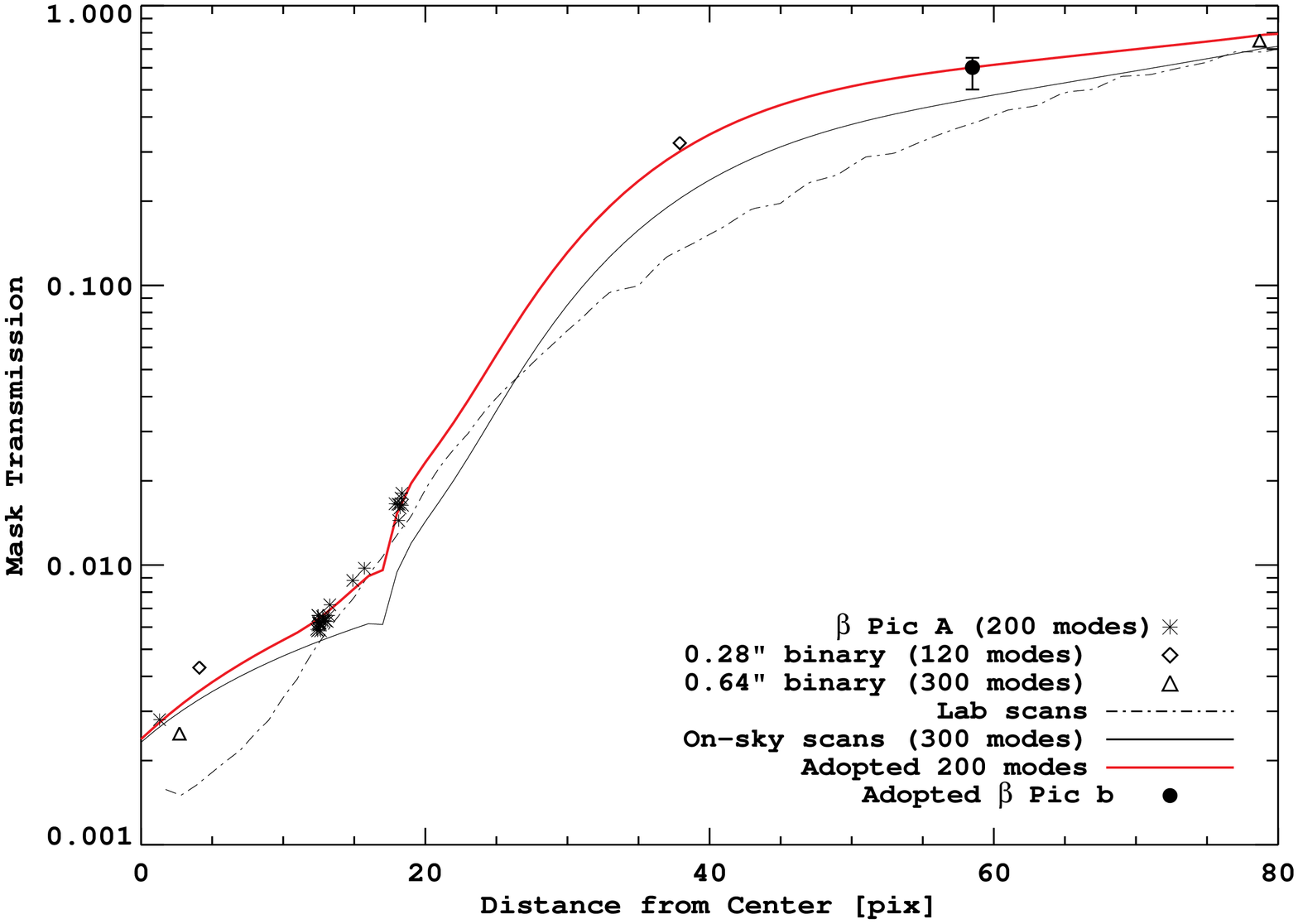}
\vspace{-.15in}
\caption{Transmission of the VisAO occulting mask. The dot-dashed line is the median smoothed profile measured in the laboratory (Strehl $>90$\%).  The solid black line is a piecewise polynomial fit to the on-sky scans with 300 modes of wavefront correction.  The individual points correspond to measurements of $\beta$ Pic A and binaries.  The red curve is our boot-strapped profile for 200 modes of wavefront correction. The separation of $\bPicb$ is $\sim59$ pixels $\approx0.47''$, which gives a mask transmission of $0.60^{+0.05}_{-0.10}$.  
\label{fig:coron_profile}}
\end{figure}

We first measured the mask transmission and PSF by scanning an artificial test source across the mask with the instrument off the telescope.  Our internal alignment source has a slightly faster f/\#, resulting in smaller FWHM (it was originally designed for the LBTAO systems), and delivers a Strehl $>90$\%.  The source was scanned along 12 different lines, spaced roughly 30 degrees apart, across the mask in the $Y_S$ filter.  We found the center of the mask by determining the x,y position which best symmetrizes all the scans simultaneously.  We measured attenuation of the mask using a 3 pixel radius photometric aperture.  The results of our laboratory scans are shown in Figure \ref{fig:coron_profile}.  The maximum attenuation is 0.0015, or ND = 2.8.  

During the first-light commissioning run (Comm1, Nov.--Dec.\ 2012), we observed a $0.28''$, $\Delta Y_S \sim 3.5$ binary both on and off the mask, providing a single transmission measurement.  The center of the mask was found using AO-off (seeing-limited) acquisition images, which show a well defined circular shadow.  This is also plotted in Figure \ref{fig:coron_profile}, and is over 200\% higher than expected based on the lab scans.  The key point is that this measurement was done with a 200 mode reconstructor, with gains applied to only the first 120 modes.

We can also use acquisition images of $\beta$ Pic A, recorded during coronagraph alignment.  These sample much closer in than the position of the planet, but provide guidance on the changes in the transmission profile due to wavefront correction.  We also include the median transmission of $\beta$ Pic A centered on the mask, which was measured using the science exposures.  During the $\beta$ Pic observations the same reconstructor was used as for the $0.28''$ binary, but gains were applied to all 200 modes.  We thus expect slightly better correction.

During our 2nd commissioning run (Comm2, Mar.--Apr.\ 2013), we scanned a star across the mask with the AO loop closed in the $Y_S$ filter.  The caveat to these measurements is that we used a 300 mode reconstructor, significantly improving wavefront correction over the 200 mode matrix in use on Comm1.  We took two scans, separated by 90 degrees, and found the best-fit center of symmetry.  We corrected for Strehl ratio variations using the un-occulted beamsplitter ghost, measured with the same 3 pixel radius aperture.  The result, also shown in Figure \ref{fig:coron_profile}, is intermediate between the lab and the $0.28''$ binary.  We also observed a $0.64''$ binary on and off the mask during Comm2, with the 300 mode reconstructor.  The mask transmission measured on this binary was higher than for both the lab and on-sky scans.  

\begin{wrapfigure}[16]{r}{2.5in}
\vspace{-.5in}
\includegraphics[width=2.in,clip=true,angle=90]{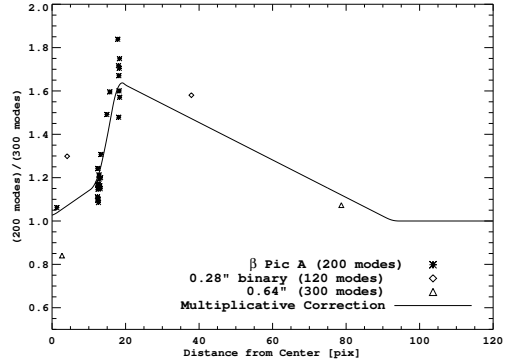}
\vspace{-.15in}
\caption{Ratio of mask transmission at lower wavefront correction quality to the transmission measured on-sky with a 300 modes.  The solid line is a piecewise linear function which we use to form a  boot-strap estimate of transmission under 200 modes of wavefront correction.    This results in the red curve shown in Figure \ref{fig:coron_profile}.
\label{fig:coron_profile_ratio}}
\end{wrapfigure}

In Figure \ref{fig:coron_profile_ratio} we show the ratio of coronagraph transmission for $\beta$ Pic A (200 modes), the $0.28''$ binary (120 modes), and the $0.64''$ binary (300 modes) to the transmission profile measured on-sky (300 modes).  To form an estimate of the complete transmission profile under 200 modes of wavefront control, we fit this ratio with a piecewise linear function, which we then multiply by the on-sky profile.  Our fit is obviously not the only functional form which could be used to describe the transmission change from 300 to 200 modes of correction, but it is the simplest estimate we can make given the data.  We note that even significant changes in the ratio close to the center result in relatively small changes in the transmission at 59 pixels, the separation of $\bPicb$, due to the contraints at wider separations.  The resultant boot-strapped profile is also shown in  \ref{fig:coron_profile}.  Using it, our adopted transmission at the location of $\bPicb$ is $0.60^{+0.05}_{-0.10}$.

\begin{figure}
\centering
\includegraphics[width=2.in,clip=true,angle=90]{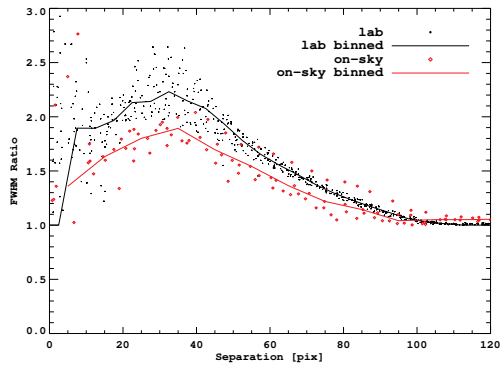}
\vspace{-.15in}
\caption{PSF shape measurements. The broad apodized transmission profile results in a PSF which varies with distance from the coronagraph center.  We plot the ratio of FWHMs of the elliptical Gaussian which best fits the PSF at each location.  On-sky, the ratio does not reach 1.0, as there is usually an elongation in the wind direction.  Correction quality also appears to have an effect on shape, as the on-sky measurements have a lower peak FWHM ratio. See also Figure \ref{fig:zooms}.
\label{fig:coron_fwhm}}
\end{figure}

The final consideration when working under the mask is that it changes the PSF, causing an elongation in the radial direction.  We measured this both in the lab and on-sky by fitting an elliptical Gaussian at each point in the scans.  The result is shown in Figure \ref{fig:coron_fwhm}, along with an on-sky image of a star at roughly the separation of $\bPicb$ shown earlier in Figure \ref{fig:zooms}.  This change in shape was taken into account when conducting photometry on the planet.

\end{document}